# Phase meter based on zero-crossing counting of digitized signals

Wataru Kokuyama,[1,a)] Hideaki Nozato[1] and Thomas R. Schibli[2,3]

[1]*National Metrology Institute of Japan (NMIJ), National Institute of Advanced Industrial Science and Technology (AIST), Tsukuba, 305-8563, Japan*
[2]*Department of Physics, University of Colorado at Boulder, Boulder, Colorado, 80309-0930, USA*
[3]*JILA, NIST and the University of Colorado, Boulder, Colorado, 80309-0440, USA*

**Abstract**

We developed a compact and easy-to-use phase meter based on a zero-crossing counting algorithm for digitized signals. Owing to the algorithm, the phase meter has low-noise and wide dynamic range. Low-noise differential phase measurements can be done for square waves ($-204 \text{ dBrad}^2/\text{Hz}$ for a 1-kHz, 1-$V_{p-p}$ signal, 10 Hz–1 kHz offset, with cross-correlation) as well as sinusoidal waves, with a measurement error of $< 1 \times 10^{-4}$ rad. We also demonstrated a direct phase measurement of an optical-beat note from a free-running laser over 10 decades (0.25 mHz–10 MHz) with a wide dynamic range of ~280 dB at 0.25 mHz. The phase meter can be an alternative for conventional phase meters and frequency counters in wide range of experiments.

## I. Introduction

The phase measurement of radio-frequency (RF) signals plays an important role in many precise electronic and optical experiments, including length metrology,[1] optical-frequency combs,[2,3] and photonic-microwave generation,[4,5] as well as precise spaceborne measurements, such as gravitational-wave detection[6,7] and mapping the Earth's gravitational field.[8] Furthermore, RF-distribution systems with precise phase characterization capability are required to ensure appropriate control of high-energy particle accelerators.[9] For such measurements, multiple types of phase measurement instruments, such as phase meters, lock-in amplifiers[10–15], frequency discriminators and phase-noise analyzers are used.

The low-noise measurement of large phase fluctuations of an optical beat note is still a challenge even with these instruments, because large phase fluctuations are beyond the measurement range of these precise instruments. For example, the software-defined radio[14] or the tracking-DDS[15] based on a digital phase-locked loop (PLL) with a numerically-controlled oscillator (NCO) can conduct precise phase measurements of optical beat notes. However, the measurement ranges of these instruments are still limited by the feedback bandwidth of the control loop of the PLL or the dynamic range of the NCO. Another example is an electrical or optical-delay line, which works as a frequency discriminator and can be used for measuring

---

a) Corresponding author. Electronic mail: wataru.kokuyama@aist.go.jp

large phase fluctuations. However, delay lines do not offer a good long-term phase stability because of the instability of the delay line itself and readout noise in frequency output.

In contrast, zero-crossing counting, on which frequency counters are based, does not suffer from the feedback bandwidth and the instability of the delay line; a frequency counter potentially mitigates the dilemma between a large measurement range and low noise. Frequency counters are extensively used to evaluate frequency stability in time and frequency metrology.[16] Some frequency counters based on the synchronous phase-reading technique,[17] the continuous time-stamping technique[18] and the linear regression of frequency data (the Ω-counter technique)[19,20] have been used experimentally and even commercialized. Additionally, a phase meter based on a principle similar to the synchronous phase-reading technique is also proposed for the Laser Interferometer Space Antenna.[21,22] However, these instruments do not achieve both low-phase-noise and sufficient measurement speed (>MHz) needed for optical beat note characterization, because they cannot use information from all zero crossings of RF signals, owing to the insufficient speed of signal processing.

To overcome these limitations, we developed a phase meter by implementing an novel simple algorithm, which we previously proposed,[23,24] on a field-programmable gate array (FPGA) with high-speed analog-to-digital converters (ADCs) to realize real-time processing of RF signals of up to ~250 MHz. Using the algorithm, the phase output is directly estimated from the zero-crossing counting and zero-crossing interpolations of an input signal, unlike time-interval measurements[25] in the frequency counters. Consequently, the output is obtained at a constant time interval, regardless of the frequency of the input signal. The information from all zero crossings is used to estimate the phase, resulting in excellent low-noise characteristics even for drifty or frequency-modulated signals. Moreover, a considerable amount of phase noise of RF signals can be precisely tracked at a bandwidth that exceeds 10 MHz.

The remainder of this study is structured as follows. In Section II, the principles of the phase-measuring algorithm are introduced and a 4-channel FPGA-phase meter is described. In Section III, the instrument is characterized in terms of its noise level, measurement error, linearity, and dynamic range; moreover, cross-correlation measurements are demonstrated to reduce the instrument's intrinsic noise level. In Section IV, the direct measurement of phase noise of a free-running optical-beat signals are described. Finally, discussions of the advantages, limitations, and potential applications of the phase meter are presented in Section V.

## II. Instrument
### A. Measurement principle

The principle of phase measurement is based on a zero-crossing counting algorithm, which was first proposed in our previous study.[23,24] In this subsection, we revisit the principle and briefly explain how the algorithm works. A detailed explanation, including an example of its calculation, is presented in Section I of the supplementary material.

This algorithm takes advantage of a combination of zero-crossing counting and interpolation using digitized data; however, it is different from previous methods that calculate frequency (or phase) from the elapsed time between zero crossings. In this technique, the phase is directly extracted from the data; consequently, the output is in a constant time interval, regardless of the frequency of the input signal. Here we express the input signal digitized by ADC as $V_i$, with an integer number of $i$. Then, we define a zero-crossing detected function as follows:

$$\lambda_i \equiv 1, \text{if } \text{sgn}_\text{M}(V_{i-1})\text{sgn}_\text{M}(V_i) = -1 \\ \equiv 0, \text{if } \text{sgn}_\text{M}(V_{i-1})\text{sgn}_\text{M}(V_i) = +1. \tag{1}$$

Here, $\text{sgn}_\text{M}(x)$ is a modified sign function, which is defined as

$$\text{sgn}_\text{M}(x) \equiv -1, \text{if } x < 0 \\ \equiv +1, \text{if } x \geq 0. \tag{2}$$

The difference from the usual sign function is $\text{sgn}_\text{M}(0) = +1$, while $\text{sgn}(0) = 0$. In a practical viewpoint, $\lambda_i$ can also be defined as $\lambda_i \equiv \text{XOR}\big(b_{2\text{C}}(V_{i-1})b_{2\text{C}}(V_i)\big)$, where $b_{2\text{C}}(x)$ means the most significant bit of $x$ in a two's complement representation. Using the zero-crossing detected function, $C_i$ is defined as the counter value, which is the number of zero crossings counted at the i-th data:

$$C_i \equiv \sum_{k=1}^{i} \lambda_k, \tag{3}$$

$F_i$ is defined as the fraction value, meaning interpolation of the zero crossing, which is calculated only at the positions of each zero crossing:

$$F_i \equiv \left(\frac{|V_{i+1}|}{|V_i| + |V_{i+1}|}\right) \lambda_{i+1} \tag{4}$$

Here we have to use $\lambda_{i+1}$ rather than $\lambda_i$. The reason for this usage is explained in Section I.A in the supplementary material. Then, the averaged phase estimator $\Phi$ (radian) can be calculated from every N samples according to Equation (5):

$$\Phi[z] = \left(\frac{\pi}{N} \sum_{i=1+(z-1)N}^{zN} (C_i + F_i)\right) + C_0, \tag{5}$$

where $z$ is an integer for indexing the averaged phase estimator. Note that the correction term $C_0$ is $\pm \pi/2$, depending on the initial sign of the input signal. This phase estimator is calculated in every N samples such that the phase output rate is 1/N of the sampling rate of ADC ($f_\text{ADC}$). Note that the signal frequency should be below $f_\text{ADC}/4$ to ensure zero crossings are detected properly.

### B. Hardware implementation

We implemented the algorithm on a commercial digitizer module (Teledyne SP Devices, ADQ14AC-4C-USB) equipped with four-channel ADCs and an FPGA (Figure 1). Phase measurement is performed for both input channels A and B, and then the phase difference is calculated by subtracting the two phase-measurement results. The same process is applied for both channels C and D.

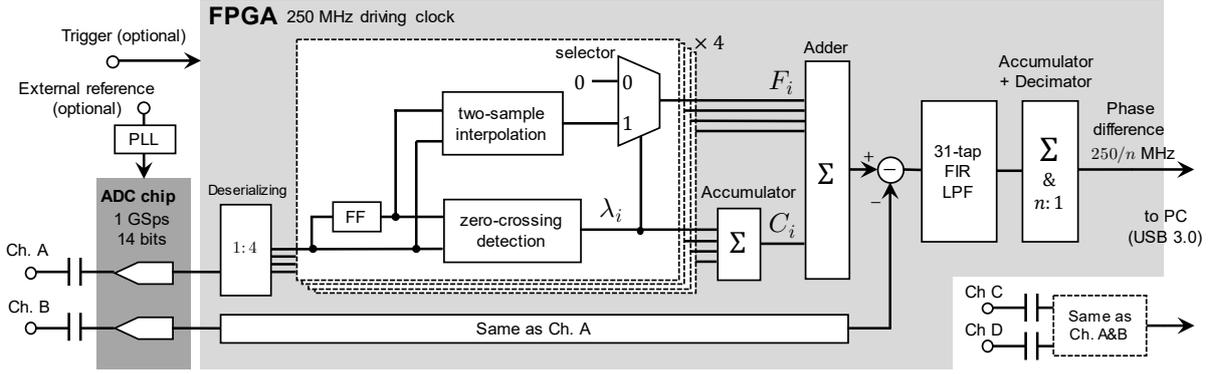

Figure 1. System block diagram of the instrument. FF: flip-flop, FIR: Finite impulse response, LPF: Low pass filter, PC: personal computer; PLL: phase-locked loop.

The input signal on channel A is digitized using a 14-bit two-channel ADC (Analog Devices, AD9680).[26] The analog inputs are AC (alternating-current) coupled (−3-dB high-pass bandwidth: 80 Hz) and have an impedance of 50 Ω. The sampling rate is 1 giga-samples per second (GSps). Note that the digitizer module's manufacturer applies nonlinearity and gain compensation technology for the ADCs. The maximum input voltage is 1.9 V peak-to-peak (+9.6 dBm for sinusoidal signals) and the input bandwidth at −1 dB is 900 MHz. However, we do not input signals over 250 MHz; therefore, the ADCs are always in Nyquist condition. Moreover, the signal-to-noise ratio (SNR), which is limited by the aperture jitter of 55 $fs_{RMS}$, is 64 dB; this corresponds to an effective number of bits of 9.6. The noise-spectral density, which is a white-noise spectrum over the Nyquist bandwidth, is −151 dBFS/Hz. Note that this is end-to-end noise; it includes noise from the analog front-end, overvoltage protection, gain compensation, and other signal shaping applied before or after ADC. As for isolation between channels, the specification of the ADC is 95 dB. Note that we selected the digitizer module with AC-coupled input because it has better SNR for input signals of >30 MHz.[27]

The external-reference signal, the frequency of which should be 10 MHz within a ±5-ppm range, can be inputted such that the ADC driving clock is synchronized with other instruments. Moreover, the start time of phase measurement can be synchronized using an external trigger signal with a resolution of 1 ns.

The digitized signal is processed in an FPGA (Xilinx, Kintex-7, XC7K325TFFG900-2), which is running on a 250-MHz driving clock. Data from the ADC is deserialized by four, such that the FPGA with a 250-MHz clock cycle can deal with the 1-GSps data. The data is then sent to two core blocks in the algorithm: zero-crossing detection and two-sample interpolation. The zero-crossing detection block generates $\lambda_i$ according to Equation (1). Then, the zero-crossing counter value $C_i$ is obtained by accumulating $\lambda_i$. The interpolation block continuously interpolates the input signal, while $F_i$ is generated via the selector driven by $\lambda_i$ (Equation (3)). The interpolation is done with a digital-signal-processing (DSP) block. Then, $C_i$ and $F_i$ are added, resulting in a single-channel phase at a rate of 250 MSps. This corresponds to the case of $N = 4$ in Equation (5). The register to keep the added value is 96-bits width in order to avoid overflow. Note that in Figure 1 some minor functions, such as the timing alignment between $C_i$ and $F_i$, are not shown for clarity.

Another phase measurement for the second input (i.e., channel B) is performed in parallel such that the phase difference is generated by subtracting the result for channel B from the result for channel A. The 250-MSps phase difference data has large fluctuations at high-frequencies due to the discontinuous increase in the phase at zero crossings (cf. Subsection II in the supplementary material). A 31-tap finite impulse response (FIR) digital low-pass filter (LPF) is applied to the phase difference to eliminate these high-frequency fluctuations. We selected a Blackman-type LPF to balance the flat gain in the passband and sufficient roll-off in the stopband. The cutoff frequency of the LPF is 16.67 MHz. All filter coefficients are 16-bit wide, and the sum of them is precisely 1, leading to 0-dB gain in a low-frequency limit. Although the logic size tends to be large, we selected FIR rather than infinite impulse response (IIR) to obtain a constant group delay across the whole frequency range. The LPF can be bypassed to avoid effects from the LPF when necessary. The phase result format is a 64-bit fixed floating point. The data is accumulated by $n:1$ to be converted to the desired sampling rates of $250/n$ MHz, where $n$ is the integer not exceeding $2.5 \times 10^7$. Thus, the final output rate, $f_{\mathrm{ADC}}/N$, is between 10 Hz and 250 MHz. Note that the averaging factor $N$ is limited to multiples of four; $N = 4n$.

Table 1 shows the total resources used in the FPGA. The values are estimated by subtracting the amount used for the peripheral (default) logic provided by the manufacturer from the total amount for the whole logic. Resources regarding signal processing, such as slices and the DSP48 slices, are extensively used. Note that a large amount of the logic is occupied by the two-sample interpolation block, which includes a 14-bit fixed-point divider function. Moreover, FIR LPF consumes a large amount of resources. We can drastically reduce this by replacing the LPF with an IIR type if we are not concerned with ripple and signal distortion in the passband. Note again that the phase meter logic for our implementation is not completely optimized in terms of the logic size and additional reduction is possible.

Table 1. Resources used for the FPGA

|  | Total available[28] | Used for the phase meter[a] | Ratio (%) |
| --- | --- | --- | --- |
| Logic cells | 326,080 | 26,357 | 8 |
| Slices | 50,950 | 35,284 | 69 |
| 36-kbit Block RAM | 445 | 26 | 6 |
| DSP48 Slices | 840 | 382 | 45 |

[a]Resource used by phase meter logic is calculated by subtracting the value of peripheral logic from that of the whole logic.

The calculated result is stored in 2-GByte peripheral memory in the digitizer module. The module is connected by a USB 3.0 link, which transfers data to a laptop personal computer (PC) at ~200 mega bytes per second. In the PC, data can be converted to power spectral density (PSD), Allan deviation[29–32], integrated timing jitter and other statistical parameters by custom-made codes using open-source numerical computation software (Scilab Enterprises, Scilab 6.0.2). Because of the limitation of the driver software in the PC, the total amount of data is limited to 76 mega samples. However, the maximum continuous measurement time becomes $7.6 \times 10^6$ s at 10 Sps, which is sufficient for most applications. We consider continuous reading from the digitizer module, which makes measurement time virtually infinite to be possible if there is a slight improvement in the driver software in the PC. Finally, the phase meter can be easily transported between labs because of the compactness of the digitizer module ($191 \times 108 \times 62$ mm$^3$, 750 g, 42 W power consumption).

The phase-measurement result has been validated by comparing it with the phase-noise standard at National Metrology Institute of Japan (NMIJ).[33] We used white-phase noise from −120 dBc/Hz (−117 dBrad$^2$/Hz) to −90 dBc/Hz (−87 dBrad$^2$/Hz) at a carrier frequency of 10 MHz. The expanded uncertainty ($k = 2$) for the −100-dBc/Hz calibration signal is 0.6 dB (at 1 Hz), 0.26 dB (at 10 Hz), and 0.22 dB (at 1 kHz, 10 kHz, and 100 kHz).[33] The two results from the phase meter and the phase-noise standard agreed within the uncertainties, showing the validity of the phase meter including the phase-measurement algorithm and self-made PC-based data analysis software for Fourier transformations. Detail of the comparison is found in Subsection IV.C of the supplementary material.

## III. Characterization
### A. Noise level

Noise, which defines how small of a signal can be detected, is a random fluctuation in a measurement that originates from internal noises in devices and/or external references. To measure the intrinsic noise level of the phase meter, two identical signals (split from the same original signal generated by a synthesizer) are sent to the two input channels of the phase meter and the PSD of the phase difference is calculated (Fig. 2). This measurement eliminates common-mode phase noise arising from the test signal source or from the sampling clock of the ADCs. We used an input power of +7 dBm for sinusoidal waves and 1 V$_{p-p}$ for square waves.

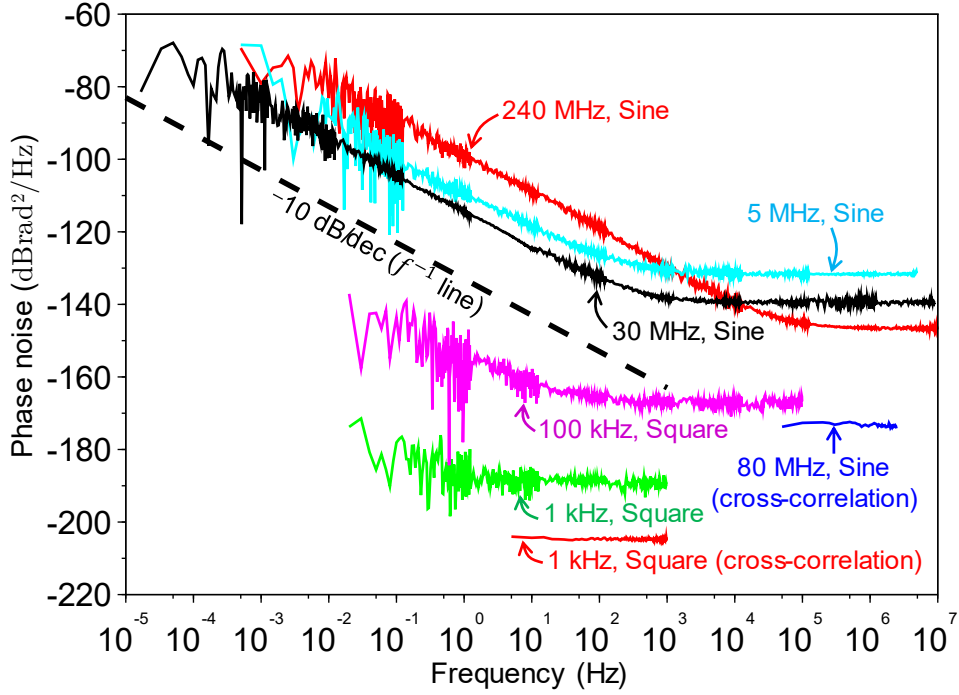

Figure 2. Noise level of the phase meter for different input-signal frequencies and different waveforms.

The noise levels are expressed in terms of two noise components (i.e., white-phase noise and flicker-phase noise) over 12 orders of frequency from $10^{-5}$ Hz to $10^7$ Hz. The white phase noise, which appears flat in the frequency domain, comes from the white-voltage noise of the ADC. We confirmed that the white-phase-noise level $S_\phi$ (rad$^2$/Hz) is determined by Equation (6):

$$S_\phi = \frac{P_\text{ND}}{P_\text{eff}} D(f_\text{sig}). \tag{6}$$

Here, $P_\text{ND}$ (dBm/Hz) is the noise-spectral density, which is determined by the SNR of the ADC. $P_\text{eff}$ (dBm) indicates the effective power of the input signal (i.e., the equivalent power of the sinusoidal signal that exhibits the same slew rate at zero crossings as that of the input signal). $f_\text{sig}$ is the input-signal frequency and $D(f_\text{sig})$ is a degradation factor, which is the ratio between $f_\text{sig}$ and the sampling rate of the ADC $f_\text{ADC}$, $D(f_\text{sig}) \equiv f_\text{ADC}/(4f_\text{sig})$. Based on Equation (6), the white noises for low-frequency square waves reach very low levels. For example, the white-phase noise for the 1-kHz square wave (slew rate: ~$3 \times 10^8$ V/s) is as low as $-190$ dBrad$^2$/Hz. This level exceeds the 50-Ω thermal-noise limit of $-178$ dBrad$^2$/Hz for a signal power of $+7$ dBm. The trick of such low phase noise is that the effective power is far larger ($+104$ dBm in this example) than the maximum input power at the ADCs; the effective power only reflects the slew rate at the zero-crossings, not the real power of the signal determined by the shape of the whole signal.

The second component is flicker-phase noise ($-10$-dB/dec dependency), which dominates at the low-offset frequency. The noise level is found to be independent of the power of the input signal, but it does depend on the input-signal frequency. We speculate that there are two noise sources: one is the flicker-noise component of the ADC-aperture jitter and the other is nonlinear conversion from the near-DC flicker-voltage noise at the ADCs. A more detailed noise analysis is found in Section III of the supplemental material.

The cross-correlation technique[34,35] can be used to improve the noise level, since the ADC noise is statistically independent between the individual ADC channels. For cross-correlation measurements, the signals are split and inputted into the four channels of the phase meter; channels C and D measure nominally the same-phase difference as channels A and B. Figure 2 presents the results with cross correlation. For 80-MHz, +7-dBm sinusoidal waves, a suppression of the ADC noise of 27 dB was achieved by averaging 25,000 traces. The level of noise suppression is limited by the common-mode phase difference ($-173$ dBrad$^2$/Hz) between the channel pairs Note that, in this plot, we used the correlation of a sufficient number of traces ($8 \times 10^8$) to identify the common-mode noise level. Using similar measurements, we find that the common-mode noise is $-204$ dBrad$^2$/Hz for a 1-kHz square-wave input signal. This phase noise corresponds to $0.3$ ps of timing jitter when integrated up to 1 kHz.

We also confirmed that suppression with cross correlation can reach at least 39 dB when we measure signals with added white-phase noise ($\sim -80$ dBrad$^2$/Hz); this level is still smaller than the expected value (47.5 dB) from the specification of channel isolation of 95 dB, according to the datasheet.[26] We consider that this is because of worse isolations for end-to-end systems when all channels A–D are active. Refer to the Subsection V.B in the supplemental material for more detail on the cross-correlation technique and performance.

### B. Error and linearity

Error refers to the difference between the actual and measured phases. We theoretically analyzed the error components of the phase-measurement principle in detail (see Section II of the supplementary material) and found that four sources of error appear in the phase measurement. Under a condition like that for an ordinary phase meter (i.e., low-drift input signals with a low measurement bandwidth), only the error arising from zero-crossing interpolation (called "zero-crossing interpolation (ZI) error") appears in the phase-measurement results among the four errors, around specific values of the input-signal frequency (so-called "singular frequencies") determined by $f_{\mathrm{ADC}}$.

Note that the four types of error are: (i) trapezoidal-approximation error, which comes from the trapezoidal approximation in the phase-estimation process; (ii) first-zero-crossing error, which comes from a slight difference in the trapezoidal approximation for the first zero-crossing; (iii) quantization-aliasing error, which comes from aliasing of the phase quantization occurring at zero crossings; and (iv) the ZI error. In most cases, these errors appear in the results for the AC components, and cause a periodic pseudo signal in the phase-measurement results.

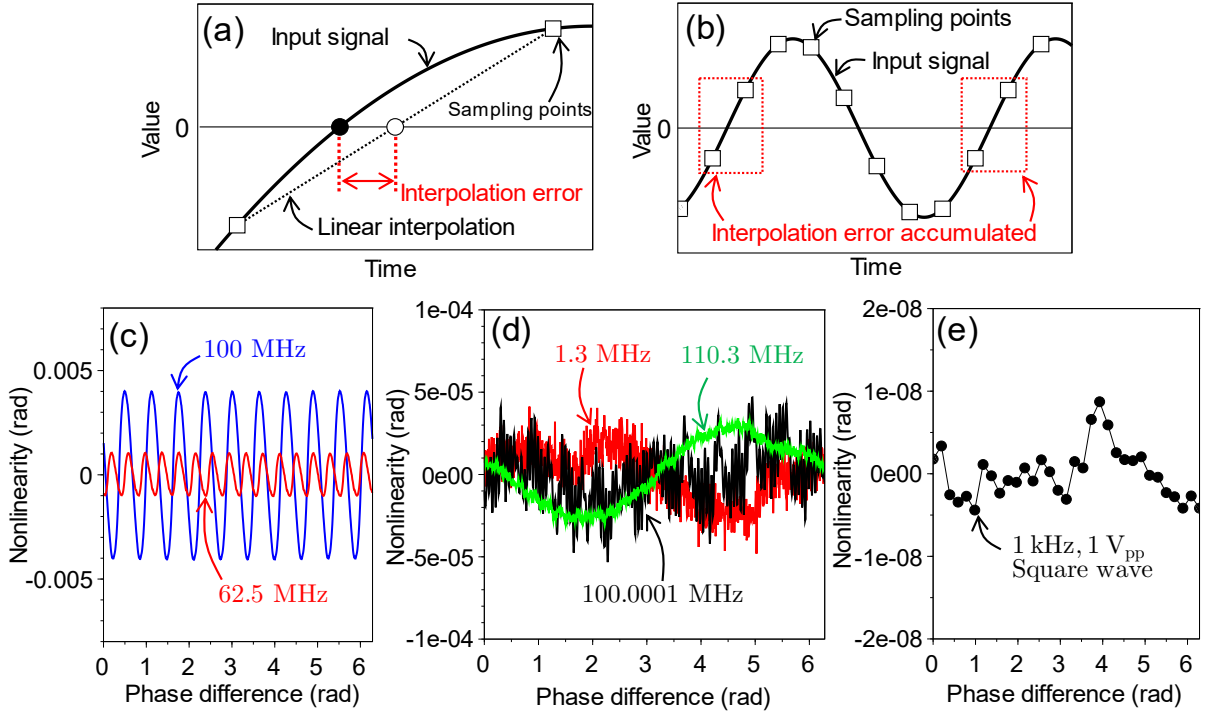

Figure 3. (a) The zero-crossing interpolation (ZI) error, (b) Accumulation mechanism of the ZI error, (c) Measured nonlinearity with input signal frequency at the singular frequency, (d) Measured nonlinearity for input frequency that are far from the singular frequency. (e) Measured nonlinearity for 1-kHz, 1-Vpp square wave.

The ZI error, which originates from the error in zero-crossing interpolation, appears at low-offset frequencies only when the input-signal frequency is around a singular value. This is graphically explained as Figure 3(a)-(b). The interpolation error essentially comes from coupling between the nonlinearity of the input signal at the zero-crossings and the sampling timing (Figure 3(a)). Hence, when the input-signal frequency and the sampling frequency are in a rational relation (i.e. the singular frequency), the interpolation error is accumulated over time, rather than it is decreased with averaging (Figure 3(b)). Quantitatively, the singular frequency is expressed as Equation (7):

$$f_{\text{sglr}} \equiv \frac{1}{2\left(s + \frac{q}{p}\right)} f_{\text{ADC}}. \tag{7}$$

The parameters $s$, $p$, and $q$ are the integers that characterize $f_{\text{sglr}}$, which obey $2 \leq s$, $1 \leq p$, $0 \leq q < p$. In addition, $p$ and $q$ are coprime. For example, when $(s, q, p) = (3, 1, 3)$, $f_{\text{sglr}} = 150$ MHz for our hardware ($f_{\text{ADC}} = 1$ GHz).

According to the detailed analysis presented in the supplementary material, the ZI error appears as a fundamental oscillation at a frequency of $2(sp + q)\delta f$ (Hz) and its harmonics, when the input-signal frequency $f_{\text{sig}}$ satisfies $f_{\text{sig}} = f_{\text{sglr}} + \delta f$. For sinusoidal input signals, the fundamental peak has an approximate amplitude of $4p^{-3} f_{\text{sig}}^{3} f_{\text{ADC}}^{-3}$ (rad) and the harmonics have a dependence of $-60$ dB/dec in the frequency domain. For example, when $f_{\text{sig}} = 150$ MHz $+ 3$ Hz, the ZI error appears as a 60-Hz pseudo signal with an amplitude of $\sim 6 \times 10^{-4}$ rad. Moreover, for $\delta f = 0$ (i.e., when the input-signal frequency is equal to the singular frequency), a DC-phase error or a secular-phase shift may occur in the results. The maximum error is also $4p^{-3}f_{\text{sig}}^3 f_{\text{ADC}}^{-3}$ (rad).

For experimental characterization of the phase meter, we measured nonlinearity, which refers to how the measurement error behaves when the true phase moves over a single cycle (0 rad to $2\pi$ rad), instead of the phase error itself. This is because a phase standard with uncertainly of $10^{-5}$-rad range does not exist and is hard to create. In particular, we use a 2-ch function generator (Keysight, 33622A) to generate two input signals, the frequency of which is slightly (10 Hz) different from each other. The phase difference fitted with $20\pi$-rad/s drift, corresponding to a 10-Hz frequency difference. Then the result is band-pass filtered with a cutoff frequencies of 5 Hz and 1 kHz to eliminate noise and drifts. The nonlinearity is then divided into 100 segments and averaged coherently for the 100 cycles to obtain the results with reduced noise.

The results at two singular frequencies (100 MHz and 62.5 MHz) are shown in Figure 3(c). The frequency of 100 MHz and 62.5 MHz corresponds to $(s, p, q) = (5, 0, 1)$ and $(8, 0, 1)$, respectively. The results are matched with theoretical estimations that can be found in Table S2 in the supplementary material; 10 cycles with peak amplitude of ~$4 \times 10^{-3}$ rad ($f_{\text{sig}} =$ 100 MHz), and 16 cycles with the peak of ~$1 \times 10^{-3}$ rad ($f_{\text{sig}} = 62.5$ MHz). In contrast, the nonlinearity for the frequency different from the singular frequency is unaffected by the ZI error (Figure S3(d)). In the plot, we confirmed that the nonlinearity of 1.3-MHz and 110.3-MHz signals are less than $10^{-4}$ rad. The levels of nonlinearities shown here are representative for all choices of input-signal frequencies we investigated; we did not find larger nonlinearities for other $f_{\text{sig}}$. We also confirmed excellent-low nonlinearity for low-frequency square waves (Figure 3(e)). The measured nonlinearity is less than $1 \times 10^{-8}$ rad, corresponding ~1.6-ps nonlinearity for the time difference in a full range of 1 ms.

If the input signal frequency is too close, or even equal, to the singular frequencies, the nonlinearity can still be avoided using different clock frequencies for the ADC. For instance, if the input signal frequency is 100 MHz, we can effectively shift the frequency by 1 ppm using a 1-ppm shifted reference clock. For such a case, the level of nonlinearity is similar to that for 100.0001 MHz (Figure 3(d)), which is comparable to other levels.

From these evaluations, we can conclude that the phase meter has a nonlinearity at a level of approximately $10^{-4}$ rad or lower for input signal frequency above ~1 MHz. Practically, the fluctuations induced by flicker noise (Section III.A) could be larger than $10^{-4}$ rad, in which case the nonlinearity would not be the dominant source of measurement uncertainty for long measurement times. Moreover, for sinusoidal input signals, there are a limited number of singular frequencies, at which the peak amplitude exceeds $1.0 \times 10^{-4}$ rad, where the peak amplitude is proportional to $f_{\text{sig}}^3 p^{-3}$. In our case, there are only 55 such singular frequencies with associated peak amplitudes over $1.0 \times 10^{-4}$ rad (see Table S2 of the supplementary material).

Note that we can only evaluate an approximate upper limit of the nonlinearity in this evaluation. This is because we need a phase standard that should have a much lower nonlinearity than $1 \times 10^{-4}$ rad to distinguish whether the function generator or phase meter is the source of the nonlinearity for this test setup. A detailed estimation of the nonlinearity is found in Section II.F of the supplementary material.

### C. Dynamic range

The dynamic range describes the difference between the maximum measurable level of the instrument and minimum measurable level determined by the instrument noise. Here, we briefly estimate the maximum measurable level of this algorithm. The condition for the phase-measuring algorithm is that zero crossings are adequately detected by digitized signals, i.e. the input-signal frequency should be in the range of $0 < f_{\text{sig}} < f_{\text{ADC}}/4$. Therefore, when the input signal has phase noise, Equation (8) should be satisfied:

$$0 < f_{\text{sig}} \pm k\sigma_f < f_{\text{ADC}}/4. \tag{8}$$

Here, $\sigma_f^2$ is the frequency variance of the signal induced by the phase noise and $k$ is a factor of the margin to avoid cycle slips. Empirically, $k = 10$ is sufficient.

For example, when a white-phase noise ($S_\phi = b_0 f^0$) is measured, the frequency variance is expressed as

$$\sigma_f^2 = \int_{f_{\text{RBW}}}^{f_{\text{C}}} f^2 S_\phi df \cong b_0 \frac{f_{\text{C}}^3}{3}, \tag{9}$$

where $f_{\text{RBW}} = 1/(\text{measurement time})$ indicates the resolution bandwidth and $f_{\text{C}}$ indicates the cutoff frequency, over which the phase noise is negligible. Therefore, the measurement condition of Equation (8) is converted to $b_0 < 3k^{-2}f_{\text{C}}^{-3}f_{\text{sig}}^2$. Here, for clarity, we assume the simplest case: $f_{\text{sig}} < f_{\text{ADC}}/8$. For example, when $f_{\text{sig}} = 100$ MHz, $f_{\text{C}} = 5$ MHz, and $k = 10$, the condition is estimated as $b_0 < -56$ dBrad$^2$/Hz. Note that the level is proportional to $f_{\text{C}}^{-3}$; when $f_{\text{C}}$ is reduced to 5 kHz, the condition becomes $b_0 < +34$ dBrad$^2$/Hz. Here, the noise level is approximately $-140$ dBrad$^2$/Hz and thus the dynamic range is more than 170 dB in this condition.

### IV. Application to dynamic phase measurement

One application that benefits from a large dynamic range is the measurement of drifty signals, such as free-running optical-beat notes. For example, here we present a direct measurement of the phase noise of a free-running monolithic mode-locked laser (MLL).[37,38]

The setup is shown in the inset of Figure 3(a). The beat note between the MLL and cavity-stabilized CW laser (1550 nm, 0.9 mW at the beat note) is detected at the 250-MHz-bandwidth photodetector. The repetition rate of the MLL is approximately 1 GHz; the averaged power, central wavelength, and linewidth of the MLL are 33 mW, 1560 nm, and 13 nm, respectively. No frequency servo is activated for the MLL; only intensity stabilization is applied. After amplification and filtering, the detected signal is inputted to channel A of the phase meter. The SNR of the signal is ~60 dB at a 100-kHz-resolution bandwidth. The signal is centered around 190 MHz, and thus a 190-MHz reference signal from a signal generator is inputted into channel B of the phase meter. The CW laser is expected to have a very high fractional stability of about $10^{-15} - 10^{-14}$, owing to the Pound–Drever–Hall (PDH) locking of a single-mode fiber laser (NKT Photonics, Koheras BASIK MIKRO E15) to an ultra-low expansion glass cavity in vacuum (manufactured by Stable Laser Systems, Inc.).

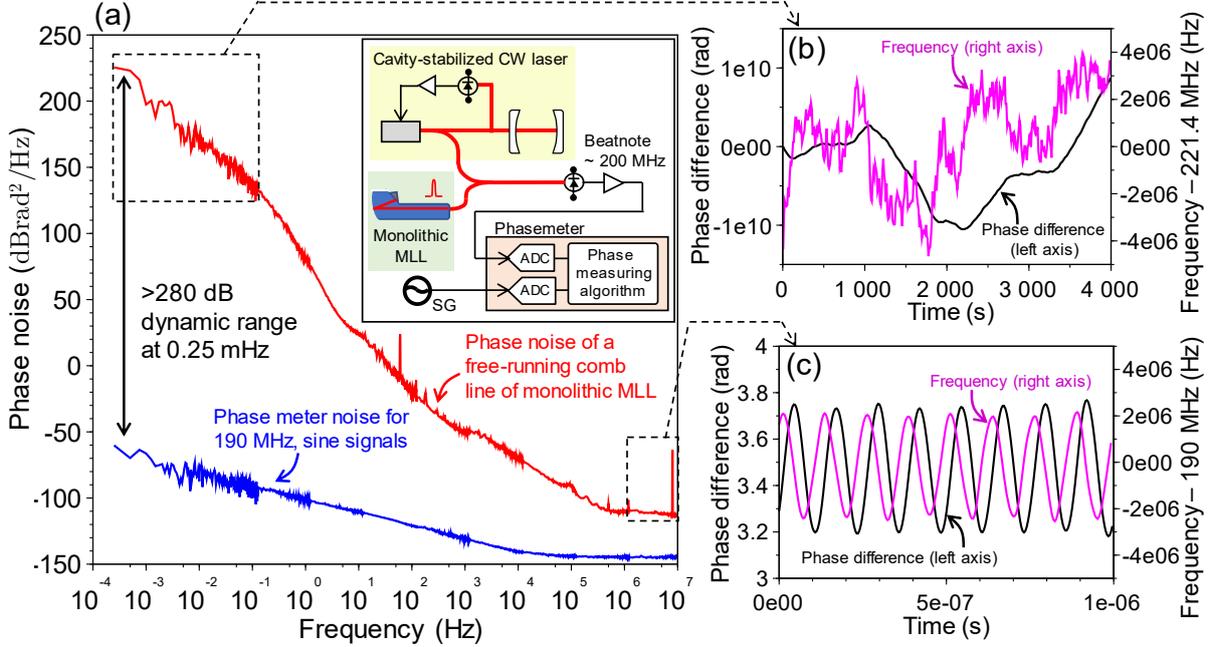

Figure 4. Direct phase-noise measurement of a free-running mode-locked laser. (a) Results and measurement setup (inset), (b) Long-term results of phase difference and frequency, (c) Short-term results of phase difference and frequency. CW: continuous wave, MLL: mode-locked laser, SG: signal generator, ADC: analog-to-digital converter.

Figure 4(a) shows the phase noise of the beat note over a wide frequency range from 0.25 mHz to 10 MHz (>10 orders of magnitude). The phase noise corresponds to the free-running phase noise of a comb mode of the MLL. In the high-frequency range above 1 MHz, this noise is limited by the beat note's SNR, whereas at frequencies below several Hz, the phase noise is dominated by the change of the repetition rate induced by the MLL's temperature fluctuation. Compared with the noise level of the phase meter, a dynamic range of ~280 dB at 0.25 mHz is achieved. Note that this is measured with a single instrument, rather than additional servo controls or external phase-locked loops such as tracking filters.

One advantage of the phase meter is that time-domain phase measurement can be conducted. Figure 4(b) shows the time evolution of the phase difference and frequency for long time scales. The frequency is calculated by taking the derivative of the phase difference. The sampling rate for the plots is 0.1 Hz. Note that the reference frequency is numerically shifted to 221.4 MHz in data analysis, just for the clarity of the plot. A phase difference up to $10^{10}$ rad is successfully tracked, corresponding to $\sim \pm 4$ MHz of frequency fluctuation. Compared to the measurement noise of $\sim 10^{-4}$ rad, this range shows a dynamic range of roughly $10^{14}$ (280 dB), agreeing with the number from the spectral domain.

At the same time, the phase meter can directly measure the phases of frequency-modulated signals that cannot be measured by other instruments. Figure 4(c) shows the phase evolution over a short-time scale. The sampling rate for the plot is 250 MHz; however, the measurement bandwidth is limited to 16.67 MHz by the FIR LPF in the phase meter. The 8-MHz phase modulation, which is used for the PDH locking of the CW laser, is clearly observed. The measured modulation depth is ~0.25 rad, which is equivalent to frequency modulation with an amplitude of ~2 MHz. Such a signal with a high modulation frequency and a large frequency deviation cannot be measured with ordinary frequency counters.

## V. Discussions

Compared to an ordinary phase measurement with I/Q (in-phase and quadrature-phase) demodulation, this technique has the benefit of a wide input-frequency range with only a small increase of the noise level. In terms of the noise level, I/Q demodulation has the best performance as the multiplication of the sine/cosine references is the best approach to extract the carrier components in an orthogonal basis. However, the input-frequency range is limited by the time constant of the demodulation. If an external PLL is used, the bandwidth is further reduced by its time constant. On the contrary, this phase meter has a wider frequency range of input signals: $f_{\text{sig}} < f_{\text{ADC}}/4$. The drawback is the increase of the noise level by the degradation factor $D(f_{\text{sig}})$, which is normally less than 20 dB.

From a sampling-theory viewpoint, zero crossings carry sufficient information about the phase noise. This is because zero crossings occur at twice the signal frequency ($2f_{\text{sig}}$); the information carried by the zero crossings has a bandwidth of $f_{\text{sig}}$ according to the sampling theorem. This equals to the Nyquist bandwidth of the phase noise, $f_{\text{sig}}$. Note that only a fraction of data from the ADC (i.e. adjacent to zero-crossings) is used to estimate the phase in the algorithm. The ratio between the used-data rate ($4f_{\text{sig}}$) and total-data rate ($f_{\text{ADC}}$) corresponds to the degradation factor, $D(f_{\text{sig}})$.

There are several measurements that can only be conducted by this phase meter. (i) Constant-rate phase measurement of square-wave signals can be done. Unlike ordinal zero-crossing-based time-interval phase/frequency measurements, this algorithm provides constant-rate, dead-time free and high-speed phase estimation with a low noise, since the algorithm directly estimates the phase from input signals. Moreover, the constant-rate phase information enables us to apply the cross-correlation technique, which can significantly lower the instrument noise level. (ii) Direct phase measurements of frequency-modulated signals are possible, as long as the zero-crossings are adequately detected. Such dynamic phase measurements may be also useful for dynamic-response analysis of VCOs, PLLs, and phase/frequency feedback-control loops. (iii) Higher-speed (>10 MHz) frequency measurements than that with ordinary frequency counters are possible, simply by taking the derivative of the estimated phase. Owing to the nature of phase measurements, such frequency measurements are automatically dead-time free, making this suitable for evaluating stable signals by calculating the Allan deviation from the measured phase evolution.

The phase meter has limitations that originates from the algorithm. First, a secular phase shift can occur in the measurement results due to ZI errors, when the input-signal frequency is equal to one of the singular frequencies. This is an intrinsic, inevitable error of the algorithm. A reference clock with a slightly shifted frequency can help to avoid this issue, by effectively shifting the input-signal frequency. Second, the phase meter cannot be used to measure multi-tone signals, but a single frequency only. In other words, the phase meter can be an alternative for frequency counters or phase meters, but not for lock-in amplifiers. Finally, phase measurements from zero crossings are sensitive to higher-order harmonics. For instance, noise around the third harmonics of the input signal ($3f_{\text{sig}}$) may be converted into low-frequency phase noise in the measurement results. Appropriate filtering of those noises around the harmonics is needed to avoid such aliasing noise.

Several improvements are discussed in the supplementary material. For example, to avoid the miscounting of the zero crossings, so-called cycle slips, tracking filters can be used prior to the phase meters. (Subsection IV.A of the supplementary material). Additionally, the input-frequency range of the phase meter can be expanded by using a prescaler, in exchange of resolution (Subsection V.C of the supplementary material). Moreover, a time-to-digital converter[36] (TDC) (rather than an ADC) architecture can be used to avoid ZI error (Subsection V.D of the supplementary material).

We are applying the phase meter to various types of measurements, especially optics experiments conducted in research labs. For example, the phase meter was already used to characterize the free-running phase noise of heterodyne beat signals obtained from optical frequency combs.[40–42] We are also applying this phase meter to a readout of a heterodyne interferometer system for vibration metrology.[43,44] We are also applying this technique to a heterodyne interferometer used for birefringence measurements of ultrathin glass[45] and picosecond-level timing delay evaluation of a pump-probe pulse laser system for thermal conductivity measurement at NMIJ.[46,47]

We believe that this instrument can be used for many applications in which conventional phase meters and frequency counters have been used. The potential applications include quartz crystal microbalance sensors,[48] quartz-based accelerometers,[49] atom interferometric gravimeters,[50] heterodyne laser interferometers with picometer-level nonlinearity,[51] evaluation of additive phase noise of optical fiber,[52] dynamic pressure measurements based on optical resonator,[53] and frequency-modulated continuous-wave light detection and ranging (FMCW-LIDAR).[54]

## Supplementary material

Please refer to the supplementary material for detailed analysis, including the derivation of measurement principle, error analysis, noise analysis and performance analysis.

**Acknowledgments**

This work was partially supported by Shizuoka prefectural government fund for promoting technology-advanced businesses (FY2015–2017) and JSPS Kakenhi 17K18421. WK would also like to thank the following researchers who provided us some insights through discussions: Hidemi Tsuchida, Sho Okubo, Keisuke Nakamura, Daisuke Akamatsu, Takehiko Tanabe, Masato Wada, Hajime Inaba, Atsushi Onae, Shinya Yanagimachi, Takashi Yagi, Tatsuya Zama, Nobuyuki Toyama, Masahiro Ishihara, Yoshiteru Kusano, Tamio Ishigami, Hiromi Mitsumori, Tomofumi Shimoda, Koichiro Hattori and Akihiro Ota (AIST), Akira Koike (System In Frontier, Inc.), Kenji Tadokoro (Mish international, Inc.), Hirokazu Ishida, Yu Morimoto and Hiroyuki Kowa (UNIOPT Co., Ltd.), Takuma Oi (Toho Mercantile Co., Ltd.), Yoshiaki Nakajima (Toho University), Kaoru Minoshima (University of Electro-communications), Kenichi Hitachi and Atsushi Ishizawa (NTT Basic Research Lab.), Shuko Yokoyama and Toshiyuki Yokoyama (Micro Optics, Co., Ltd.), Haochen Tian (Tianjin University), Seiji Kawamura (Nagoya University), Koji Arai (LIGO/Caltech), Kenji Numata (NASA/GSFC), Yuta Michimura, Masaki Ando, and Mamoru Endo (The University of Tokyo). Manoj Kalubovilage (University of Colorado Boulder) helped us to take the data presented in Section IV. WK has personal financial interests through patents[24,39] (Assignee: National Institute of Advanced Industrial Science and Technology).

**Data availability**

The data that support the findings of this study are available from the corresponding author upon reasonable request.

# Supplementary material: Phase meter based on zero-crossing counting of digitized signals

Wataru Kokuyama,[1,a)] Hideaki Nozato[1] and Thomas R. Schibli[2,3]

[1]*National Metrology Institute of Japan (NMIJ), National Institute of Advanced Industrial Science and Technology (AIST), Tsukuba, 305-8563, Japan*
[2]*Department of Physics, University of Colorado at Boulder, Boulder, Colorado, 80309-0930, USA*
[3]*JILA, NIST and the University of Colorado, Boulder, Colorado, 80309-0440, USA.*

**Abstract**

This supplementary material provides a detailed analysis of the phase-measuring algorithm, which was recently proposed by the authors. Despite its importance, such detailed analysis has not yet been published. Section I presents a quantitative derivation of the measurement principle. We algebraically and quantitatively prove why phase measurement can be conducted using Equation (5) in the main text. We also explain how the phase measurement algorithm works using a detailed example. Section II shows a detailed error analysis. All sources of error in the phase measuring algorithm are identified and classified into four categories, and the characteristics of the four errors are analyzed. We conclude that, for low-drift input signals with low measurement bandwidth, only an error from zero-crossing interpolation appears in the measurement results only when the input signal frequency is near singular frequencies, calculated through the sampling rate of the analog-to-digital converter (ADC). Moreover, the nonlinearity of the phase meter is analyzed, and it is within $1 \times 10^{-4}$ rad. Section III shows a detailed noise analysis. Measurement of phase difference is limited by two noises, white phase noise and flicker phase noise. White phase noise originates from the white noise in the analog-to-digital (A/D) conversion process, whereas flicker phase noise originates from the aperture jitter and nonlinearity of the ADC. Section IV presents the performance of the phase meter. We show that cycle slips, from which conventional frequency counters also suffer, appears when either the signal-to-noise ratio (SNR) or slew rate of the input signal is low. We create a theoretical or empirical model on the rate of cycle slips and confirmed correlations with measurement or simulation. We also measured the effect of amplitude-modulation to phase-modulation (AM-PM) conversion. Furthermore, we validate the measurement of the phase meter with the national phase noise standard of Japan. Section V introduces some techniques to improve the phase meter. The techniques include offset compensation, cross-correlation, expansion of input frequency range, and elimination of zero-crossing interpolation error.

---

a) Corresponding author. Electronic mail: wataru.kokuyama@aist.go.jp

# Table of Contents

# Notation and Acronyms

| Item | Description | Unit | Ref. | Page |
|---|---|---|---|---|
| $t$ | Time | sec | — | S5 |
| $S(t)$ | Input analog signal for the phase meter | V | (S1) | S5 |
| $\phi(t)$ | Phase of the input signal | rad | (S1) | S5 |
| $\Phi[z]$ | Averaged phase estimator | rad | (S2) | S5 |
| $z$ | Index for $\Phi[z]$ | — | (S2) | S5 |
| $T$ | Measurement interval for $\Phi[z]$ | sec | (S2) | S5 |
| $p(t)$ | Normalized phase | No dim. | (S3) | S6 |
| $\tau_j$ | the timing of the zero-crossings of the input signal | sec | — | S6 |
| $j$ | Index used for zero-crossings | — | — | S6 |
| $q(t)$ | Phase estimation function | No dim. | (S6) | S7 |
| $t_{\text{ADC}}$ | Sampling interval of ADC | sec | — | S8 |
| $V_i$ | Sampled data by ADC for the input signal $S(t)$ | V | — | S8 |
| $i$ | Index used for variables with same sampling rate as digitized data ($f_{\text{ADC}}$) | — | — | S8 |
| $N_j$ | the number of the sampled data just after the zero-crossing | — | — | S8 |
| $Q(t)$ | Windowed phase estimation function | No dim. | (S10) | S8 |
| $w(t)$ | Window function | No dim. | (S11) | S9 |
| $U(t)$ | Step function | No dim. | (S12) | S9 |
| $Q_i$ | Instantaneous phase estimator | rad | (S14) | S9 |
| $\lambda_i$ | Zero-crossing detection function | — | (S19) | S9 |
| $C_i$ | Counting value | — | (S22) | S10 |
| $F_i$ | Fraction value | — | (S25) | S10 |
| $B_i$ | Boxcar filter coefficients to obtain final measurement result | — | (S27) | S11 |
| $C_0$ | Constant added for $Q_i$ | — | (S26) | S11 |
| $\delta\tau_j$ | Perturbation of zero-crossing timing | sec | — | S11 |
| $\delta\Phi$ | Perturbation of the averaged phase estimator | rad | (S30) | S11 |
| $L$ | Number of zero-crossings in time range of measurement interval | — | — | S11 |
| TA error | Trapezoidal Approximation error | rad | — | S17 |
| FZ error | First Zero-crossing error | rad | — | S17 |
| QA error | Quantization-Aliasing error | rad | — | S17 |
| ZI error | Zero-crossing Interpolation error | rad | — | S18 |
| $q_{\text{ERR}}(t)$ | Error of phase estimation function $q(t)$ | — | — | S24 |
| $H_w(f)$ | Transfer function of $w(t)$ | — | (S64) | S24 |
| $\delta f$ | Detuning frequency | Hz | (S66) | S24 |
| $f_{\text{sglr}}$ | Singular frequency | Hz | (S68) | S24 |

## Notation and Acronyms (continued)

| Item | Description | Unit | Ref. | Page |
|---|---|---|---|---|
| $f_{\text{fund}}$ | Fundamental frequency (frequency of fundamental peak) | Hz | (S70) | S25 |
| $A_{\text{fund,(rad)}}$ | Amplitude of the fundamental peak | rad | (S72) | S25 |
| $\delta f_{\text{MAX}}$ | Maximum detuning frequency | Hz | (S73) | S26 |
| $A_{\text{MAX,(rad)}}$ | maximum amplitude associated to $\delta f_{\text{MAX}}$ | rad | (S74) | S26 |
| $f_{\text{MAX}}$ | Maximum input signal frequency free from QA error | Hz | (S77) | S28 |
| $f_{\text{CMBW}}$ | Critical measurement bandwidth | Hz | (S80) | S28 |
| $\gamma(t)$ (ZIEF) | Zero-crossings interpolation error function (See also Figure S12) | rad | — | S29 |
| VS-ZIEF | Virtually-sampled ZIEF | rad | — | S29 |
| deltaZIEF | Sum of delta functions converted from VS-ZIEF | rad | — | S29 |
| W-ZIEF | Windowed ZIEF, which is convolution of deltaZIEF and window function $w(t)$ | rad | — | S29 |
| $Q_{i,\text{ERR}}$ | ZI error in instantaneous phase estimator $Q_i$ | rad | — | S29 |
| $\gamma_0(t)$ (oneZIEF) | One-cycle Zero-crossing Interpolation Error Function | rad | (S87) | S32 |
| $\phi_t$ | $2\pi f_{\text{sig}} t$ | rad | (S90a) | S33 |
| $\phi_{\text{nyq}}$ | $2\pi f_{\text{sig}} t_{\text{ADC}}/2$ | rad | (S90b) | S33 |
| $\gamma_{\text{MAX}}$ | The maximum value of oneZIEF | rad | — | S33 |
| $C_l$ | Fourier expansion coefficient of oneZIEF | rad | (S98) | S37 |
| $S_{\text{CLK}}$ | Phase noise of sampling clock | $\text{rad}^2/\text{Hz}$ | (S109) | S56 |
| $S_{\phi,\text{CLK}}$ | Sampling clock (phase noise of the phase measurement result induced by the sampling clock) | $\text{rad}^2/\text{Hz}$ | (S109) | S56 |
| $\Delta f_{\text{sig}}$ | Frequency difference between two input signals in two channels | Hz | — | S56 |
| SG | Signal generator | — | — | S57 |
| $\sigma_V$ | Root mean square value of ADC noise | V | (S111) | S58 |
| $P_{\text{ND}}$ | Power spectral density of ADC noise | W/Hz | — | S58 |
| $\sigma_t$ | Root mean square of timing noise at zero-crossings | s | (S112) | S58 |
| $\delta\Phi$ | Root mean square value of phase estimator | rad | (S113) | S59 |
| $S_\phi$ | Power spectrum density of the measured phase noise | $\text{rad}^2/\text{Hz}$ | (S114) | S59 |
| $P_{\text{eff}}$ | Effective power of input signal | W | (S116) | S59 |
| $D(f_{\text{sig}})$ | Degradation factor | — | (S117) | S59 |
| $\mathcal{L}_{\text{(dBc/Hz)}}$ | Single sideband phase noise | dBc/Hz | (S124) | S61 |
| $\overline{V_i}$ | Sampled data without noise (used for explanation of cycle slips) | V | — | S68 |
| $N(x)$ | Normal distribution | — | (S133) | S69 |
| $C_N(x)$ | Cumulative distribution function for normal distribution | — | (S134) | S69 |
| SR | Slip rate | $\text{s}^{-1}$ | (S139) | S69 |
| $f_{\text{NBW}}$ | Noise bandwidth | Hz | — | S71 |
| $P_{\text{noise}}$ | Integrate noise power over noise bandwidth | W | (S142) | S74 |

# I. Measurement principle

In our previous work[1,2], we presented an intuitive and graphical explanation to show how our proposed digital phase-measuring algorithm works. In this section, we prove algebraically and quantitatively why phase measurement can be done with Equation (5) (see main text) or Equation (S28) below. All sources of error are identified and analyzed in detail in Section II by using the quantitative derivation.

## I.A. Measurement principle
### 1. Derivation

We start with continuous time. Discrete time, which must be used to describe digitized signals, will be considered later. We assume a periodic input signal,

$$S(t) \equiv P(\phi(t)), \tag{S1}$$

where $\phi(t)$ is a monotonically increasing phase; $P(x)$ is a periodic function with a period of $2\pi$, and $P(x) = 0$ at $x = n\pi$. For example, a pure sinusoidal signal without any phase fluctuation $S(t)$ is expressed exactly as $S(t) = \sin(2\pi f_{\text{sig}} t)$, where $f_{\text{sig}}$ is the input signal frequency. In other words, $\phi(t) = 2\pi f_{\text{sig}} t$, and $P(x) = \sin(x)$. Generally, the input signal may have phase/frequency noise or fluctuations; therefore, this general expression is used. The goal of the phase measuring algorithm is to estimate the phase $\phi(t)$ or equivalent estimator from the digitized input signal. Note that in this expression, amplitude fluctuation is not considered, and the signal amplitude is normalized. What we want to derive is a phase estimator with a given constant interval of time $T$. The most appropriate phase estimator should be averaged phase $\Phi$ over the time interval, i.e.,

$$\Phi[1] \equiv \frac{1}{T} \int_0^T \phi(t) dt$$

$$\Phi[2] \equiv \frac{1}{T} \int_T^{2T} \phi(t) dt$$

$$\vdots$$

$$\Phi[z] \equiv \frac{1}{T} \int_{(z-1)T}^{zT} \phi(t) dt, \tag{S2}$$

where z is an integer ($1 \leq z$).

In Figure S1, we describe how the phase estimator is derived from digitized data. First, the phase of the input signal $\phi(t)$ is approximated with the quantized phase estimation function $q(t)$ by using the timing of zero-crossings. Next, we carry out a convolution of $q(t)$ with a window function to obtain the windowed phase estimation function $Q(t)$. We then apply sampling to $Q(t)$ at the analog to digital converter (ADC) sampling rate. In this step, a linear interpolation is used to estimate the zero-crossing timing. Finally, we derive the averaged phase estimator by applying averaging (moving average filter) and decimation. Note that the definitions of functions and variables presented in Figure S1 appear later in this subsection.

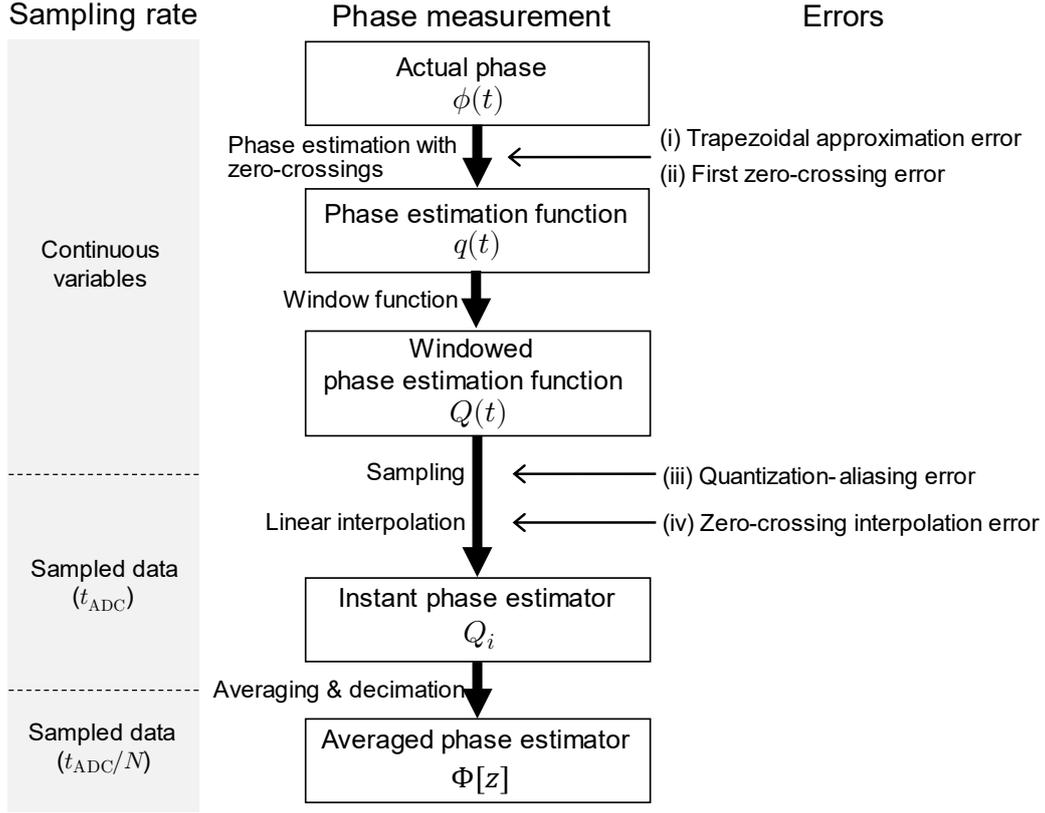

Figure S1. Overview of the phase measuring algorithm. The phase estimation sequence is explained in the center. In the left column, the associated sampling rate is shown. On the right side, errors occurring at certain steps of the phase measurement are marked.

To simplify the calculation, we use a normalized dimensionless phase,

$$p(t) \equiv \frac{\phi(t)}{\pi}, \tag{S3}$$

as follows. For the first step, we define $\tau_j$ $(1 \leq j)$ as the timing of the zero-crossings of the input signal. By definition, $p(\tau_j) = j$ because $\sin\left(p(\tau_j)\right) = 0$ at every zero-crossing. In addition, we use Equation (S4) for simplicity,

$$p_0 \equiv p(0), \tag{S4}$$

where $p_0$ is an initial value of $p(t)$. Although the actual initial phase is in the range of $-1 \leq p_0 < 1$, for clarity, we assume that $0 \leq p_0 < 1$ without losing any generality. This is justified if we add the initial phase offset of $+\pi$ (for normalized phase $p(t)$, $+1$) when the actual initial phase is below zero. If the initial phase offset applied, the result should be compensated with this offset as follows:

$$\Phi[z] \to \Phi[z] - \pi. \tag{S5}$$

Next, we define a phase estimation function $q(t)$ that approximates $p(t)$ by using quantization when of zero-crossings occur, as given by Equation (S6),

$$q(t) \equiv \frac{1}{2}, \quad (0 \leq t < \tau_1)$$
$$\equiv 1 + \frac{1}{2}, (\tau_1 \leq t < \tau_2)$$
$$\vdots$$
$$\equiv j + \frac{1}{2} \; (\tau_j \leq t < \tau_{j+1}); \tag{S6}$$

$q(t)$ is shown as the thin line in Figure S2.

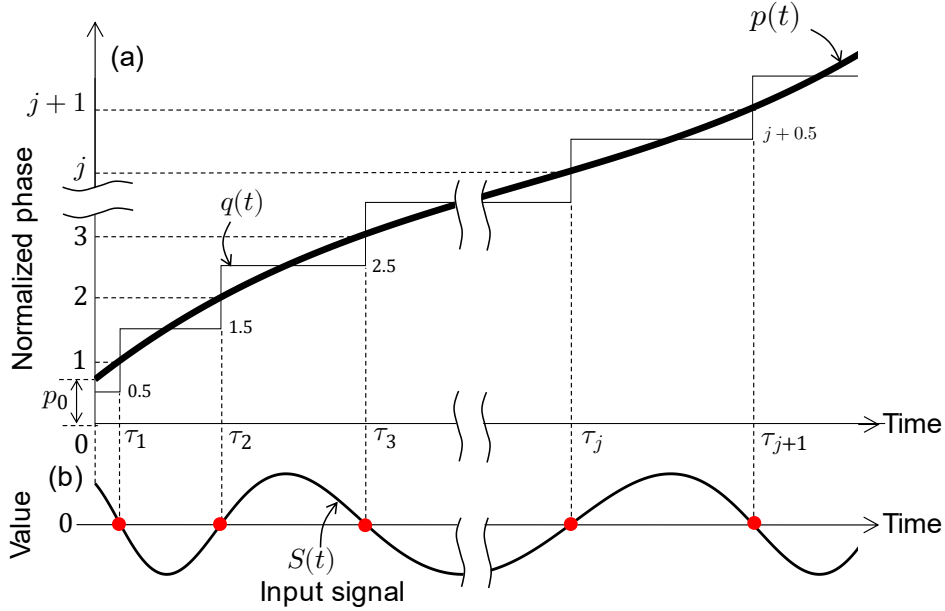

Figure S2. (a) Normalized phase evolution of the input signal and (b) corresponding input signal. Filled circles show the positions of zero-crossings.

This approximation function uses zero-crossings as markers to indicate when the input signal reaches an integer multiple of π. It is justified because the integration of $q(t)$ is equal to the trapezoidal approximation of $p(t)$ with two points $(\tau_j, j)$ and $(\tau_{j+1}, j+1)$ as follows:

$$\int_{\tau_j}^{\tau_{j+1}} p(t)dt \cong \frac{1}{2}(\tau_{j+1} - \tau_j)(j + j + 1) = \left(j + \frac{1}{2}\right)(\tau_{j+1} - \tau_j) = \int_{\tau_j}^{\tau_{j+1}} q(t)dt \tag{S7}$$

In the aforementioned transformation, we assume continuous time. However, for digitized signals, time should be discrete. Hereafter, we consider how $q(t)$ can be expressed by digitized data from the ADC. Let us consider the region around the j-th zero-crossing as shown in Figure S3.

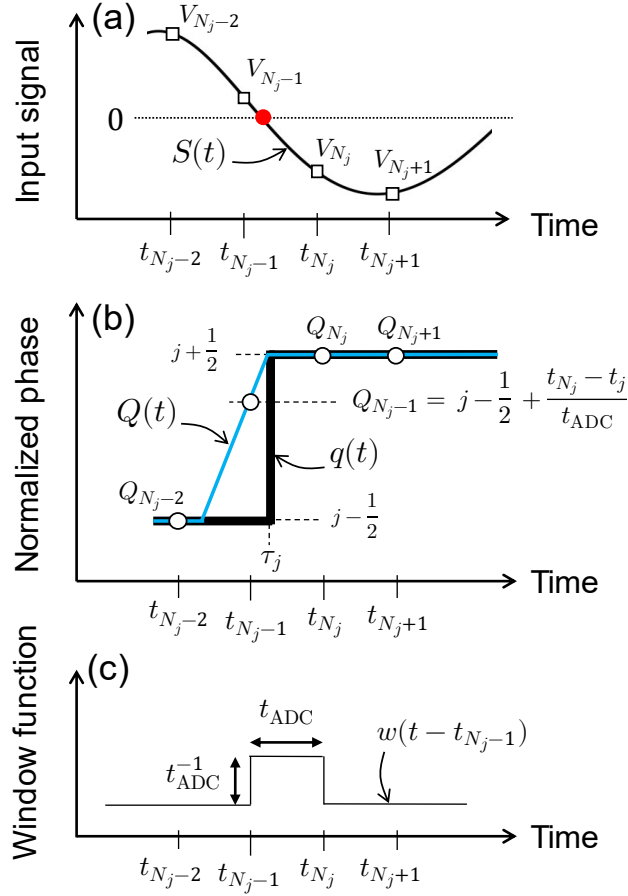

Figure S3. Around the j-th zero-crossing (a) input signal $S(t)$, digitized data $V_i$, and the zero-crossing timing (filled red circle); (b) the phase estimation function $q(t)$, window phase estimation function $Q(t)$, and instantaneous phase estimator $Q_i$; (c) example of the window function $w(t - t_{N_j-1})$.

Here, the sampling interval of the ADC is defined as $t_{\text{ADC}}$, i.e., sampling frequency $f_{\text{ADC}} = t_{\text{ADC}}^{-1}$, thus resulting in the ADC output value to be $V_i$, as shown in Figure S3(a). We also assume that at time $t = 0$, the first sample $V_1$ is obtained, and at time $t = (i-1)t_{\text{ADC}}$, the i-th sample is obtained:

$$V_i = S((i-1)t_{\text{ADC}}). \tag{S8}$$

The time at which the i-th sample is obtained is then defined as:

$$t_i = (i-1)t_{\text{ADC}}. \tag{S9}$$

We define $N_j$ to be the number of sampled data just after the zero-crossing. In other words, j-th zero-crossings occur between the data at $N_j - 1$ and at $N_j$ as also shown in Figure S3(a).

Then, $q(t)$ is the continuous phase estimation function shown in Figure S3(b). To consider how $q(t)$ can be expressed with discrete values, we carry out $q(t)$ sampling in two steps; (i) a window function is applied to $q(t)$, and then, (ii) the windowed function at the time interval $t_{\text{ADC}}$ is sampled. Note that the window function works as an anti-alias filter for sampling. For the first step (i), to sample $q(t)$, we apply a convolution with the window function according to Equation (S10),

$$Q(t) \equiv \int_0^\infty q(t')w(t' - t)dt'. \tag{S10}$$

The window function is defined as a rectangular window (see Figure S3(c)), as
$$w(t) \equiv \frac{1}{t_{\text{ADC}}} \left( U(t) - U(t - t_{\text{ADC}}) \right), \tag{S11}$$
where the step function $U(t)$ is given as:
$$\begin{aligned} U(t) &= 0, & (t < 0) \\ &= 1 & (t \geq 0). \end{aligned} \tag{S12}$$
Note that $w(t)$ is normalized by the factor $1/t_{\text{ADC}}$ to maintain unity gain:
$$\int_0^\infty w(t)\,dt = 1. \tag{S13}$$
The windowed phase estimation function $Q(t)$ is also a continuous function albeit with a much shallower slope than that of $q(t)$, as shown in Figure S3(b). For the second step (ii), $Q(t)$ is sampled according to Equation (S14).
$$Q_i = Q(t_i) \tag{S14}$$
Here, the $Q_i$ sequence represents the instantaneous phase estimator at a rate of $t_{\text{ADC}}$. Next, we consider how $Q_i$ can be generated from $V_i$. As shown in Figure S3(b), $Q_{N_j-2}$ and $Q_{N_j}$, which are not at time $t_{N_j-1}$, can be simply derived as follows:
$$Q_{N_j-2} = \int_{t_{N_j-2}}^{t_{N_j-1}} q(t')dt' = j - \frac{1}{2} \tag{S15}$$
$$Q_{N_j} = \int_{t_{N_j}}^{t_{N_j+1}} q(t')dt' = j + \frac{1}{2} \tag{S16}$$
When the sampling time is $t_{N_j-1}$,
$$Q_{N_j-1} = \int_{t_{N_j-1}}^{t_{N_j}} q(t')dt' = \left(j - \frac{1}{2}\right) + \frac{t_{N_j} - \tau_j}{t_{\text{ADC}}}. \tag{S17}$$
Equations (S15)–(S17) show that $Q_i$ has three components, the counting-up $C_i$, the additional $F_i$ that appears only at $t_{N_j-1}$, and a constant $C_0$:
$$Q_i = C_i + F_i + C_0. \tag{S18}$$
To explicitly express them, a zero-crossing detected function is defined as follows.
$$\begin{aligned} \lambda_i &\equiv 1, \text{if } \text{sgn}_{\text{M}}(V_{i-1})\text{sgn}_{\text{M}}(V_i) = -1 \\ &\equiv 0, \text{if } \text{sgn}_{\text{M}}(V_{i-1})\text{sgn}_{\text{M}}(V_i) = +1. \end{aligned} \tag{S19}$$

Here, $\text{sgn}_{\text{M}}(x)$ is a modified sign function, which is defined as
$$\begin{aligned} \text{sgn}_{\text{M}}(x) &\equiv -1, \text{if } x < 0 \\ &\equiv +1, \text{if } x \geq 0. \end{aligned} \tag{S20}$$
The difference from the usual sign function is $\text{sgn}_{\text{M}}(0) = +1$, while $\text{sgn}(0) = 0$. In a practical viewpoint, $\lambda_i$ can also be defined as
$$\lambda_i \equiv \text{XOR}\big(b_{2\text{C}}(V_{i-1}) b_{2\text{C}}(V_i)\big), \tag{S21}$$
where $b_{2\text{C}}(x)$ means the most significant bit of $x$ when two's complement is used to represent the digitized value. Using $\lambda_i$, the counting-up $C_i$ is given as,

$$C_i \equiv \sum_{k=1}^{i} \lambda_k, \tag{S22}$$

where $C_k$ is the counter started at the beginning of the measurement. We call $C_i$ the counting value. As for the additional value $F_i$, there is no exact information on the actual timing of the zero-crossing $\tau_j$ because only digitized data are available. Therefore, we use linear interpolation instead, as shown in Figure S4.

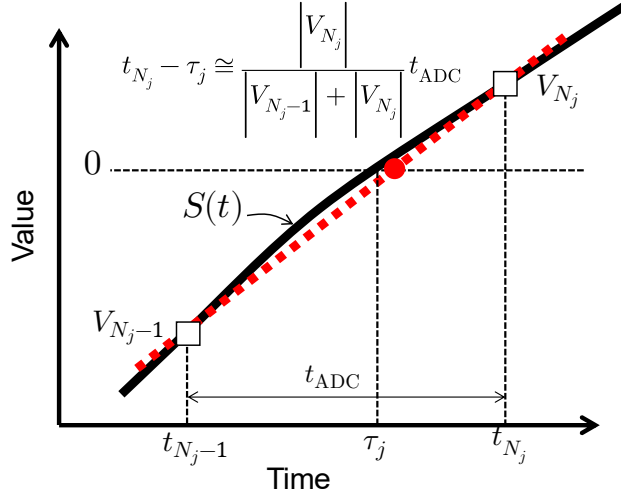

Figure S4. Linear interpolation of the timing of the j-th zero-crossings. The two white squares show the sampling timing. The thick line is the actual signal $S(t)$, and the thick red dotted line between the two squares indicates the linear interpolation.

In this interpolation, the actual zero-crossing timing $\tau_j$ can be approximated as shown in Figure S4:

$$\tau_j \cong t_{N_j} - \frac{|V_{N_j}|}{|V_{N_j-1}| + |V_{N_j}|} t_{\text{ADC}}. \tag{S23}$$

Then,

$$\frac{t_{N_j} - \tau_j}{t_{\text{ADC}}} \cong \frac{|V_{N_j}|}{|V_{N_j-1}| + |V_{N_j}|}. \tag{S24}$$

This approximation has error, which is discussed later in Section II.E. To generalize it, $F_i$ can be defined at every sampling timing including the positions of each zero crossing, as given by Equation (S25):

$$F_i \equiv \left( \frac{|V_{i+1}|}{|V_i| + |V_{i+1}|} \right) \lambda_{i+1}, \tag{S25}$$

which we call the fraction value. Note that $\lambda_{i+1}$ is used in this definition instead of $\lambda_i$, so that $F_i$ is assigned to the position $t_{N_j-1}$. Note that although Figure S4 only shows rising zero crossing, the same result is derived for falling zero crossings. For the third component, the constant in Equation (S18), we include the compensation term from Equation (S5), which is $-1$ depending on the initial phase of the input signal. The compensation can be calculated easily by using the first value $V_1$; therefore, the constant $C_0$ should be:

$$C_0 = -\frac{\pi}{2}, \quad \text{if } V_1 \leq 0$$
$$= +\frac{\pi}{2}, \quad \text{if } V_1 > 0. \tag{S26}$$

Note that the instantaneous phase estimator $Q_i$ can be used as the phase information sampled with frequency $f_{\text{ADC}}$. As expressed in Equation (S2), we seek a measured phase with a given constant interval time of $T$, $\Phi[z]$. To obtain it, we use a boxcar filter (or moving average and decimation) $B_i$ for simplicity.

$$B_i[z] \equiv \frac{1}{N}, \quad (1 + (z-1)N < i < zN)$$
$$\equiv 0, \quad (\text{others}) \tag{S27}$$

Then, we derive

$$\Phi[z] \equiv \pi \sum_{i=1}^{\infty} Q_i B_i[z]$$
$$= \left( \frac{\pi}{N} \sum_{i=1+(z-1)N}^{zN} (C_i + F_i) \right) + C_0, \tag{S28}$$

which appears as Equation (5) in the main text.

Here, the averaged phase estimator expresses the average of the phase over the time interval $Nt_{\text{ADC}}$. Therefore, the corresponding time stamp for $\Phi[z]$ should be as follows:

$$t[z] = \frac{1}{Nt_{\text{ADC}}} \int_{(z-1)Nt_{\text{ADC}}}^{zNt_{\text{ADC}}} t\,dt$$
$$= \left( z - \frac{1}{2} \right) Nt_{\text{ADC}}. \tag{S29}$$

Equation (S29) means that the time stamp should not be the beginning of the time interval, $(z-1)Nt_{\text{ADC}}$, nor should it be at the end of the time interval, $zNt_{\text{ADC}}$.

### 2. Response to perturbations

For error analysis and noise analysis later, we consider perturbation of the zero-crossing timing here. Based on Equations (S24) and (S25), when the zero-crossing timing $\tau_j$ is perturbated as $\tau_j \to \tau_j + \delta\tau_j$, the fraction value $F_{N_j}$ decreases like $F_{N_j} \to F_{N_j} - \delta\tau_j/t_{\text{ADC}}$. Thus, perturbation of the averaged phase estimator $\delta\Phi$ is expressed as

$$\delta\Phi = \frac{\pi}{N} \sum_{j}^{L} \delta F_{N_j}$$
$$= -\frac{\pi}{T} \sum_{j}^{L} \delta\tau_j, \tag{S30}$$

where $L$ is the number of zero-crossings in the range of time $T$. This relation (Equation (S30)) means that the perturbation of the measured phase $\delta\Phi$ can be expressed with measurement time $T$ and the zero-crossing timing perturbations. We consider one specific case as an example. Assume the input signal is sinusoidal, $S(t) = \sin 2\pi f_{\text{sig}} t$, and the phase is perturbed by $\delta\Phi$. Then, considering that the number of zero-crossings is determined as $L = 2f_{\text{sig}}T$, the estimated phase shift from Equation (S30) becomes $\delta\Phi = -2\pi f_{\text{sig}}\delta\tau$. On the other hand, the actual perturbation of the zero-crossing timing is by definition $\delta\tau = -\delta\Phi/2\pi f_{\text{sig}}$, which agrees with the result from Equation (S30).

## I.B. Calculation example

For the reader's convenience, we present an example of the algorithm being used in a calculation. First, consider an input signal as follows:

$$S(t) = \sin(2\pi f_{\text{sig}} t + \phi_0), \tag{S31}$$

where $t$, $f_{\text{sig}}$, and $\phi_0$ are time in ns, signal frequency in GHz, and initial phase in rad, respectively. As an example, we choose $f_{\text{sig}} = 0.11$ (GHz) and $\phi_{\text{ofs}} = 0.6$ (rad). Therefore, the true phase is expressed as

$$\phi(t) = 0.22\pi t + 0.6. \tag{S32}$$

In Figure S5, the example values are shown for $0 \leq t \leq 30$. Note that the first sample $V_1$ is obtained at time $t_1 = 0$, and the N-th sample is obtained at the time $t_N = (N-1)$.

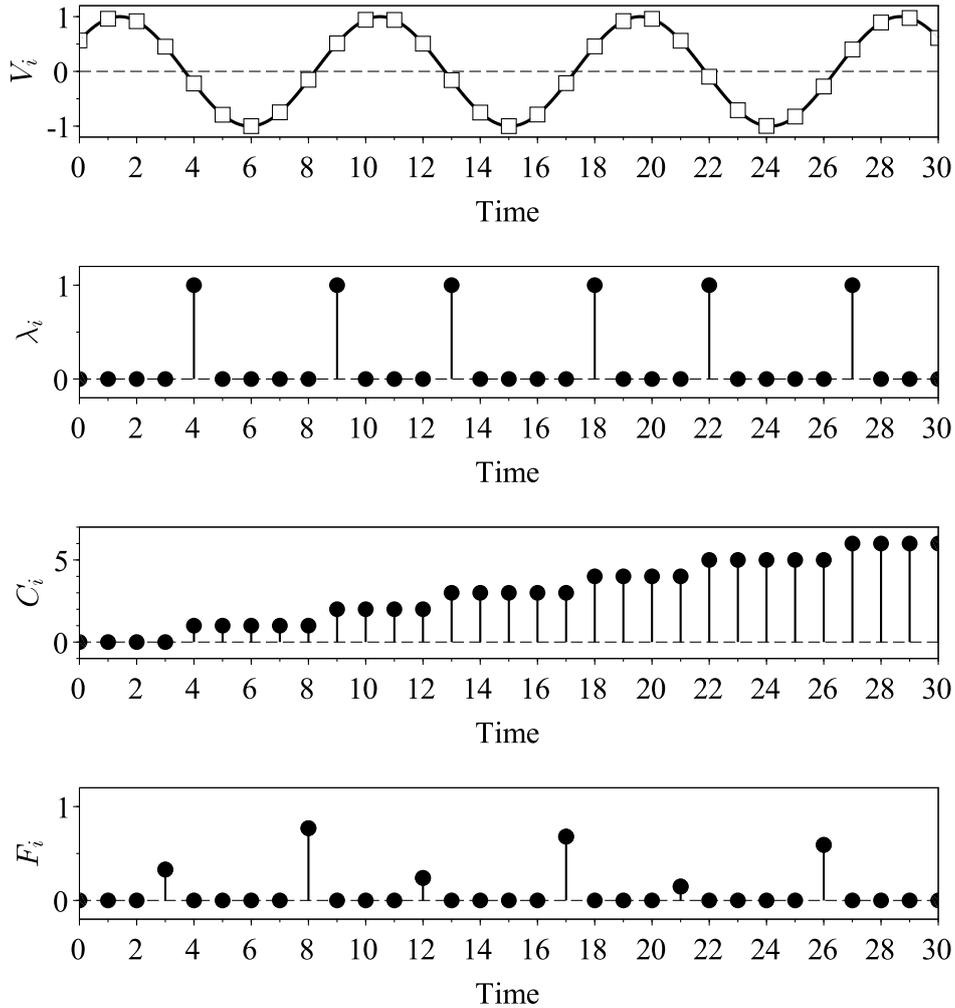

Figure S5. An example of phase calculation with the algorithm: (a) input signal (line) and digitized data (square); (b) the zero-crossing detected function $\lambda_i$; (c) counter value $C_i$; (d) fraction value $F_i$.

Then, the zero-crossing detected function $\lambda_i$ is calculated, as shown in Figure S5(b), and the counter value is obtained, as shown in Figure S5(c), according to Equation (S22). The fraction value $F_i$ should be calculated according to the definition in Equation (S25); the resulting $F_i$ are plotted in Figure S5(d). For reference, the data can be expressed as follows:

$$\begin{pmatrix} t_i \\ V_i \\ C_i \\ F_i \end{pmatrix} = \begin{pmatrix} 0 & 1 & 2 & 3 & 4 & 5 & \cdots \\ 0.565 & 0.961 & 0.917 & 0.451 & -0.221 & -0.792 & \cdots \\ 0 & 0 & 0 & 0 & 1 & 1 & \cdots \\ 0 & 0 & 0 & 0.329 & 0 & 0 & \cdots \end{pmatrix}. \tag{S33}$$

The average factor $N$ can be chosen so that the phase can be obtained at the desired sample rate. For example, here $N = 10$. By applying Equation (S26), the correction factor becomes $R = +\pi/2$, because $V_1 \geq 0$ in this case. Then, the phase result is obtained from Equation (S28):

$$\Phi[1] = \left(\frac{\pi}{10}\sum_{i=1}^{10}(C_i + F_i)\right) + \frac{\pi}{2},$$
$$\vdots$$
$$\Phi[z] = \left(\frac{\pi}{10}\sum_{i=10z-9}^{10z}(C_i + F_i)\right) + \frac{\pi}{2}. \tag{S34}$$

Since the beginning of the time range is $t = 10z - 10$, and the beginning of the next time range is $t = 10z$, the range of the index is $10z - 9 \leq i \leq 10z$, and the corresponding time range should be $10z - 10 \leq t < 10z$. The corresponding time stamp according to Equation (S29) is:

$$t[1] = 5,$$
$$\vdots \tag{S35}$$
$$t[z] = 5z.$$

Then, the phase and corresponding time stamp should be as follows:

$$\begin{pmatrix} \Phi[1] & \Phi[2] & \Phi[3] & \cdots \\ t[1] & t[2] & t[3] & \cdots \end{pmatrix} = \begin{pmatrix} 4.115 & 10.970 & 17.825 & \cdots \\ 5 & 10 & 15 & \cdots \end{pmatrix}. \tag{S36}$$

On the other hand, integrating Equation (S32) over the time range $10z - 10 < t < 10z$ results in the theoretical value $\overline{\Phi}[z]$,

$$\overline{\Phi}[z] = \frac{1}{10}\int_{t=10z-10}^{t=10z} \phi(t)dt \tag{S37}$$
$$= 2.2\pi z + (0.6 - 1.1\pi),$$

for the same time stamp. Then,

$$\begin{pmatrix} \overline{\Phi}[1] & \overline{\Phi}[2] & \overline{\Phi}[3] & \cdots \\ t[1] & t[2] & t[3] & \cdots \end{pmatrix} = \begin{pmatrix} 4.056 & 10.967 & 17.879 & \cdots \\ 5 & 10 & 15 & \cdots \end{pmatrix}. \tag{S38}$$

Therefore, the estimation error $\Delta\Phi[z] \equiv \Phi[z] - \overline{\Phi}[z]$ for this example is

$$\Delta\overline{\Phi}[z] = \begin{pmatrix} 0.059 & 0.003 & -0.053 & \cdots \end{pmatrix}. \tag{S39}$$

For comparison, we also estimate the phase using conventional I/Q (in-phase/quadrature signals) demodulation:

$$\Phi_{\text{Dem}}[z] = \arctan\left(\frac{\sum_{i=10z-9}^{10z} V_i \sin(2\pi f t_i)}{\sum_{i=10z-9}^{10z} V_i \cos(2\pi f t_i)}\right) + 10(z-1)2\pi f \tag{S40}$$

$$= (4.026 \quad 10.965 \quad 17.977 \quad \cdots)$$

Then,

$$\Phi_{\text{Dem}}[z] - \overline{\Phi}[z] = (-0.030 \quad -0.002 \quad 0.098 \quad \cdots), \tag{S41}$$

thus showing that the error in the phase measured by both methods, i.e., using Equation (S36) and Equation (S38), are comparable. Note that the I/Q demodulation uses a known signal frequency ($f = 0.11$). Generally, in an I/Q demodulation process, the signal frequency should be known roughly within the measurement bandwidth $f_{\text{MBW}} \sim (2Nt_{\text{ADC}})^{-1}$ prior to calculation. On the other hand, the phase measuring algorithm does not require any prior information about the input signal frequency.

We present an example of our implementation. Figure S6(a) shows the test setup. We use test signals from the function generator (Keysight 33622A) with two phase-locked outputs. A 100-MHz sinusoidal signal with phase modulation (5 kHz, 180 deg$_{0\text{-p}}$, sinusoidal) is the input for Ch. A of the phase meter. For Ch. B of the phase meter, we use a 100-MHz sinusoidal signal without phase modulation. The signal power for both channels is +7 dBm. We set three test points in this example. One is the raw signal from the ADCs (point $\alpha$ in Figure S6(b)), and the second is the phase difference (point $\beta$ in Figure S6(b)), which is a 250-MHz sampling rate. The third is picked up after LPF in the FPGA (point $\gamma$ in Figure S6(b)). The LPF is a Blackman-type, and its 3-dB cutoff frequency is 16.667 MHz.

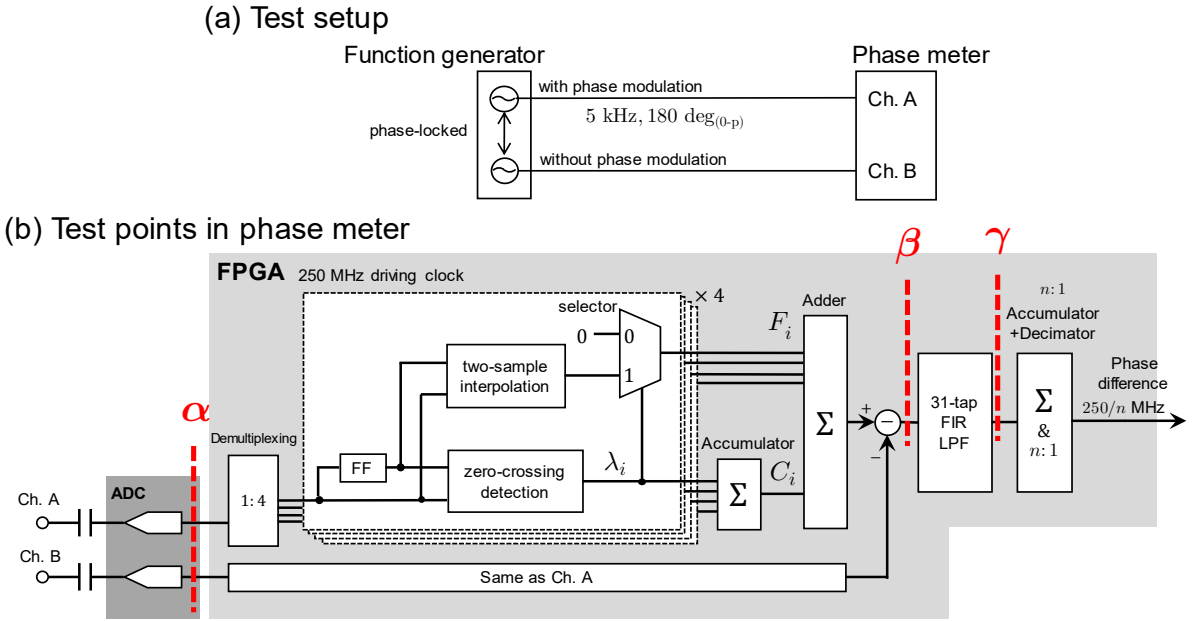

Figure S6. (a) Test setup, (b) Test points in the phase meter. Three test points ($\alpha, \beta, \gamma$) correspond to data in Figure S7.

Figure S7 presents the results. Raw data from the ADC is shown only in $2 \times 10^{-7}$ seconds; therefore, the phase difference is almost constant in this time window. Figure S7(b) shows the phase difference before and after LPF. Here, for clarity, the data after LPF is shifted backward to compensate for the group delay of the filter, equaling $16/(2.5 \times 10^8) = 6.4 \times 10^{-8}$ seconds.

Before LPF, the phase estimator with a 250-MHz sampling rate has a large dispersion of $\sim\pm 1$ rad (see also the inset of Figure S7(b)), which is reasonable. The sampling rate of the phase estimator (250 MHz) is larger than the rate of zero-crossings (200 MHz), which is twice the signal frequency. In terms of power spectral density (PSD), this dispersion corresponds to large peaks in the high-frequency region, which is above 10 MHz in this example (the black curve in Figure S7(c)). These high-frequency peaks in Figure S7(c) originate from an error called the quantization-aliasing error, which is described in Section II.D. We retrieve the accurate phase differences by eliminating the high-frequency components with the LPF (the red line in Figure S7(b)). The red curve excellently matches with the given 5-kHz, sinusoidal, and $\pm\pi$-rad phase modulation. Moreover, the spectral shape between 10 kHz and 10 MHz is flat, i.e., white phase noise, reflecting that the noise level is uncontaminated by spurious.

Here, we would like to state that the $n\colon 1$ accumulator and decimator (the last most right part of FPGA in Figure S6(b)), is not an optimal selection because the boxcar filter does not have an optimally steep response in its rejection band. In other words, to achieve the best rejection performance for the high-frequency components, we should apply other processing such as the cascaded digital LPF, instead of the $n\colon 1$ accumulator and decimator.

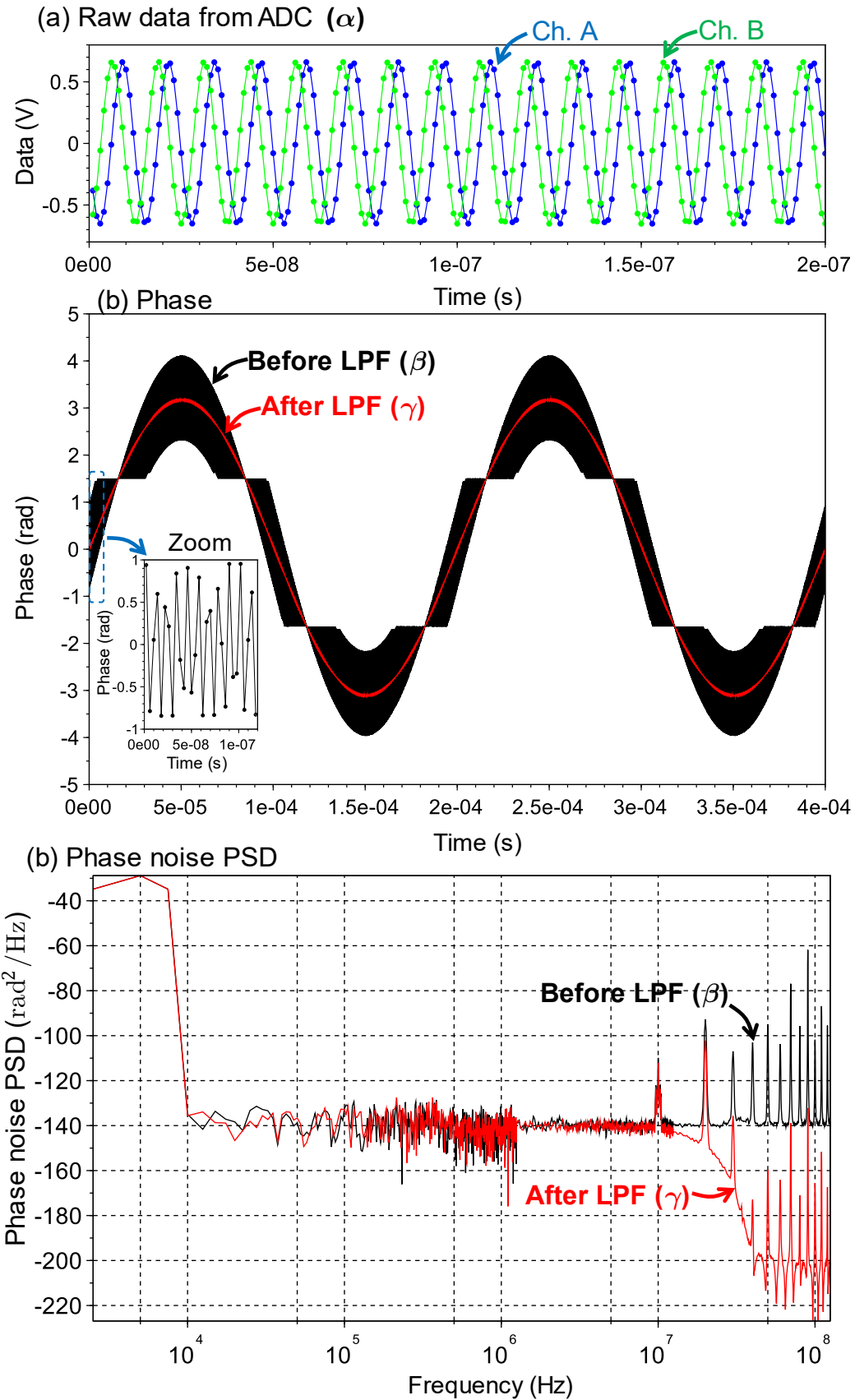

Figure S7. Results for the test setup shown in Figure S6. (a) Raw data from the two ADCs (b) Phase difference before and after LPF (c) Phase noise power spectral density (PSD) for the two-phase differences

## II. Error analysis

Generally, an error refers to the difference between what is measured and what the actual value is. In this section, we analyze what type of error components is present in the phase measuring algorithm. We also estimate how large the components are and when the errors arise in the measurement.

### II.A. Overview: classification of errors

In terms the phase measuring algorithm error, four sources of error appear in the phase measurement principle, which are listed in Figure S1. In the Subsections II.B–II.D, we quantitatively analyze components of the following errors in detail.

1. Trapezoidal approximation error (TA error). This error comes from the trapezoidal approximation in the phase estimation process. When the actual (normalized) phase $p(t)$ is approximated with $q(t)$ in Equation (S6), trapezoidal approximation is applied at zero-crossing timing. The approximation process is justified because the integration of $q(t)$ is equal to the trapezoidal integration in Equation (S7). This approximation has no error when the input signal exhibits a constant frequency; the phase increases exactly linearly. However, when the frequency of the input signal changes over time, the approximation has error. In the Subsection II.B, we evaluate that this error is proportional to $\dot{f}_{\text{sig}} f_{\text{sig}}^{-2}$, where $f_{\text{sig}}$ is the signal frequency, and $\dot{f}_{\text{sig}}$ denotes frequency drift rate.

2. First zero-crossing error (FZ error). Before the first zero-crossing, the approximation is slightly different from the trapezoidal approximation after the second zero-crossing (Figure S2). This difference results in changes of the integration of $p(t)$ and $q(t)$, thus leading to error (in addition to TA error), which we call first zero-crossing error. In Subsection II.C, we analyze the effect of this error and prove that it effectively has no influence on the measurement.

3. Quantization–aliasing error (QA error). This error comes from a coupling of quantization error and aliasing. The phase estimation function $q(t)$ has a quantization error, which occurs at a period twice the signal frequency, i.e., the rate of zero-crossings. Here the actual phase is defined as a continuous variable, whereas the data rate from the ADC is limited to a finite number, $f_{\text{ADC}}$. Therefore, the instantaneous phase estimator $Q_i$ is obtained by sampling, shown in Figure S1; aliasing peaks may occur in low-frequency components because the quantization error has a spectrum spreading to a frequency higher than $f_{\text{ADC}}$. We call the error associated with these aliasing peaks quantization–aliasing (QA) error.

Since QA error comes from sampling at the ADC and cannot be removed with a low-pass filter (anti-aliasing filter), it is a fundamental systematic error of the phase measuring algorithm. This error is analyzed in Subsection II.D, and we show that it appears when the signal frequency is around specific frequencies. Additionally, we prove that the error, in practice, does not play a role when the measurement bandwidth is appropriately limited.

4. Zero-crossing interpolation error (ZI error). ZI error comes from interpolation, i.e., a coupling between the sampling timing and nonlinearity of the input signal. Timing of the zero-crossings is approximated with linear interpolation according to Equation (S23), which causes timing estimation errors. As shown in section I, those timing errors lead to the phase error.

We analyze ZI error as follows. First, we introduce a zero-crossing interpolation error function (ZIEF) that expresses how much error occurs at a certain zero-crossing timing. The ZIEF becomes large when the input signal around the zero-crossing is highly nonlinear and/or when the input signal frequency is high. By using the ZIEF, the error induced by the zero-crossing interpolation can be expressed as sampling of the ZIEF at the frequency of zero-crossings, which we refer to as virtual sampling. Next, we use frequency-domain analysis, similar to that used to analyze QA error, to convert it to the error of the instantaneous phase estimator. The error obtained from virtual sampling is converted to a sum of delta functions, upon which a window function is applied; then, the sum is resampled at $f_{\text{ADC}}$. In the resampling, ZI error becomes present in low-offset frequencies of the measured phase due to aliasing, which is why ZI error occurs around specific frequencies, called singular frequencies. Note that this is the same for QA error. ZI error also consists of a fundamental frequency and its higher order harmonics. However, the peaks have −60 dB/dec frequency dependence; those of QA error have −20 dB/dec slope.

We must be careful because ZI error appears in the DC (0 Hz) band when the input signal frequency is the same as singular frequencies. In other words, ZI error may lead to a secular phase shift in the measurement at singular frequencies. In Subsection II.E, we analyze this error and how much it appears in the measurement.

### II.B. Trapezoidal approximation (TA) error

TA error is an error caused by trapezoidal approximation that is used in the derivation of the measurement principle shown in Section I. In particular, phase quantization in Equation (S7) causes errors on the phase estimation function $q(t)$ from the normalized phase $p(t)$. The approximation has no error when the phase $p(t)$ increases linearly, which means that the input signal frequency is constant. When the signal has a frequency drift, i.e., $\frac{d^2}{dt^2} p(t) \neq 0$, the error becomes nonzero.

Hereafter, we estimate the amount of error. Consider the integration of $p(t)$ in the range of $\tau_j \leq t < \tau_{j+1}$,

$$\int_{\tau_j}^{\tau_{j+1}} p(t) dt. \tag{S42}$$

Here, $p(\tau_j) = j$ and $p(\tau_{j+1}) = j+1$ by definition. To estimate approximate error $\delta p$, we carry out a second-order expansion of $p(t)$:

$$p(t) = \left( j + \frac{t - \tau_j}{\tau_{j+1} - \tau_j} \right) + C_2 (t - \tau_j)(t - \tau_{j+1}). \tag{S43}$$

where $C_2$ is a coefficient for the second order that we assume is constant. Here, the first term represents the applied approximation. We consider the input signal frequency $f_{\text{sig}}$ in the time range of $\tau_j \leq t < \tau_{j+1}$,

$$f_{\text{sig}} = \frac{1}{2\pi}\frac{d\phi}{dt} = \frac{1}{2}\frac{dp}{dt}. \tag{S44}$$

Taking the second derivative of Equation (S43), we derive $C_2$ as follows:

$$C_2 = \frac{1}{2}\frac{d^2 p}{dt^2} = \dot{f}_{\text{sig}}, \tag{S45}$$

and by considering the relation $p(\tau_j) = j$ and $p(\tau_{j+1}) = j+1$, $f_{\text{sig}}$ can also be approximated to

$$f_{\text{sig}} \simeq \frac{1}{2}\frac{p(\tau_{j+1}) - p(\tau_j)}{\tau_{j+1} - \tau_j} = \frac{1}{2(\tau_{j+1} - \tau_j)}. \tag{S46}$$

Taking the second order term of Equation (S43), the approximation error per unit time is given by:

$$\frac{1}{\tau_{j+1} - \tau_j}\int_{\tau_j}^{\tau_{j+1}} C_2(t-\tau_j)(t-\tau_{j+1})dt = -\frac{C_2}{6}(\tau_{j+1}-\tau_j)^2 \simeq -\frac{1}{24}\frac{\dot{f}_{\text{sig}}}{f_{\text{sig}}^2}. \tag{S47}$$

Therefore, TA error $\delta\Phi_{\text{TA}}$ for the certain measurement time from $t_{\text{start}}$ to $t_{\text{end}}$ can be roughly estimated, by integrating Equation (S47),

$$\begin{aligned}\delta\Phi_{\text{TA}} = \pi\delta p &\simeq \frac{\pi}{t_{\text{end}} - t_{\text{start}}}\int_{t_{\text{start}}}^{t_{\text{end}}}\left(-\frac{1}{24}\frac{\dot{f}_{\text{sig}}}{f_{\text{sig}}^2}\right)dt \\ &= \frac{\pi}{24}\left(\frac{f_{\text{sig(end)}}^{-1} - f_{\text{sig(start)}}^{-1}}{t_{\text{end}} - t_{\text{start}}}\right),\end{aligned} \tag{S48}$$

which means the measured phase error $\delta\Phi_{\text{TA}}$ is proportional to the mean change rate of the period of the input signal. When the frequency drift rate is sufficiently low and constant, the error can be expressed as follows:

$$\delta\Phi_{\text{TA}} \sim -\frac{\pi}{24}\frac{\dot{f}_{\text{sig}}}{f_{\text{sig}}^2}. \tag{S49}$$

For example, when $f_{\text{sig}} = 10$ MHz and $\dot{f}_{\text{sig}} = 1$ kHz/s, this error contributes by $\delta\Phi \sim -1.3 \times 10^{-12}$ rad, which is sufficiently small compared to other measurement errors and noise.

This error has the worst value when the period of the input signal changes greatly over time. For example, when $f_{\text{sig(end)}} = 10$ MHz, $f_{\text{sig(end)}} = 20$ MHz, and $t_{\text{end}} - t_{\text{start}} = 1$ μs, i.e., mean frequency drift rate is $10^{13}$ Hz/s, the phase error for 1-μs sampling is approximately:

$$\delta\Phi_{\text{TA}} \simeq -\frac{\pi}{540} \sim -0.006 \text{ (rad)}. \tag{S50}$$

Note that in 1 μs sampling, the signal creates roughly 30 zero-crossings, and the averaged phase $\Phi$ becomes $\sim 15\pi$, which is about 8000 times larger than the TA error $\delta\Phi_{\text{TA}}$.

## II.C. First zero-crossing (FZ) error

FZ error is an error occurring only in the first measured phase, $\Phi[1]$. As shown in Figure S2, the actual phase of the input signal is approximated with the phase estimation function $q(t)$. After the first zero-crossing, $\tau_1 \leq t$, the phase estimation function expresses TA, as described in Equation (S7). However, the approximation is slightly different before the first zero-crossing, $0 \leq t < \tau_1$. Figure S8 shows this time range, which is extracted from Figure S2.

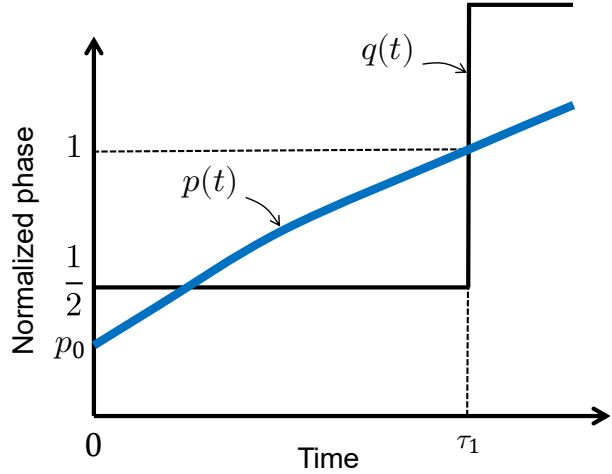

Figure S8. The actual normalized phase $p(t)$ and estimation function $q(t)$ before the first zero-crossing.

For TA, the integration of $p(t)$ should be performed as follows:
$$\int_0^{\tau_1} p(t)\, dt \cong \frac{1}{2}(1 + p_0)\tau_1. \tag{S51}$$
However, the phase estimation function is
$$\int_0^{\tau_1} q(t)\, dt = \frac{1}{2}\tau_1. \tag{S52}$$
Equation (S51) and Equation (S52) are different because the initial phase cannot be known based on data before the first zero-crossing only. The error contribution $\delta p[1]$ is
$$\delta p[1] = \int_0^{\tau_1} \left(q(t) - p(t)\right) dt \cong \frac{1}{2} p_0 \tau_1, \tag{S53}$$
though this depends on the initial phase itself; assuming constant frequency, the initial phase can be expressed as:
$$\tau_1 \sim \frac{1}{2 f_{\text{sig}}}(1 - p_0), \tag{S54}$$
where $f_{\text{sig}}$ is the frequency of the input signal. Then,
$$\delta p[1] \sim \frac{1}{4 f_{\text{sig}}} p_0 (1 - p_0). \tag{S55}$$
Considering that the initial phase $p_0$ moves in the range of $0 \leq p_0 < 1$, in the worst case, the error occurs when $p_0 = 1/2$:

$$(\delta p[1])_{\text{worst}} \sim \frac{1}{16 f_{\text{sig}}}. \tag{S56}$$

The first phase result, $\Phi[1]$, is the average of the instantaneous phase over $0 \leq t < T$. Therefore, the error in $\Phi[1]$ for the worst case is

$$(\delta \Phi_{\text{FZ}}[1])_{\text{worst}} \sim \frac{\pi}{16 f_{\text{sig}} T}. \tag{S57}$$

On the other hand, the approximate value of $\Phi[1]$ is expressed by using signal frequency $f_{\text{sig}}$,

$$\Phi[1] \sim \frac{\pi}{T} \int_0^T p(t)\, dt \cong \frac{\pi}{T} \int_0^T 2 f_{\text{sig}} T\, dt = \pi f_{\text{sig}} T. \tag{S58}$$

We can express the mean number of zero-crossings $n$ in the time range of $0 \leq t < T$ as

$$n \sim 2 f_{\text{sig}} T. \tag{S59}$$

Finally, we derive

$$\left(\frac{\delta \Phi_{\text{FZ}}[1]}{\Phi[1]}\right)_{\text{worst}} \sim \frac{1}{16 f_{\text{sig}}^2 T^2} \sim \frac{1}{4 n^2}. \tag{S60}$$

This evaluation indicates that FZ error causes, on average, less than ca. 1-% relative error when the interval time for the measured phase is 2.5 times the signal period, i.e., $T = 2.5/f_{\text{sig}}$, hence $n \sim 5$. Note that $\Phi[z]$ ($z \geq 2$) is not effected by FZ error. Substituting with measurement bandwidth $f_{\text{MBW}} = 1/(2T)$, this can also be expressed as

$$\left(\frac{\delta \Phi_{\text{FZ}}[1]}{\Phi[1]}\right)_{\text{worst}} \sim \left(\frac{f_{\text{MBW}}}{2 f_{\text{sig}}}\right)^2. \tag{S61}$$

### II.D. Quantization–aliasing (QA) error
#### 1. Error estimation using frequency-domain analysis

QA error is an intrinsic error in the phase measuring algorithm. Originally generated from the phase quantization process at zero-crossings, QA error is converted into the measurement bandwidth by aliasing owing to the ADC sampling.

Figure S9 shows an overview of how the error appears in the results. Here, we assume the input frequency is constant, i.e., $f_{\text{sig}}$, for simplicity. The frequency-domain analysis is conducted as follows: (i) the error induced by the quantization process is expressed as a continuous sawtooth wave in the time domain (Figure S9(a)), (ii) by taking a Fourier transformation of the error, the frequency-domain representation of the error becomes a fundamental frequency and its higher order harmonics (Figure S9(b)), (iii) a window function $w(t)$ (Figure S9(c)) is applied as in Equation (S10) to obtain a windowed error function (Figure S9(d)), (iv) the windowed error function is sampled at $f_{\text{ADC}}$. In this process, QA error may appear in low-frequency components due to aliasing when the input signal frequency is near specific frequencies (Figure S9(e)), and (v) the instantaneous phase is filtered (low-pass) and decimated to obtain the phase, resulting in limited bandwidth from $f_{\text{ADC}}/2$ to $f_{\text{MBW}}$, which we define as the measurement bandwidth. Similarly, the error is also limited its bandwidth (Figure S9(f)). Here, $f_{\text{MBW}} = f_{\text{ADC}}/2N$, and we assume that the measurement bandwidth is sufficiently lower than the signal frequency, i.e., $f_{\text{MBW}} \ll f_{\text{sig}}$.

Practical errors are also induced by the filtering and decimation process, which is shown in Figure S9(e) to Figure S9(f). In the previous paragraph, we assume the perfect attenuation for anti-aliasing filtering before sampling at $2f_{\text{MBW}}$. However, in practice, the stopband attenuation may not be perfect, and it may cause an additional aliasing peak in the final phase, which is not considered here.

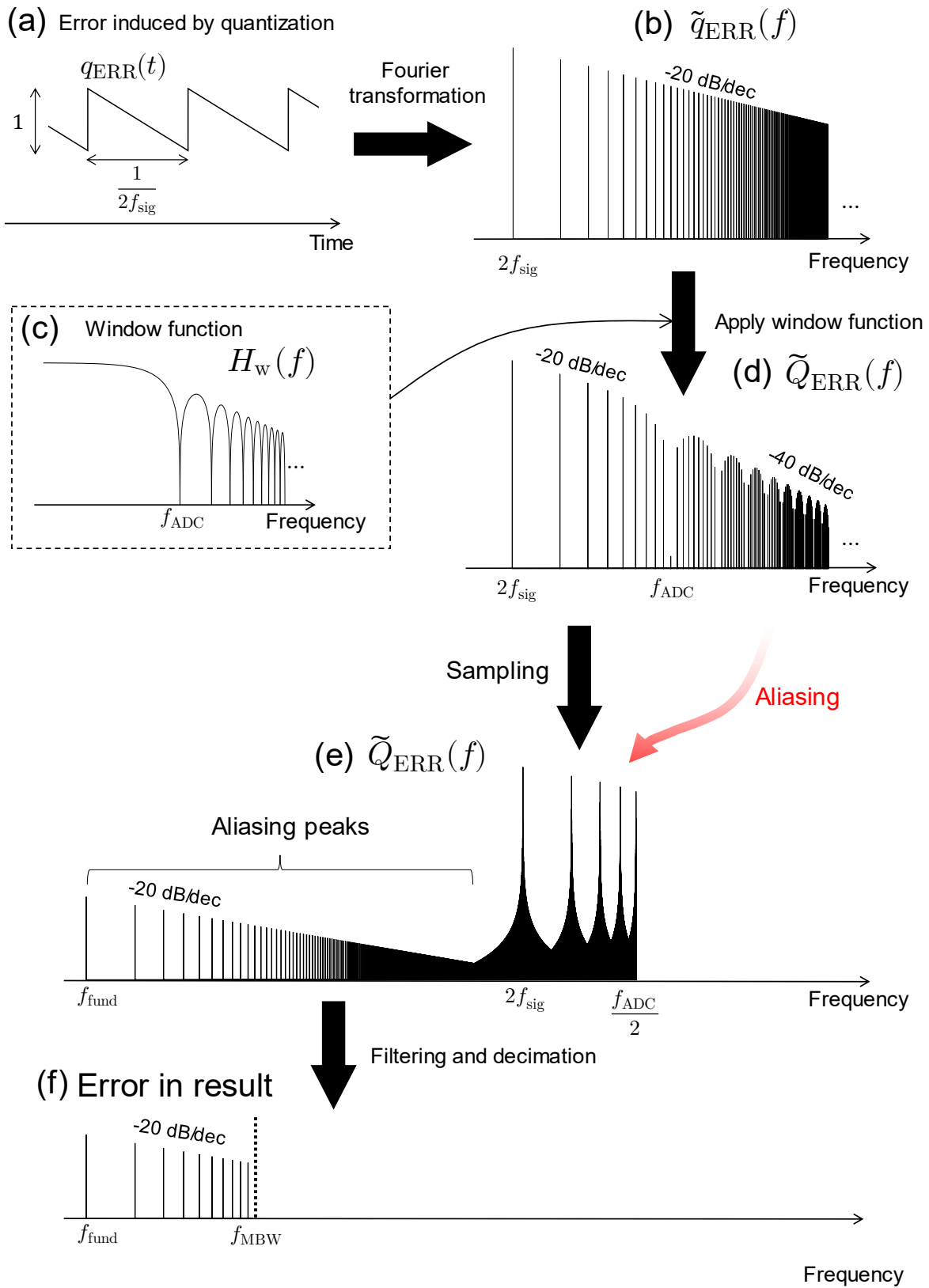

Figure S9. Overview of how quantization–aliasing error is introduced into the measurement: (a) error of phase estimation function $q_{\mathrm{ERR}}(t)$ in the time domain; (b) frequency-domain expression of $q_{\mathrm{ERR}}(t)$; (c) transfer function $H_w(f)$ for the window function; (d) spectrum of the error in the windowed phase estimation function; (e) spectrum of the error in the instantaneous phase estimator $Q_i$; (f) spectrum of the error in the measured phase $\Phi[z]$. $f_{\mathrm{sig}}$, $f_{\mathrm{ADC}}$, $f_{\mathrm{MBW}}$, and $f_{\mathrm{fund}}$ denote the input signal frequency, the sampling frequency of the ADC, the measurement bandwidth, and the fundamental frequency of the error, respectively. For this plot, $f_{\mathrm{sig}} = 50.01$ MHz, $f_{\mathrm{ADC}} = 1$ GHz, and $f_{\mathrm{MBW}} = 2$ MHz; $f_{\mathrm{fund}}$ is calculated to 200 kHz. Except in (a), all scales are logarithmic.

The error, $q_{\text{ERR}}(t)$, is the difference between the phase estimation function $q(t)$ and the actual (normalized) phase $p(t)$. As shown in Figure S2, $q(t)$ is a step-wise function, and $p(t)$ is a smooth function. Therefore, the error $q_{\text{ERR}}(t)$ is expressed as a sawtooth wave in the time domain. Its peak-to-peak amplitude is 1, and the frequency is $2f_{\text{sig}}$ (Figure S9(a)). Considering the Fourier expansion of sawtooth waves, $q_{\text{ERR}}(t)$ can be expressed as follows:

$$q_{\text{ERR}}(t) = \frac{1}{\pi} \sum_{k=1}^{\infty} \frac{(-1)^k \sin(2\pi k 2 f_{\text{sig}} t)}{k}. \tag{S62}$$

Therefore, the frequency domain representation of $q_{\text{ERR}}(t)$, which we name $\tilde{q}_{ERR}(f)$, is a summation of the signals of $2f_{\text{sig}}$ and its higher order harmonics (Figure S9(b)):

$$\tilde{q}_{ERR}(f) = \frac{1}{\pi} \sum_{k=1}^{\infty} \frac{(-1)^k}{k} \delta(f - 2k f_{\text{sig}}). \tag{S63}$$

The peaks have $-20$-dB/dec power dependency because the amplitude of the k-th peak is inversely proportional to the integer $k$. Next, the window function in Equation (S11) is applied prior to sampling. Since the width of the window is $t_{\text{ADC}} = f_{\text{ADC}}^{-1}$, as shown in Figure S3(c), the transfer function of $w(t)$ can be written as:

$$H_w(f) = \frac{\left|\sin\left(\frac{\pi f}{f_{\text{ADC}}}\right)\right|}{\frac{\pi f}{f_{\text{ADC}}}}. \tag{S64}$$

Figure S9(c) shows the gain of the frequency response of $H_w(f)$. Zeros are found when $f = m f_{\text{ADC}}$, where $m$ is an integer. Convolution is then applied, resulting in the spectrum of the error of the windowed phase estimation function,

$$\tilde{Q}_{\text{ERR}}(f) = H_w(f) \tilde{q}_{\text{ERR}}(f), \tag{S65}$$

as shown in Figure S9(d). Next, when $\tilde{Q}_{\text{ERR}}(f)$ is sampled with the rate $f_{\text{ADC}}$, aliasing occurs. When we consider only low frequencies, i.e., below the measurement bandwidth, $f \lesssim f_{\text{MBW}}$, only frequency components around integer multiples of $f_{\text{ADC}}$ appear as aliasing peaks. In other words, aliasing peaks appear only if the input signal frequency is near a singular frequency:

$$f_{\text{sig}} = f_{\text{sglr}} \pm \delta f. \tag{S66}$$

Here, $\delta f$ is a detuning frequency, which is assumed to be smaller than the signal frequency, $\delta f \ll f_{\text{sig}}$. The singular frequency should satisfy the relation with $f_{\text{ADC}}$ using two integers, $k$ and $p$, as follows:

$$2k f_{\text{sglr}} = p f_{\text{ADC}}. \tag{S67}$$

For convenience, we set the integer $k$ as $k = sp + q$, where the two integers $s$ and $q$ are defined as $s \equiv \lfloor k/p \rfloor$ and $q \equiv k - p \lfloor k/p \rfloor$. Here, $p$ and $q$ are coprime. Then, Equation (S67) can be rewritten as:

$$f_{\text{sglr}} = \frac{1}{2\left(s + \frac{q}{p}\right)} f_{\text{ADC}}, \tag{S68}$$

where the following conditions are true for $s$, $p$, and $q$:

$$2 \leq s \tag{S69a}$$

$$1 \leq p \tag{S69b}$$
$$0 \leq q < p. \tag{S69c}$$

Note that $q$ can be zero only if $p = 1$ to avoid double counting. Note that the parameter $s$ cannot be 1 because the zero-crossings can only be detected properly when $f_{\text{sig}} < f_{\text{ADC}}/4$. Equation (S68) indicates that the $(sp+q)$-th peak of $\tilde{Q}_{\text{ERR}}(f)$ is near the p-th multiple of $f_{\text{ADC}}$, inducing an aliasing peak at the frequency of

$$f_{\text{fund}} \equiv 2(sp+q)\delta f, \tag{S70}$$

which we name the fundamental frequency. We proceed to evaluate how large the aliasing peak is. From Equation (S63), the amplitude of the $(sp+q)$-th peak of $\tilde{q}_{\text{ERR}}(f)$ is $1/(sp+q)\pi$. The gain of the window function can be estimated with the first order approximation around the frequency of $2(sp+q)f_{\text{sig}}$.

$$H_w(2(sp+q)f_{\text{sig}}) \cong 2(sp+q)\delta f \left(\frac{\mathrm{d}}{\mathrm{d}f}H_w(f)\right)_{f \cong pf_{\text{ADC}}}$$
$$= \frac{2(sp+q)\delta f}{pf_{\text{ADC}}} \tag{S71}$$

The amplitude at the fundamental peak (in the unit of radian) then becomes

$$A_{\text{fund,(rad)}} = \pi \frac{1}{(sp+q)\pi} H_w(2(sp+q)f_{\text{sig}}) = \frac{2\delta f}{pf_{\text{ADC}}}. \tag{S72}$$

The fundamental peak amplitude is proportional to the detuning frequency and is also proportional to $p^{-1}$. It is also readily found that there are higher order harmonic peaks at $f = mf_{\text{fund}}$, and their amplitude becomes $1/m$, i.e., they have $-20$-dB/dec dependency. Figure S10 shows an example of peaks induced by this error. Three cases are shown for different detuning frequencies. We also plot phase noise, which consists of white phase noise and flicker phase noise.

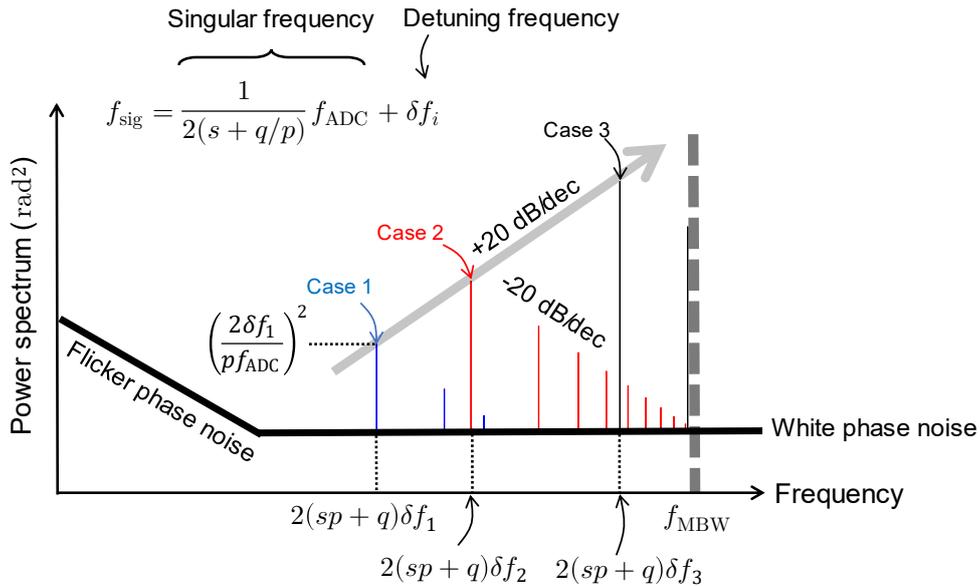

Figure S10. Quantization–aliasing error spectra dependency on detuning frequency. The fundamental frequency and its amplitude are proportional to the detuning frequency (fundamental power has +20 dB/dec dependency). When $p\delta f > f_{\text{MBW}}$, the error is filtered by the bandwidth-limitation (anti-aliasing) filter just before the results are obtained.

## 2. Upper limit on detuning frequency

The QA error does not appear in the result when the fundamental frequency is higher than the measurement bandwidth, i.e., when $f_{\text{MBW}} < \delta f_{\text{MAX}}$, because the frequency component beyond the measurement bandwidth should be sufficiently eliminated by the anti-aliasing filter before obtaining the final measured phase. In other words, the detuning frequency has an upper limit with respect to QA error.

The upper limit can be derived as follows. Figure S10 shows how QA errors are distributed in phase noise power spectrum. As the detuning frequency increases, the fundamental frequency increases proportionally according to Equation (S70). At the same time, the fundamental peak increases with 20 dB/dec dependency. The maximum detuning frequency $\delta f_{\text{MAX}}$ is obtained as

$$\delta f_{\text{MAX}} = \frac{f_{\text{MBW}}}{2(sp+q)}, \quad (S73)$$

and the maximum amplitude associated with $\delta f_{\text{MAX}}$ becomes

$$A_{\text{MAX,(rad)}} = \frac{f_{\text{MBW}}}{p(sp+q)f_{\text{ADC}}}. \quad (S74)$$

Considering that the higher order harmonic peaks have a $-20$ dB/dec dependency, we approximate the total error magnitude with $A_{\text{MAX,(rad)}}$. Using Equations (S68) and (S74), the distribution of maximum error amplitude over the input signal frequency is obtained, as shown in Figure S11. Examples for sets of singular frequency, maximum detuning, and maximum amplitude are listed in Table S1. To calculate these values, we used $f_{\text{ADC}} = 1$ GHz, and $f_{\text{MBW}} = 500$ kHz and only present results in the range of $p(sp+q) < 15$, which corresponds to an amplitude greater than $0.4 \times 10^{-4}$ rad. Note that the root mean squared (RMS) noise amplitude with a white phase noise level of $-145$ dBrad$^2$/Hz becomes $0.4 \times 10^{-4}$ rad. The upper limit of the input signal frequency is $f_{\text{ADC}}/4$, which corresponds to $(s, q, p) = (2, 0, 1)$. Upon combining Equations (S68) and (S74), we obtain the following relation:

$$A_{\text{MAX,(rad)}} = \frac{f_{\text{MBW}}}{2p^2 f_{\text{ADC}}^2} f_{\text{sig}}, \quad (S75)$$

which indicates that the maximum amplitude is proportional to $f_{\text{sig}}$ when $f_{\text{MBW}}$ is fixed. Equation (S75) also shows that the factor is proportional to $p^{-2}$; corresponding sequences of $p$ are found in Figure S11.

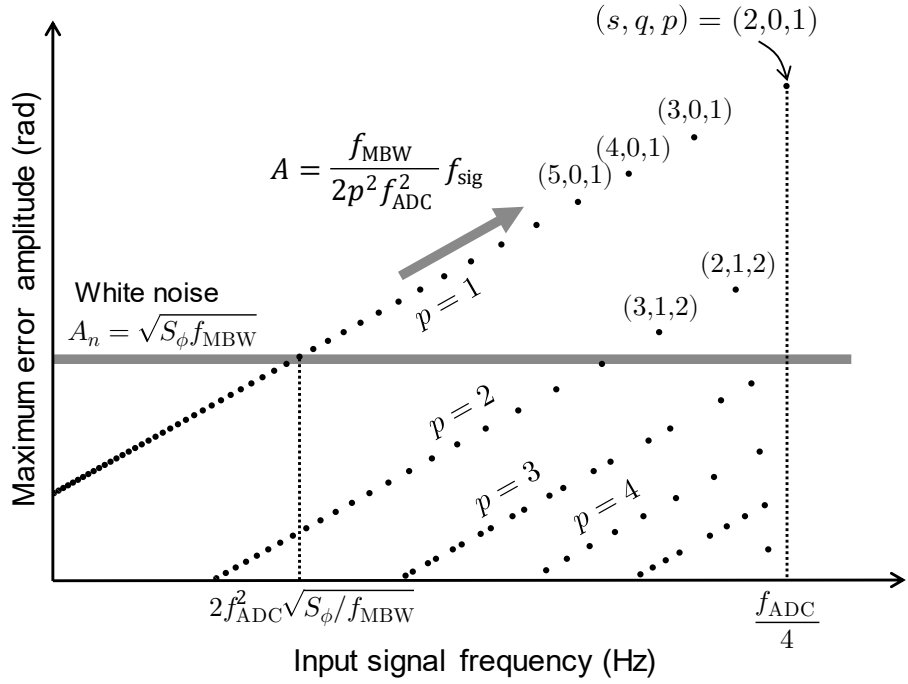

Figure S11. Distribution of the maximum error amplitude over the input signal frequency. Both axes are logarithmic.

Table S1. A partial list of singular frequencies and their corresponding maximum detuning and maximum amplitudes. The values were calculated using $f_{\text{ADC}} = 1$ GHz and $f_{\text{MBW}} = 500$ kHz. For simplicity, only results in the range of $p(sp+q) < 15$ are shown. Note that the RMS noise amplitude for the $f_{\text{MBW}}$ is about $0.4 \times 10^{-4}$ rad when the white phase noise level is $-145$ dBrad$^2$/Hz.

| $s$ | $q$ | $p$ | $2(sp+q)$ | Singular frequency $f_{\text{sglr}}$ (MHz) $\dfrac{f_{\text{ADC}}}{2(s+q/p)}$ | Maximum amplitude $A_{\text{MAX,(rad)}}$ ($\times 10^{-4}$ rad) $\dfrac{f_{\text{MBW}}}{f_{\text{ADC}}}\dfrac{1}{p(sp+q)}$ | Maximum detuning $\delta f_{\text{MAX}}$ (kHz) $\dfrac{f_{\text{MBW}}}{2(sp+q)}$ |
|---|---|---|---|---|---|---|
| 2 | 0 | 1 | 2 | 250.0000 | 2.50 | 125.0 |
| 2 | 1 | 2 | 10 | 200.0000 | 0.50 | 50.0 |
| 3 | 0 | 1 | 3 | 166.6667 | 1.67 | 83.3 |
| 3 | 1 | 2 | 14 | 142.8571 | 0.36 | 35.7 |
| 4 | 0 | 1 | 4 | 125.0000 | 1.25 | 62.5 |
| 5 | 0 | 1 | 5 | 100.0000 | 1.00 | 50.0 |
| 6 | 0 | 1 | 6 | 83.3333 | 0.83 | 41.7 |
| 7 | 0 | 1 | 7 | 71.4286 | 0.71 | 35.7 |
| 8 | 0 | 1 | 8 | 62.5000 | 0.63 | 31.3 |
| 9 | 0 | 1 | 9 | 55.5556 | 0.56 | 27.8 |
| 10 | 0 | 1 | 10 | 50.0000 | 0.50 | 25.0 |
| 11 | 0 | 1 | 11 | 45.4545 | 0.45 | 22.7 |
| 12 | 0 | 1 | 12 | 41.6667 | 0.42 | 20.8 |
| 13 | 0 | 1 | 13 | 38.4615 | 0.38 | 19.2 |
| 14 | 0 | 1 | 14 | 35.7143 | 0.36 | 17.9 |

### 3. Practical limit by the measurement bandwidth

Equation (S72) indicates that the measured phase is effectively free from QA error when the measurement bandwidth is less than a critical value. As shown in Figure S10, the measured phase contains noise, and QA error is below the noise level when the detuning frequency is low. Here, we consider only white phase noise, which is generally sufficient to estimate the noise above several kHz. The RMS amplitude of noise $A_{\text{noise}}$ is given by

$$A_{\text{noise}} = \sqrt{S_\phi f_{\text{MBW}}}. \tag{S76}$$

By applying Equations (S72) and (S76) and taking the worst case ($p=1$), the relation $A_{\text{MAX,(rad)}} < A_{\text{noise}}$; therefore, the noise amplitude exceeds the error amplitude, becoming

$$f_{\text{sig}} < f_{\text{MAX}} \equiv 2 f_{\text{ADC}}^2 \sqrt{\frac{S_\phi}{f_{\text{MBW}}}}. \tag{S77}$$

Equation (S77) shows that the frequency is free from QA error up to a maximum frequency ($f_{\text{MAX}} \propto f_{\text{MBW}}^{-\frac{1}{2}}$). In Subsection III.C, the white phase noise is expressed by Equation (S118), signifying that the white noise power spectrum is proportional to $f_{\text{sig}}^{-1}$ as follows:

$$S_\phi = \frac{P_{\text{ND}}}{P_{\text{eff}}} \left( \frac{f_{\text{ADC}}}{4 f_{\text{sig}}} \right). \tag{S78}$$

$P_{\text{ND}}$ and $P_{\text{eff}}$ are the noise density of the ADC (W/Hz) and effective signal power (W), respectively. Both are defined again later in Subsection III.C. Now, by using Equation (S78), Equation (S77) can be rewritten as:

$$f_{\text{MAX}}^3 = \frac{4 f_{\text{ADC}}^5}{f_{\text{MBW}}} \frac{P_{\text{ND}}}{P_{\text{eff}}}. \tag{S79}$$

Then, we can derive the critical measurement bandwidth by applying $f_{\text{MAX}} = f_{\text{ADC}}/4$ to (S79),

$$f_{\text{CMBW}} = 256 f_{\text{ADC}}^2 \frac{P_{\text{ND}}}{P_{\text{eff}}}. \tag{S80}$$

The critical measurement bandwidth indicates that if the measurement bandwidth is less than $f_{\text{CMBW}}$, the RMS amplitude of this error is always less than that of the white noise, regardless of the input signal frequency.

For example, taking the actual parameters of our hardware, $\log_{10} P_{\text{ND}} = -154$ dBFS/Hz, $10 \log_{10} P_{\text{eff}} = -2$ dBFS, and $f_{\text{ADC}} = 1$ GHz. Then, $f_{\text{CMBW}}$ becomes approximately 160 kHz. Note that $f_{\text{CMBW}} \propto P_{\text{eff}}^{-1}$, which means that the higher the effective power $P_{\text{eff}}$ is, the smaller $f_{\text{CMBW}}$ becomes due to lower white noise levels.

### II.E. Zero-crossing interpolation (ZI) error
#### 1. Analysis overview

ZI error is caused by the error at the linear interpolations when estimating the timing of zero-crossings in Equation (S23). In this subsection, we analyze how much error appears and how it affects the phase measurement. It is clear that this error originates from two coupled features; one is how nonlinear the input signal is at zero-crossings, and the second is the relative position between the sampling timing and zero-crossing timings. To analyze this complicated coupling, we carried out the procedure presented in Figure S12.

In Figure S12(a), we assume that an input signal $S(t)$ with a constant input signal frequency of $f_{\text{sig}}$ is sampled with the frequency of the ADC ($f_{\text{ADC}}$). Then, zero-crossings appear at the frequency $2f_{\text{sig}}$. We express how the error arises at zero-crossings by introducing a ZIEF (Figure S12(b)). The ZIEF shows the amount of phase error the estimated phase contains. It is calculated from the relation between the zero-crossing timing and sampling timing. The error induced by zero-crossing interpolation can be expressed by sampling the ZIEF at the zero-crossing timing (filled circle in Figure S12(b)), which is called virtual sampling. From it, we obtain the error at the rate of zero-crossings ($2f_{\text{sig}}$) (filled circles in Figure S12(b)), which is the virtually-sampled ZIEF (VS-ZIEF).

To convert the VS-ZIEF to the error of the instantaneous phase estimator, $Q_{i,\text{ERR}}$, which should be at the rate of $f_{\text{ADC}}$, we applied the sub-steps, as shown in Figure S12(b)–(e). First, we convert the VS-ZIEF to a summation of delta functions (Figure S12(c)), which we call deltaZIEF. Then, we take a convolution of deltaZIEF with the window function, using a time constant of $t_{\text{ADC}}$ by using the window function defined in Equation (S11). We call this convolution the windowed ZIEF (W-ZIEF) (Figure S12(d)). Finally, we resample the W-ZIEF at a sampling rate of $f_{\text{ADC}}$ to obtain the error of the instantaneous phase estimator, $Q_{i,\text{ERR}}$, which is the ZI error in the instantaneous phase estimator $Q_i$ (Figure S12(e)). The $Q_{i,\text{ERR}}$ can also be converted to the error in the measured phase by applying a low-pass filter and decimations to obtain the error on the measured phase in the measurement bandwidth (not shown in Figure S12).

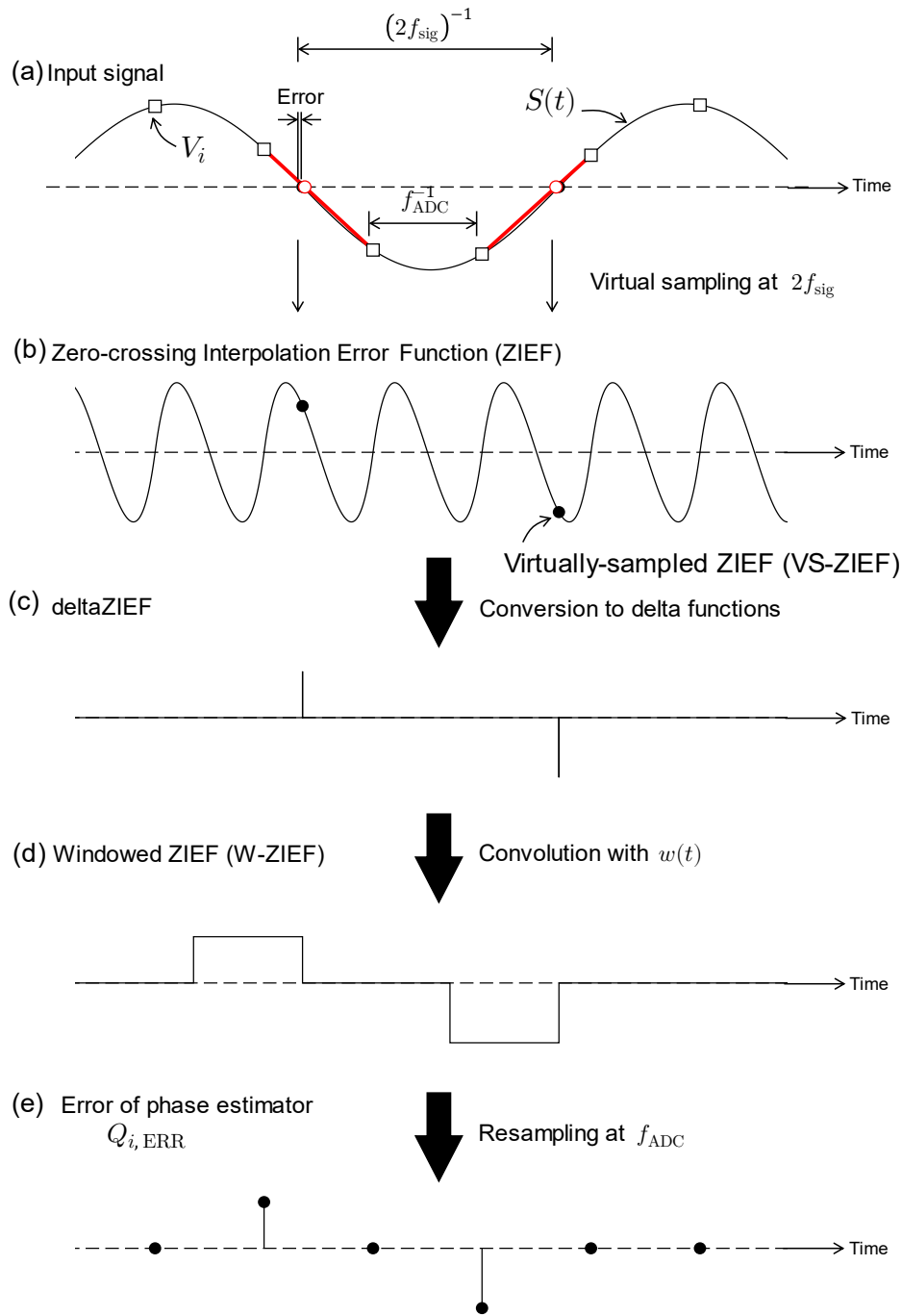

Figure S12. Introducing the zero-crossing interpolation error function (ZIEF) to analyze the zero-crossing interpolation error: (a) the input signal $S(t)$, sampling at the frequency of $f_{\text{ADC}}$, and linear interpolations; (b) ZIEF and its virtual sampling; (c) the error on the instantaneous phase estimator $Q_{i,\text{ERR}}$, which is resampled from the virtually-sampled data in (b).

## *2. Zero-crossing interpolation error function (ZIEF)*

To explicitly express the ZIEF, we start with a one-cycle zero-crossings interpolation error function (oneZIEF) represented by $\gamma_0(t)$. The oneZIEF is defined within only one sampling interval of the ADC, i.e., the oneZIEF is defined as a continuous variable defined in the range of

$$-\frac{t_{\text{ADC}}}{2} \leq t \leq \frac{t_{\text{ADC}}}{2}. \tag{S81}$$

We define the value $\gamma_0(t)$ as the phase error when the sampling timings are $t = \pm t_{\text{ADC}}/2$ and the actual zero-crossing is at $t$. The relation is graphically presented in Figure S13. In Case 1, the actual zero-crossing timing is $t = t_1$, and the corresponding interpolation error is $\gamma_0(t_1)$. Note that $\gamma_0(t)$ is given in phase, not in time, hence $\gamma_0(t)$ is multiplied by the input signal frequency as $\gamma_0(t_1) = 2\pi f_{\text{sig}} \delta t(t_1)$. In addition, both boundaries of this function are zero:

$$\gamma_0\left(-\frac{t_{\text{ADC}}}{2}\right) = \gamma_0\left(\frac{t_{\text{ADC}}}{2}\right) = 0, \tag{S82}$$

since the sampling position is the same as the zero-crossing timing in this case. Therefore, the oneZIEF has a cyclic boundary condition, and it can be expanded to a continuous function, the ZIEF, without losing continuity.

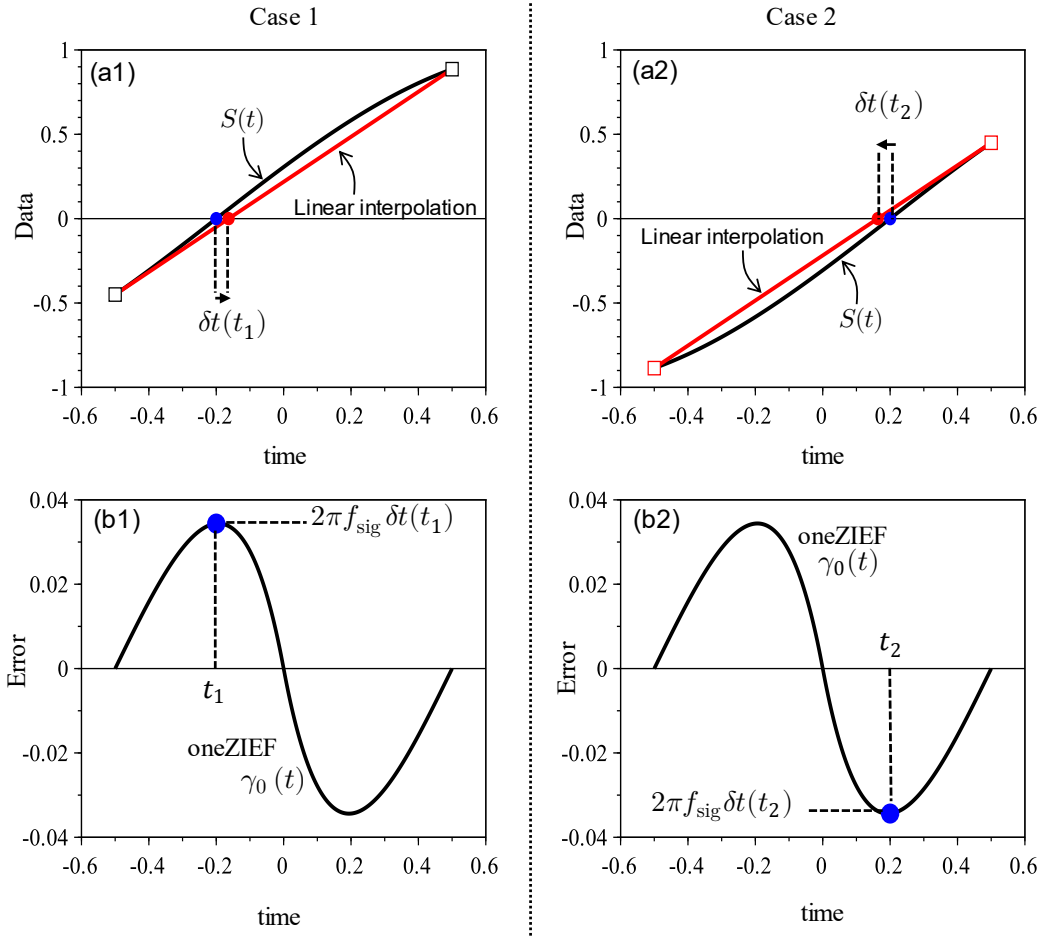

Figure S13. The one-cycle zero-crossing interpolation error function (oneZIEF) ((b1) and (b2)). The input signal $S(t)$ and its linear interpolation are shown in (a1) and (a2). Case 1 (left) and Case 2 (right) feature different zero-crossing timings. When the actual zero-crossing timing is $t$, the estimated timing is $t + \gamma_0(t)/2\pi f_{\text{sig}}$.

When considering how the oneZIEF expressed with input signals, we note that the same oneZIEF is obtained for both rising and falling edges if both edges are axisymmetric to each other around the zero-crossings, i.e.,

$$S_{\text{Rising}}(t) = -S_{\text{Falling}}(t). \tag{S83}$$

Here, $S_{\text{Rising}}(t)$ and $S_{\text{Falling}}(t)$ are the signal for the rising and falling edges, respectively. In this case, Case 1, $|A| = -A = -S_{\text{Rising}}(-t - \frac{t_{\text{ADC}}}{2})$ and $|B| = B = S_{\text{Rising}}(-t + \frac{t_{\text{ADC}}}{2})$, and for falling edges, $|A| = A = S_{\text{Falling}}(-t - \frac{t_{\text{ADC}}}{2})$ and $|B| = -B = S_{\text{Falling}}(-t + \frac{t_{\text{ADC}}}{2})$. Thus,

$$\begin{aligned}
\left(\frac{|B|}{|A|+|B|}\right)_{\text{Rising}} &= \left(\frac{B}{-A+B}\right)_{\text{Rising}} \\
&= \frac{S_{\text{Rising}}\left(-t + \frac{t_{\text{ADC}}}{2}\right)}{-S_{\text{Rising}}\left(-t - \frac{t_{\text{ADC}}}{2}\right) + S_{\text{Rising}}\left(-t + \frac{t_{\text{ADC}}}{2}\right)} \\
&= \frac{-S_{\text{Falling}}\left(-t + \frac{t_{\text{ADC}}}{2}\right)}{S_{\text{Falling}}\left(-t - \frac{t_{\text{ADC}}}{2}\right) - S_{\text{Falling}}\left(-t + \frac{t_{\text{ADC}}}{2}\right)} \\
&= \left(\frac{-B}{A-B}\right)_{\text{Falling}} \\
&= \left(\frac{|B|}{|A|+|B|}\right)_{\text{Falling}}.
\end{aligned} \tag{S84}$$

Therefore, we only need to consider rising edges. If the condition in Equation (S82) is not satisfied, the oneZIEF can be considered separately for each type of edge, and they can be expanded to ZIEF after they are joined. The linear interpolation becomes

$$\frac{|B|}{|A|+|B|} = \frac{S\left(-t + \frac{t_{\text{ADC}}}{2}\right)}{S\left(-t + \frac{t_{\text{ADC}}}{2}\right) - S\left(-t - \frac{t_{\text{ADC}}}{2}\right)}, \tag{S85}$$

and, the estimated zero-crossing timing $t_{\text{est}}$ determined by the linear interpolation is

$$\begin{aligned}
t_{\text{est}} &= \frac{t_{\text{ADC}}}{2} - t_{\text{ADC}} \frac{|B|}{|A|+|B|} \\
&= -\frac{t_{\text{ADC}}}{2} \frac{S\left(-t + \frac{t_{\text{ADC}}}{2}\right) + S\left(-t - \frac{t_{\text{ADC}}}{2}\right)}{S\left(-t + \frac{t_{\text{ADC}}}{2}\right) - S\left(-t - \frac{t_{\text{ADC}}}{2}\right)}.
\end{aligned} \tag{S86}$$

By using the actual zero-crossing timing $t$, the oneZIEF $\gamma_0(t)$ should be

$$\begin{aligned}
\gamma_0(t) &= 2\pi f_{\text{sig}}(t_{\text{est}} - t) \\
&= 2\pi f_{\text{sig}}\left(-t - \frac{t_{\text{ADC}}}{2} \frac{S\left(-t + \frac{t_{\text{ADC}}}{2}\right) + S\left(-t - \frac{t_{\text{ADC}}}{2}\right)}{S\left(-t + \frac{t_{\text{ADC}}}{2}\right) - S\left(-t - \frac{t_{\text{ADC}}}{2}\right)}\right).
\end{aligned} \tag{S87}$$

This is a generalized expression of the oneZIEF. From the Equation (S81), $-t \pm \frac{t_{\text{ADC}}}{2}$ is in the range from $-t_{\text{ADC}}$ to $t_{\text{ADC}}$, which indicates that the shape of the input signal $S(t)$ should be known in the range of $\pm t_{\text{ADC}}$ around the zero-crossings to calculate $\gamma_0(t)$. Practically, $S(t)$ should be measured or estimated with higher time resolution than $t_{\text{ADC}}$. Possible ways to carry this out include (i) measuring the input signal by using an equivalent-time sampling mode of a digital oscilloscope, (ii) using a high-speed signal acquisition instrument, or (iii) calculating the shape of the signal theoretically.

### 3. ZIEF for sinusoidal input signals

Here, we focus on sinusoidal input signals, i.e., we assume the input signal to be
$$S(t) = \sin(2\pi f_{\text{sig}} t). \tag{S88}$$
Using the additive theorem of trigonal functions, we obtain the following relations:
$$S\left(-t + \frac{t_{\text{ADC}}}{2}\right) + S\left(-t - \frac{t_{\text{ADC}}}{2}\right) = 2\sin(-2\pi f_{\text{sig}} t)\cos(\pi f_{\text{sig}} t_{\text{ADC}}) \\ = -2\sin(\phi_t)\cos(\phi_{\text{nyq}}), \tag{S89a}$$

$$S\left(-t + \frac{t_{\text{ADC}}}{2}\right) - S\left(-t - \frac{t_{\text{ADC}}}{2}\right) = 2\cos(\phi_t)\sin(\phi_{\text{nyq}}), \tag{S89b}$$

using the following definitions:
$$\phi_t \equiv 2\pi f_{\text{sig}} t \tag{S90a}$$
$$\phi_{\text{nyq}} \equiv 2\pi f_{\text{sig}} \frac{t_{\text{ADC}}}{2}. \tag{S90b}$$
Note that from Equation (S81), the range of $\phi_t$ is limited to
$$-\phi_{\text{nyq}} \leq \phi_t \leq \phi_{\text{nyq}}. \tag{S91}$$
Applying Equations (S89a) and (S89b) to Equation (S87), the oneZIEF is derived for sinusoidal signals without approximation.
$$\gamma_0(t) = -\phi_t + \frac{\phi_{\text{nyq}}}{\tan \phi_{\text{nyq}}} \tan \phi_t \tag{S92}$$

In Figure S14(a), oneZIEFs having different $f_{\text{sig}}$ are plotted. For example, $t_{\text{ADC}} = 1$ ns and $f_{\text{sig}}$ is 50 MHz, 100 MHz, and 200 MHz, while the maximum input frequency is $t_{\text{ADC}}^{-1}/4 = 250$ MHz. From Equation (S92), it is clear that the oneZIEFs are odd functions, i.e., symmetric around $t = 0$. In addition, the function $\gamma_0(t)$ has two peaks; $\gamma_{\text{MAX}}$ represents the maximum value. Figure S14(b) shows a plot of $\gamma_{\text{MAX}}$ versus $f_{\text{sig}}$. As shown in Equation (S97), $\gamma_{\text{MAX}}$ is proportional to $f_{\text{sig}}^3$ when $f_{\text{sig}} \ll f_{\text{ADC}}/4$. In particular, for the parameters chosen for this example,
$$\gamma_{\text{MAX}} \cong 4.0 \times 10^{-9} \times \left(\frac{f_{\text{sig}}}{1 \text{ MHz}}\right)^3. \tag{S93}$$
Considering that the flicker phase noise limits the phase fluctuation on the order of about $10^{-4}$ rad, this error does not effectively worsen the measurement accuracy below ca. 30 MHz.

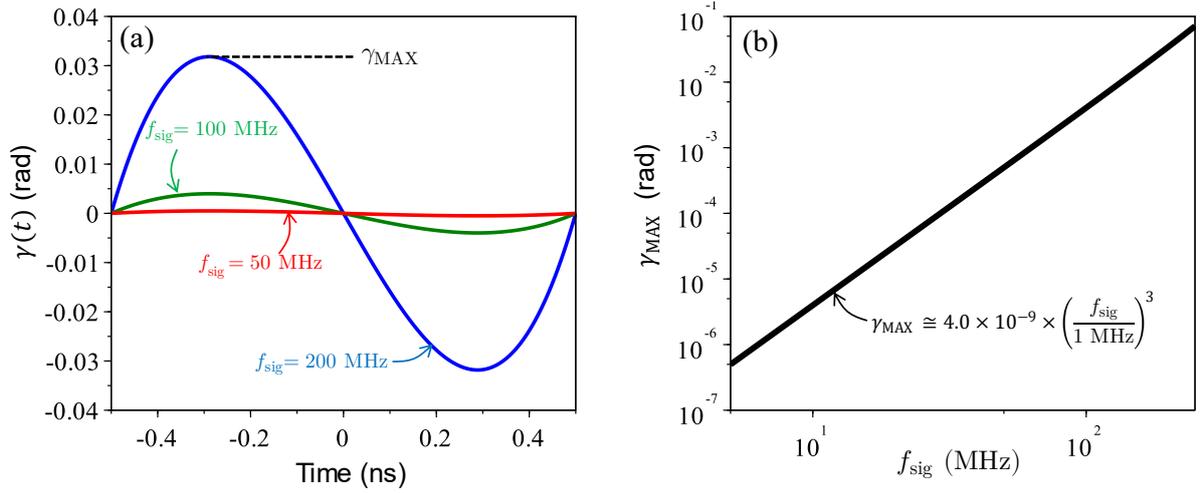

Figure S14. (a) The one-cycle zero-crossing interpolation error function (oneZIEF) for different input signal frequencies given sinusoidal input signals; (b) the maximum value of the oneZIEF versus input frequency.

### *4. Approximation of oneZIEF for low input frequencies*

When $f_{\text{sig}}$ is sufficiently lower than $f_{\text{ADC}}$, an analytical expression of the oneZIEF can be obtained, and we can approximate the $\gamma_{\text{MAX}}$ values according to $\phi_{\text{nyq}} \ll \pi$ and $\phi_t \ll \pi$. Then, the following approximations are used for a tangent function to the third order.

$$\tan \phi_t \cong \phi_t + \frac{\phi_t^3}{3} \tag{S94a}$$

$$\frac{1}{\tan \phi_{\text{nyq}}} \cong \frac{1}{\phi_{\text{nyq}}} - \frac{\phi_{\text{nyq}}}{3} \tag{S94b}$$

Applying Equations (S94a) and (S94b) to (S92), $\gamma_0(t)$ can be expressed as a third-order polynomial.

$$\begin{aligned}\gamma_0(t) &\cong -\phi_t + \frac{\phi_{\text{nyq}}}{\tan \phi_{\text{nyq}}} \tan \phi_t \\ &\cong \frac{1}{3}\phi_t(\phi_t + \phi_{\text{nyq}})(\phi_t - \phi_{\text{nyq}}) \\ &= \frac{(2\pi f_{\text{sig}})^3}{3} t \left(t + \frac{t_{\text{ADC}}}{2}\right)\left(t - \frac{t_{\text{ADC}}}{2}\right)\end{aligned} \tag{S95}$$

Note that this approximation is also sufficiently appropriate for rough estimations when $f_{\text{sig}} \sim f_{\text{ADC}}/4$. According to numerical calculation, the approximation (Equation (S95)) and exact solution (Equation (S92)) differ by less than 30%, even if $f_{\text{sig}} = 250\,\text{MHz} = f_{\text{ADC}}/4$. When $f_{\text{sig}} < 100\,\text{MHz}$, the difference becomes less than 5%. By using Equation (S95), $\gamma_{\text{MAX}}$ can be estimated analytically by solving $\frac{d}{dt}\gamma_0(t) = 0$; then, the extrema can be derived as

$$t = \pm \frac{t_{\text{ADC}}}{\sqrt{12}}, \tag{S96}$$

and the maximum is given as follows:

$$\gamma_{\text{MAX}} = \frac{(2\pi f_{\text{sig}} t_{\text{ADC}})^3}{36\sqrt{3}} \cong 4.0 \times \left(\frac{f_{\text{sig}}}{f_{\text{ADC}}}\right)^3. \tag{S97}$$

Note that $\gamma_{\text{MAX}}$ is expressed in radians. This result agrees with that of Figure S14(b). It is also notable that $\gamma_{\text{MAX}}$ depends only on the input signal frequency, and it does not depend on the power of the input signal.

### *5. Estimation of ZI error using frequency-domain analysis*

To analyze how the error presents in the measurement, we carry out a similar frequency-domain analysis as done for QA error. The overview is presented in Figure S15. A Fourier transformation is applied to the ZIEF, as shown in Figure S15(a). The peaks in the frequency domain consist of the fundamental frequency, $f_{\text{ADC}} = t_{\text{ADC}}^{-1}$, and its higher order harmonics because the ZIEF is a periodic function with a frequency of $f_{\text{ADC}}$. Here, we assume sinusoidal input signals, leading to −60 dB/dec dependency in the peaks, as shown later. Next, the ZIEF is sampled at $2f_{\text{sig}}$, which represents virtual sampling (Figure S15(b)). In this process, the peaks are transferred to the low-frequency range of the VS-ZIEF though aliasing.

To convert the VS-ZIEF to the error of the instantaneous phase estimator, which has a sampling rate of $f_{\text{ADC}}$, additional steps are carried out for resampling. The process in the time domain is presented in Figure S12(c)–(e). First, the VS-ZIEF is converted to deltaZIEF. Then, high frequencies above $f_{\text{sig}}$ in the spectral domain of the deltaZIEF become repeated peaks, as shown in Figure S15(c), due to the nature of delta functions. Next, the window function is applied to the deltaZIEF, and W-ZIEF (Figure S15(d)) is obtained. Since the time constant of the window function is $t_{\text{ADC}}$, and its frequency response is a low-pass filter (Equation (S64)), the W-ZIEF has a cutoff frequency of $f_{\text{ADC}}$. Note that resampling at the frequency of $f_{\text{ADC}}$ does not significantly affect the spectral shape below $f_{\text{sig}}$ because $H_w(\omega)$ exhibits low frequency dependence below $f_{\text{sig}}$. Then, the W-ZIEF is sampled at $f_{\text{ADC}}$ again to obtain error in the instantaneous phase estimator, as shown in Figure S15(e). In this resampling process, high frequency peaks above $f_{\text{sig}}$ appear below $f_{\text{ADC}}$ due to the repeated aliasing. Finally, filtering and decimation are applied to obtain the measured phase. The error peaks remain in the measurement bandwidth, which is shown in Figure S15(f).

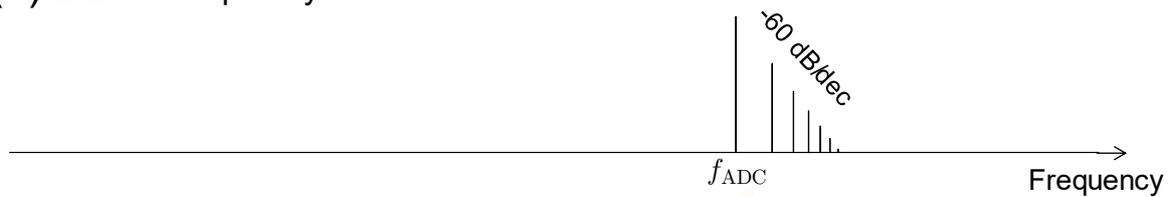
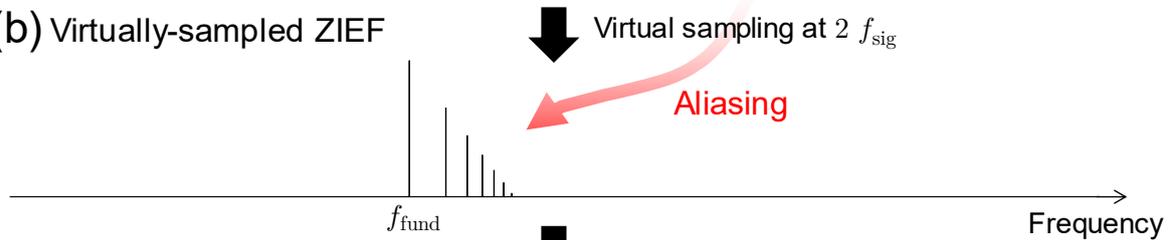
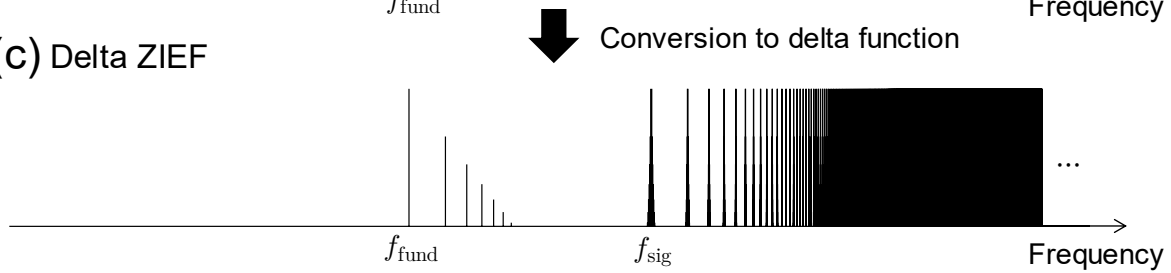
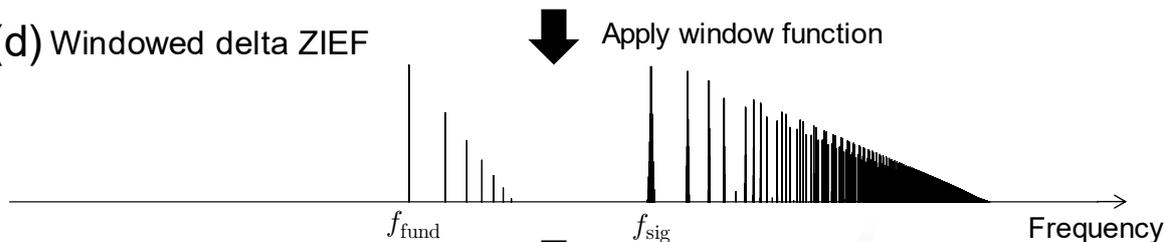
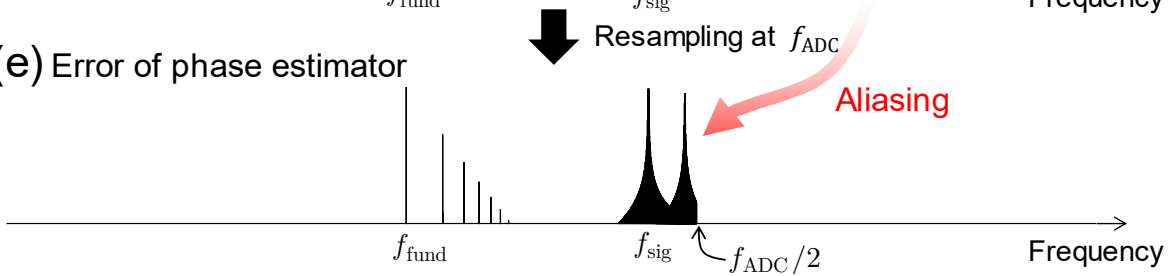
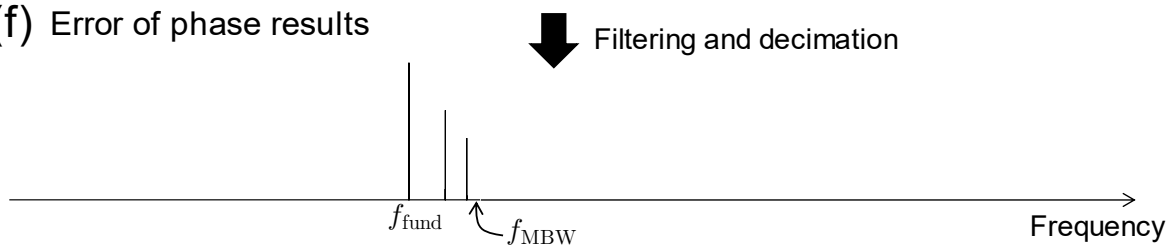

Figure S15. Overview of how the zero-crossing interpolation (ZI) error is introduced into the measurement, using a sinusoidal input signal: (a) the frequency domain representation for the ZIEF. The peaks consist of the fundamental frequency, $f_{\mathrm{ADC}} = t_{\mathrm{ADC}}^{-1}$, and its higher order harmonics. The power depends on the frequency with $-60$ dB/dec; (b) virtually-sampled ZIEF (VS-ZIEF). After effective sampling at a rate of $2f_{\mathrm{sig}}$, the peaks in the ZIEF are aliased and appear in at frequencies below $f_{\mathrm{sig}}$; (c) the deltaZIEF converted from the VS-ZIEF; (d) windowed ZIEF; (e) QA error on the instantaneous phase estimator. After resampling at $f_{\mathrm{ADC}}$ and filtering and decimation at $2f_{\mathrm{MBW}}$, the error peaks remain in the measurement bandwidth; (f) error in the measured phase. For all plots, both axes are logarithmic.

The Fourier transformation of the ZIEF can be calculated by taking a Fourier expansion of the oneZIEF since the ZIEF is the periodic expansion of the oneZIEF. Thus, we consider the Fourier expansion as follows.

$$\gamma_0(t) = \sum_{l=1}^{\infty} C_l \sin(2\pi l f_{\mathrm{ADC}} t) \tag{S98}$$

We only use sine expansion because the oneZIEF is an odd function. Here, $C_l$ denotes the Fourier expansion coefficient, which can be derived by integration, according to Equation (S97).

$$C_l = \frac{1}{t_{\mathrm{ADC}}/2} \int_{-t_{\mathrm{ADC}}/2}^{t_{\mathrm{ADC}}/2} \gamma_0(t) \sin(2\pi l f_{\mathrm{ADC}} t) \, dt. \tag{S99}$$

Using the well-known integration formula, we derive

$$\int_{-t_{\mathrm{ADC}}/2}^{t_{\mathrm{ADC}}/2} t \sin(2\pi l f_{\mathrm{ADC}} t) \, dt = -\frac{1}{2f_{\mathrm{ADC}}^2} \frac{(-1)^l}{l\pi} \tag{S100a}$$

$$\int_{-t_{\mathrm{ADC}}/2}^{t_{\mathrm{ADC}}/2} t^3 \sin(2\pi l f_{\mathrm{ADC}} t) \, dt = \frac{1}{8f_{\mathrm{ADC}}^4} \left( -\frac{1}{l\pi} + \frac{6}{(l\pi)^3} \right)(-1)^l. \tag{S100b}$$

By using Equations (S95), (S100a), and (S100b), Equation (S99) can be directly integrated to obtain $C_l$ as

$$C_l = \frac{4 f_{\mathrm{sig}}^3}{f_{\mathrm{ADC}}^3} \frac{(-1)^l}{l^3}, \tag{S101}$$

thus enabling the amplitude of the $l$-th peak to be expressed (Figure S15(b)). $C_l$ is proportional to $l^{-3}$, i.e., the frequency dependency is $-60$ dB/dec in signal power representation. Moreover, $C_l$ is also proportional to $(f_{\mathrm{sig}}/f_{\mathrm{ADC}})^3$.

Next, we focus on how the error peaks appear in the measured phase when the input signal frequency is around the singular frequencies and when the detuning frequency changes. As done in the previous Subsection II.D, here, we also use the Equations (S66) and (S68) with singular frequency $f_{\mathrm{sglr}}$ and detuning frequency $\delta f$. Here, the singular frequencies are characterized by the integers $(s, q, p)$ following the conditions in Equations (S69a)–(S69c), i.e.,

$$f_{\mathrm{sig}} = \frac{1}{2\left(s + \frac{q}{p}\right)} f_{\mathrm{ADC}} + \delta f. \tag{S102}$$

By aliasing the first peak of the ZIEF in Figure S15(b), fundamental frequency is induced, $f_{\mathrm{fund}} = 2(sp + q)\delta f$ (Equation (S70)).

The dependency of the ZIEF on the detuning frequency $\delta f$ around a singular frequency is presented in Figure S16. Note that the vertical axis is not PSD but the power spectrum, i.e., the unit is $\mathrm{rad}^2$. Three cases, Cases 1–3, are shown, each corresponding to a different detuning frequency. Here, the integers $(s, q, p)$ characterize the singular frequency as given in Equation (S102). As with QA error, ZI error also disappears from the results when $f_{\mathrm{fund}}$ exceeds the measurement bandwidth, $f_{\mathrm{MBW}}$, which is the result of filtering. Thus, the maximum detuning frequency, $\delta f_{\mathrm{MAX}}$, is also expressed by Equation (S73), $\delta f_{\mathrm{MAX}} = f_{\mathrm{MBW}}/(2(sp + q))$. Considering the aliasing process from the ZIEF to the VS-ZIEF, the fundamental amplitude is found to be invariant to the detuning frequency.

$$A_{\text{fund,(rad)}} = \frac{4f_{\text{sig}}^3}{p^3 f_{\text{ADC}}^3} \tag{S103}$$

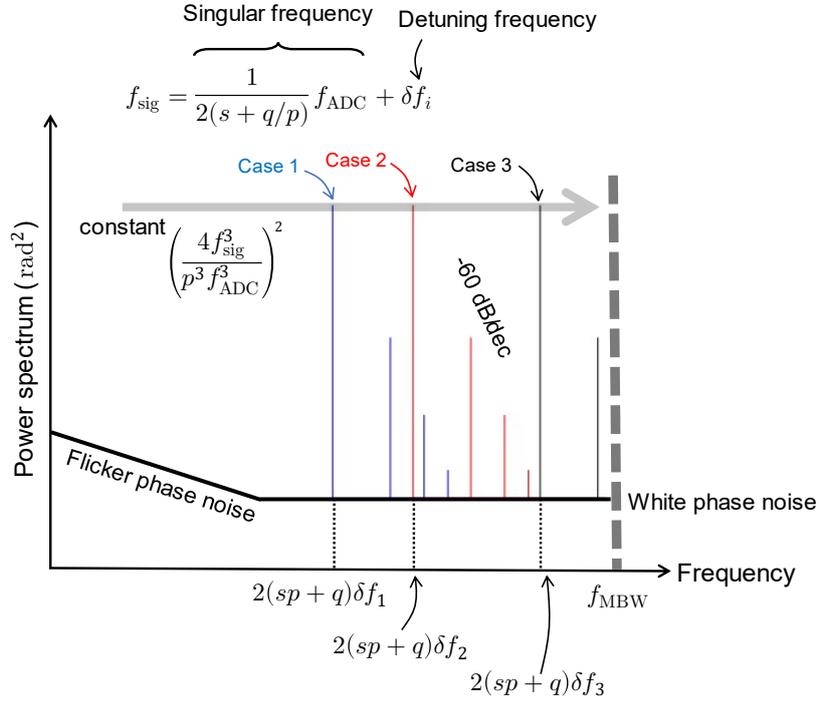

Figure S16. The dependency of the zero-crossing interpolation error on the detuning frequency around a singular frequency. The horizontal axis is frequency, and the vertical axis is the error spectral power (not PSD); both axes are logarithmic. Cases 1–3 correspond to different detuning frequencies $\delta f_i$. $f_{\text{sig}}$ and $f_{\text{ADC}}$ denote input signal frequency and sampling frequency at the ADC, respectively. The integers $(s, q, p)$ characterize the singular frequency.

The amplitude of ZI error remains constant regardless of the detuning frequency, whereas QA error decreases its amplitude when the detuning frequency becomes small, as shown in Equation (S72). This result demonstrates that ZI error causes a DC phase error (or phase shift) when the detuning frequency is zero, indicating that the input signal frequency is exactly at the singular frequency. The DC phase error can be between zero and the maximum error, $\gamma_{\text{MAX}}$, depending on the relative timing (or relative phase) between the input signal and sampling clock of the ADC.

### 6. Analysis on singular frequencies

In this subsection, we analyze how the singular frequencies are distributed among the whole input frequency range and investigate the associated error amplitudes. Figure S17 shows singular frequencies for sinusoidal input signals when $f_{\text{ADC}} = 1$ GHz. We use $f_{\text{MBW}} = 500$ kHz and $S_\phi = -145$ dBrad$^2$/Hz as realistic parameters. The fundamental error amplitudes are obtained by numerical calculation. In particular, we integrated the right-hand-side of Equation (S99) numerically by applying Equation (S92). In Figure S17, each dot represents a singular frequency. Sequences proportional to $f_{\text{sig}}^3 p^{-3}$, according to Equation (S75), are found. In Figure S15, the thick horizontal line denotes $0.4 \times 10^{-4}$ rad, which corresponds to the RMS noise amplitude for the considered hardware. The five largest peaks are associated with $(s, q, p) = (2, 0, 1), (3, 0, 1), (4, 0, 1), (2, 1, 2),$ and $(5, 0, 1)$.

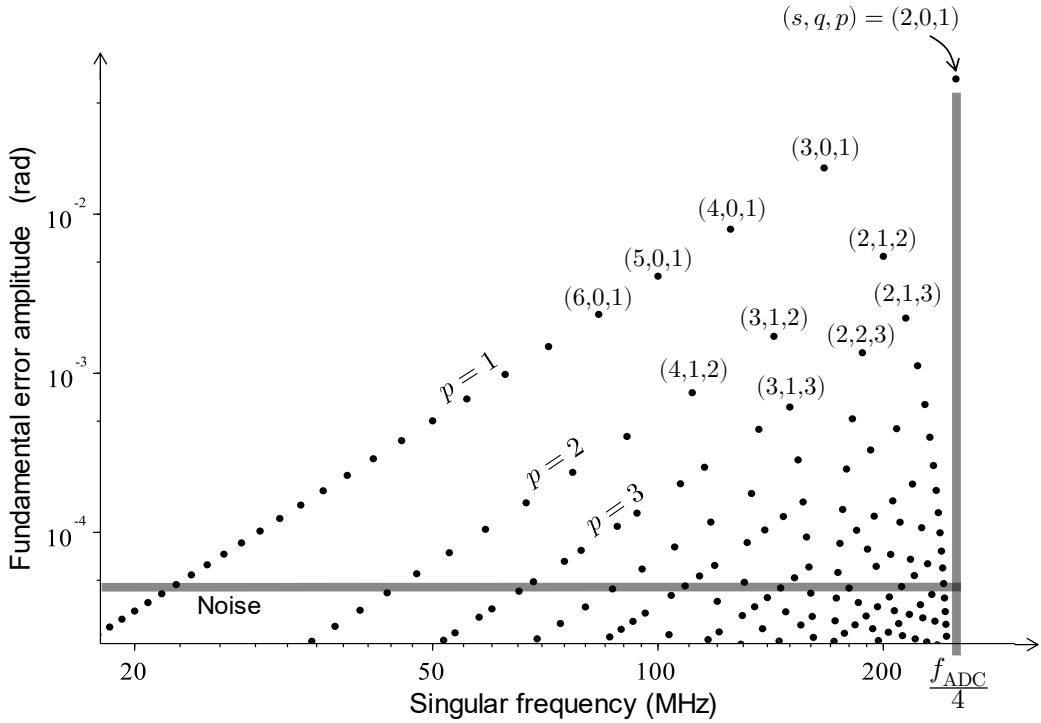

Figure S17. The zero-crossing interpolation error dependency on the detuning frequency around a singular frequency. Each dot is a singular frequency. The maximum input frequency $f_{\text{ADC}}/4$ is also shown (thick vertical line). Typical noise amplitude of our hardware (parameters: $f_{\text{ADC}} = 1$ GHz, $f_{\text{MBW}} = 500$ kHz, and $S_\phi = -145$ dBrad$^2$/Hz) is about $0.4 \times 10^{-4}$ rad (thick horizontal line).

Table S2 lists every singular frequency above the RMS noise level of $0.4 \times 10^{-4}$ rad. Note that only the maximum detuning, $\delta f_{\text{MAX}}$, depends on $f_{\text{MBW}}$. The 11 singular frequencies with peak amplitudes larger than 1 mrad are marked in bold for clarity.

As an example, we explain the case for $(s, q, p) = (2, 2, 3)$. The singular frequency for this parameter set is 187.5 MHz, and the corresponding peak amplitude is $13.5 \times 10^{-4}$ rad. The maximum detuning frequency is 31.3 kHz, indicating that ZI error should be considered only when the input signal frequency is within 187.5 MHz $\pm$ 31.3 kHz for the given $f_{\text{MBW}}$ of 500 kHz. Frequency factor refers to the ratio of the fundamental peak and detuning frequency. In this example, the fundamental peak appears at a frequency of $16 \times \delta f$. When the detuning frequency is zero, i.e., when the input signal frequency is exactly 187.5 MHz, there is a DC phase error up to a peak amplitude of $13.5 \times 10^{-4}$ rad. The magnitude of the DC phase error depends on the relative phase between the input signal and the ADC sampling clock.

Note that the measured phase will be virtually free from the ZI error when the input signal frequency is below a critical signal frequency. The critical frequency is related to flicker phase noise because flicker phase noise normally determines the RMS noise amplitude for long measurements.

Table S2. A list of singular frequencies for the zero-crossing interpolation error and the corresponding maximum amplitude and maximum detuning. Input signals are sinusoidal, $f_{\text{ADC}} = 1$ GHz, and $f_{\text{MBW}} = 500$ kHz. Large singular frequencies with more than 1 mrad peak amplitude are shown in bold.

| $s$ | $q$ | $p$ | Singular frequency $f_{\text{sglr}}$ (MHz) | Peak amplitude ($\times 10^{-4}$ rad) | Maximum detuning $\delta f_{\text{MAX}}$ (kHz) | Frequency factor |
|---|---|---|---|---|---|---|
| | | | $\dfrac{f_{\text{ADC}}}{2(s + q/p)}$ | Numerically calculated from (S92) | $\dfrac{f_{\text{MBW}}}{2(sp + q)}$ | $2(sp + q)$ |
| **2** | **0** | **1** | **250. 000 00** | **708.0** | **125.0** | **4** |
| 2 | 1 | 13 | 240. 740 74 | 0.5 | 9.3 | 54 |
| 2 | 1 | 12 | 240. 000 00 | 0.6 | 10.0 | 50 |
| 2 | 1 | 11 | 239. 130 43 | 0.8 | 10.9 | 46 |
| 2 | 1 | 10 | 238. 095 24 | 1.0 | 11.9 | 42 |
| 2 | 1 | 9 | 236. 842 11 | 1.3 | 13.2 | 38 |
| 2 | 1 | 8 | 235. 294 12 | 1.8 | 14.7 | 34 |
| 2 | 1 | 7 | 233. 333 33 | 2.6 | 16.7 | 30 |
| 2 | 2 | 13 | 232. 142 86 | 0.4 | 8.9 | 56 |
| 2 | 1 | 6 | 230. 769 23 | 4.0 | 19.2 | 26 |
| 2 | 2 | 11 | 229. 166 67 | 0.6 | 10.4 | 48 |
| 2 | 1 | 5 | 227. 272 73 | 6.4 | 22.7 | 22 |
| 2 | 2 | 9 | 225. 000 00 | 1.1 | 12.5 | 40 |
| **2** | **1** | **4** | **222. 222 22** | **11.2** | **27.8** | **18** |
| 2 | 3 | 11 | 220. 000 00 | 0.5 | 10.0 | 50 |
| 2 | 2 | 7 | 218. 750 00 | 2.0 | 15.6 | 32 |
| 2 | 3 | 10 | 217. 391 30 | 0.7 | 10.9 | 46 |
| **2** | **1** | **3** | **214. 285 71** | **22.2** | **35.7** | **14** |
| 2 | 4 | 11 | 211. 538 46 | 0.5 | 9.6 | 52 |
| 2 | 3 | 8 | 210. 526 32 | 1.2 | 13.2 | 38 |
| 2 | 2 | 5 | 208. 333 33 | 4.5 | 20.8 | 24 |
| 2 | 3 | 7 | 205. 882 35 | 1.6 | 14.7 | 34 |
| 2 | 4 | 9 | 204. 545 45 | 0.7 | 11.4 | 44 |
| **2** | **1** | **2** | **200. 000 00** | **54.3** | **50.0** | **10** |
| 2 | 5 | 9 | 195. 652 17 | 0.6 | 10.9 | 46 |
| 2 | 4 | 7 | 194. 444 44 | 1.3 | 13.9 | 36 |
| 2 | 3 | 5 | 192. 307 69 | 3.3 | 19.2 | 26 |
| 2 | 5 | 8 | 190. 476 19 | 0.8 | 11.9 | 42 |
| **2** | **2** | **3** | **187. 500 00** | **13.5** | **31.3** | **16** |
| 2 | 5 | 7 | 184. 210 53 | 1.0 | 13.2 | 38 |
| 2 | 3 | 4 | 181. 818 18 | 5.2 | 22.7 | 22 |
| 2 | 7 | 9 | 180. 000 00 | 0.4 | 10.0 | 50 |
| 2 | 4 | 5 | 178. 571 43 | 2.5 | 17.9 | 28 |
| 2 | 5 | 6 | 176. 470 59 | 1.4 | 14.7 | 34 |
| 2 | 6 | 7 | 175. 000 00 | 0.9 | 12.5 | 40 |
| 2 | 7 | 8 | 173. 913 04 | 0.6 | 10.9 | 46 |
| **3** | **0** | **1** | **166. 666 67** | **195.3** | **83.3** | **6** |
| 3 | 1 | 8 | 160. 000 00 | 0.4 | 10.0 | 50 |
| 3 | 1 | 7 | 159. 090 91 | 0.6 | 11.4 | 44 |
| 3 | 1 | 6 | 157. 894 74 | 0.9 | 13.2 | 38 |
| 3 | 1 | 5 | 156. 250 00 | 1.6 | 15.6 | 32 |
| 3 | 1 | 4 | 153. 846 15 | 2.9 | 19.2 | 26 |
| 3 | 2 | 7 | 152. 173 91 | 0.5 | 10.9 | 46 |
| **3** | **1** | **3** | **150. 000 00** | **6.1** | **25.0** | **20** |
| 3 | 2 | 5 | 147. 058 82 | 1.3 | 14.7 | 34 |
| 3 | 3 | 7 | 145. 833 33 | 0.4 | 10.4 | 48 |
| **3** | **1** | **2** | **142. 857 14** | **17.1** | **35.7** | **14** |

Table S2 (continued)

| s | q | p | Singular frequency $f_{\text{sglr}}$ (MHz) | Fundamental peak amplitude $A_{\text{fund,(rad)}}$ ($\times 10^{-4}$ rad) | Maximum detuning $\delta f_{\text{MAX}}$ (kHz) | Frequency factor |
|---|---|---|---|---|---|---|
| 3 | 3 | 5 | 138. 888 89 | 1.0 | 13.9 | 36 |
| 3 | 2 | 3 | 136. 363 64 | 4.4 | 22.7 | 22 |
| 3 | 3 | 4 | 133 .333 33 | 1.8 | 16.7 | 30 |
| 3 | 4 | 5 | 131. 578 95 | 0.9 | 13.2 | 38 |
| 3 | 5 | 6 | 130. 434 78 | 0.5 | 10.9 | 46 |
| **4** | **0** | **1** | **125. 000 00** | **80.5** | **62.5** | **8** |
| 4 | 1 | 5 | 119. 047 62 | 0.6 | 11.9 | 42 |
| 4 | 1 | 4 | 117. 647 06 | 1.2 | 14.7 | 34 |
| 4 | 1 | 3 | 115. 384 62 | 2.6 | 19.2 | 26 |
| 4 | 2 | 5 | 113. 636 36 | 0.5 | 11.4 | 44 |
| 4 | 1 | 2 | 111. 111 11 | 7.6 | 27.8 | 18 |
| 4 | 3 | 5 | 108. 695 65 | 0.5 | 10.9 | 46 |
| 4 | 2 | 3 | 107. 142 86 | 2.0 | 17.9 | 28 |
| 4 | 3 | 4 | 105. 263 16 | 0.8 | 13.2 | 38 |
| 4 | 4 | 5 | 104. 166 67 | 0.4 | 10.4 | 48 |
| **5** | **0** | **1** | **100. 000 00** | **40.8** | **50.0** | **10** |
| 5 | 1 | 4 | 95. 238 10 | 0.6 | 11.9 | 42 |
| 5 | 1 | 3 | 93. 750 00 | 1.3 | 15.6 | 32 |
| 5 | 1 | 2 | 90. 909 09 | 4.0 | 22.7 | 22 |
| 5 | 2 | 3 | 88. 235 29 | 1.1 | 14.7 | 34 |
| 5 | 3 | 4 | 86. 956 52 | 0.4 | 10.9 | 46 |
| **6** | **0** | **1** | **83. 333 33** | **23.5** | **41.7** | **12** |
| 6 | 1 | 3 | 78. 947 37 | 0.8 | 13.2 | 38 |
| 6 | 1 | 2 | 76. 923 08 | 2.4 | 19.2 | 26 |
| 6 | 2 | 3 | 75. 000 00 | 0.7 | 12.5 | 40 |
| **7** | **0** | **1** | **71. 428 57** | **14.7** | **35.7** | **14** |
| 7 | 1 | 3 | 68. 181 82 | 0.5 | 11.4 | 44 |
| 7 | 1 | 2 | 66. 666 67 | 1.5 | 16.7 | 30 |
| 7 | 2 | 3 | 65. 217 39 | 0.4 | 10.9 | 46 |
| 8 | 0 | 1 | 62. 500 00 | 9.8 | 31.3 | 16 |
| 8 | 1 | 2 | 58. 823 53 | 1.0 | 14.7 | 34 |
| 9 | 0 | 1 | 55. 555 56 | 6.9 | 27.8 | 18 |
| 9 | 1 | 2 | 52. 631 58 | 0.7 | 13.2 | 38 |
| 10 | 0 | 1 | 50. 000 00 | 5.0 | 25.0 | 20 |
| 10 | 1 | 2 | 47. 619 05 | 0.5 | 11.9 | 42 |
| 11 | 0 | 1 | 45. 454 55 | 3.8 | 22.7 | 22 |
| 11 | 1 | 2 | 43. 478 26 | 0.4 | 10.9 | 46 |
| 12 | 0 | 1 | 41. 666 67 | 2.9 | 20.8 | 24 |
| 13 | 0 | 1 | 38. 461 54 | 2.3 | 19.2 | 26 |
| 14 | 0 | 1 | 35. 714 29 | 1.8 | 17.9 | 28 |
| 15 | 0 | 1 | 33. 333 33 | 1.5 | 16.7 | 30 |
| 16 | 0 | 1 | 31. 250 00 | 1.2 | 15.6 | 32 |
| 17 | 0 | 1 | 29. 411 77 | 1.0 | 14.7 | 34 |
| 18 | 0 | 1 | 27. 777 78 | 0.9 | 13.9 | 36 |
| 19 | 0 | 1 | 26. 315 79 | 0.7 | 13.2 | 38 |
| 20 | 0 | 1 | 25. 000 00 | 0.6 | 12.5 | 40 |
| 21 | 0 | 1 | 23. 809 52 | 0.5 | 11.9 | 42 |
| 22 | 0 | 1 | 22. 727 27 | 0.5 | 11.4 | 44 |
| 23 | 0 | 1 | 21. 739 13 | 0.4 | 10.9 | 46 |

### II.F. Nonlinearity of the phase meter

Nonlinearity refers to how the measurement error behaves when the true phase moves over a signal cycle, i.e., from 0 rad to $2\pi$ rad. Practically we can experimentally confirm the performance of the phase meter in the nonlinearity, not in the phase measurement error, since appropriate phase standard is not available. In addition, for the readout of a heterodyne interferometer, nonlinearity rather than phase error should be minimized because it corresponds with the cyclic displacement error, a critical error source in displacement measurements, such as positioning stages.

From test measurements presented in this subsection, we conclude that the phase meter has a nonlinearity lower than approximately $10^{-4}$ rad for input signal frequency above ~1 MHz, when the input signal frequency is different from the singular frequency. Practically, the fluctuation induced by the flicker noise (Section III.D) is typically larger than $10^{-4}$ rad, and thus the nonlinearity does not contribute as the main source of measurement uncertainty in the long-term measurement. In contrast, when the input signal frequency is at the singular frequency, the nonlinearity ranging from $10^{-4}$ rad to $7 \times 10^{-2}$ rad occurs, according to Table S2.

### *1. Nonlinearity at the singular frequencies*

Among the four errors, we must consider only the ZI error for low-drift input signals with low measurement bandwidth, which are realistic conditions for precise low-throughput measurements (Subsection II.H.2). Hence, the ZI error appears in the measurement results only when the input signal frequency is near the singular frequencies listed in Table S2:

$$r|f_{\text{sig}} - f_{\text{sglr}}| < f_{\text{MBW}}. \tag{S104}$$

Here, $f_{\text{sglr}}$ is the singular frequency near the input signal frequency and $r$ is the frequency factor associated with the singular frequency (Table S2).

$$r \equiv 2(sp + q) \tag{S105}$$

Hence, when the input signal frequency is 100 MHz + 1 Hz, for example, the fluctuation from the nonlinearity will appear as peaks of 10 Hz and its harmonics. As such fluctuations can be removed by applying sufficient rejection with LPF over 10 Hz, we must consider nonlinearity from ZI error only when the input signal frequency is equal to the singular frequencies.

We measured the nonlinearity for the actual setup as shown in Figure S18(a). We use a 2-ch function generator (Keysight 33622A) to generate two input signals, the frequency of which is slightly different from each other. We particularly used $f_{\text{sig}} + 10$ (Hz) for Ch. A and $f_{\text{sig}}$ (Hz) for Ch. B. The power of both the signals is +7 dBm. The phase difference is recorded by the phase meter at 10-kHz sampling rate and the measurement time length is 10 s. The phase difference is fitted with $20\pi$-rad/s drift, corresponding to a 10-Hz frequency difference. Then the result is applied with a band-pass filter having a cutoff frequency of 5 Hz and 1 kHz to eliminate noise and drifts. The nonlinearity is then divided into 100 segments and averaged coherently for the 100 cycles from 0 to $2\pi$ to obtain the results. A 10-MHz frequency reference output from the function generator is input into the phase meter so that the clocks of the two instruments are phase-locked and the input signal matches the singular frequencies.

The curves in Figure S18(b1)-(b6) show the results for the six different input signal frequencies: 100 MHz $(s, p, q) = (5, 0, 1)$, 93.75 MHz $(5, 1, 3)$, 75 MHz $(6, 2, 3)$, 62.5 MHz $(8, 0, 1)$, 50 MHz $(10, 0, 1)$ and 31.25 MHz $(16, 0, 1)$. The fundamental peak amplitudes and the frequency factors associated with these frequencies are found in Table S2; e.g., ~$4 \times 10^{-3}$ rad and $r = 10$ for $f_{\text{sig}} = 100$ MHz. The results are matched with theoretical estimation; for instance, Figure S18(b1) shows the nonlinearity of a 10-cycle with a peak of ~$4 \times 10^{-3}$ rad, corresponding with the theoretical estimation (shown in broken red lines). In the case of $f_{\text{sig}} = 62.5$ MHz (Figure S18(b4)), the peak is ~$1 \times 10^{-3}$ rad and the cycle is 16. In Figure S18(b2), (b3) and (b6), fluctuations due to low-frequency noise is also observed.

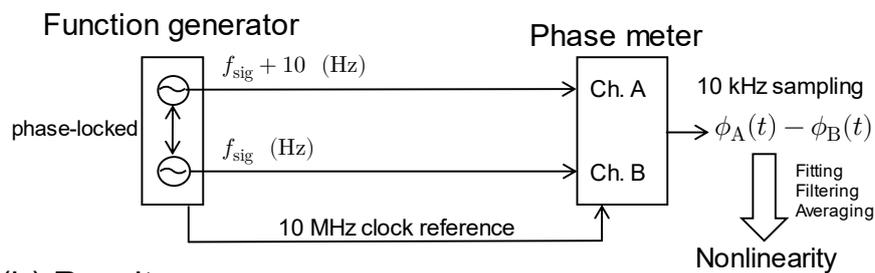

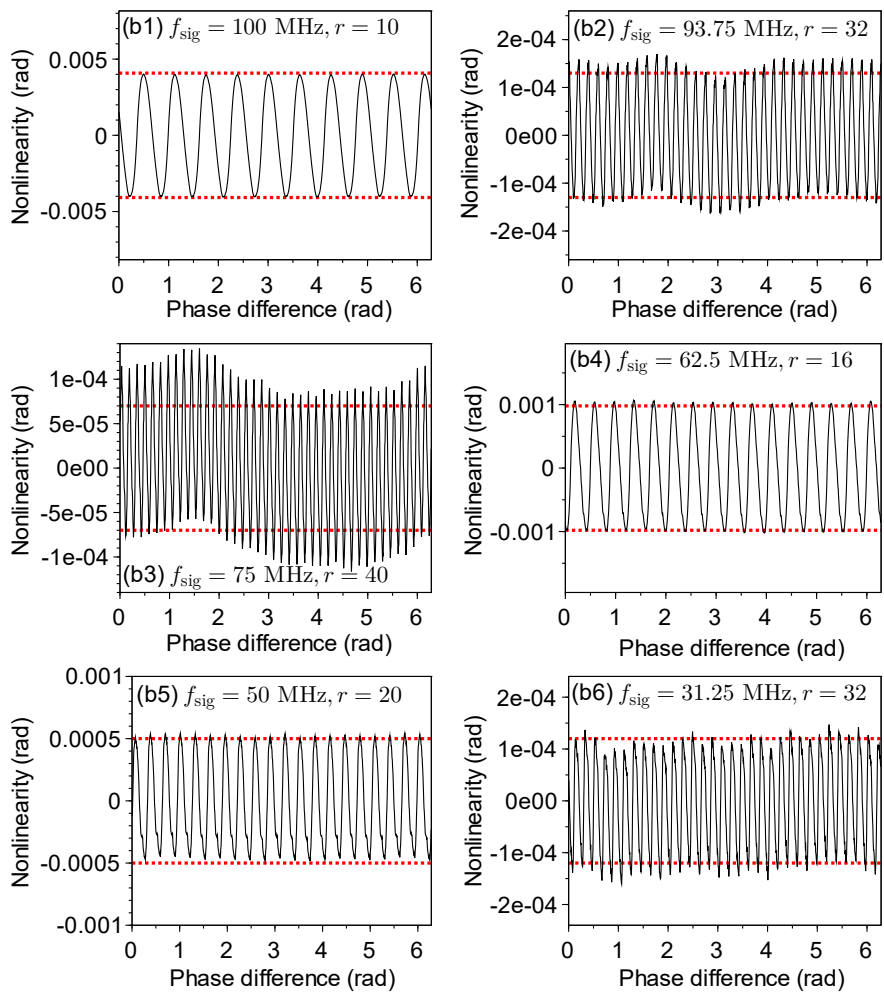

Figure S18. Measurement of nonlinearity at singular frequencies. (a) Measurement setup, (b) Results

### *2. Nonlinearity not at the singular frequencies*

In contrast, the nonlinearity for the frequency different from the singular frequency is unaffected by the ZI error; the nonlinearity is determined by the imperfection of the phase meter, other than the phase measuring algorithm. In Figure S19(a), we experimentally evaluated nonlinearity for 14 different input signal frequencies from 1.3 MHz to 110.3 MHz and confirmed that the nonlinearity is less than $10^{-4}$ rad for these cases. The irrational frequency of 31.41592659 MHz also shows a small nonlinearity (Figure S19(b)). As long as we evaluated nonlinearity in different $f_{\text{sig}}$, we confirmed similar levels of nonlinearity.

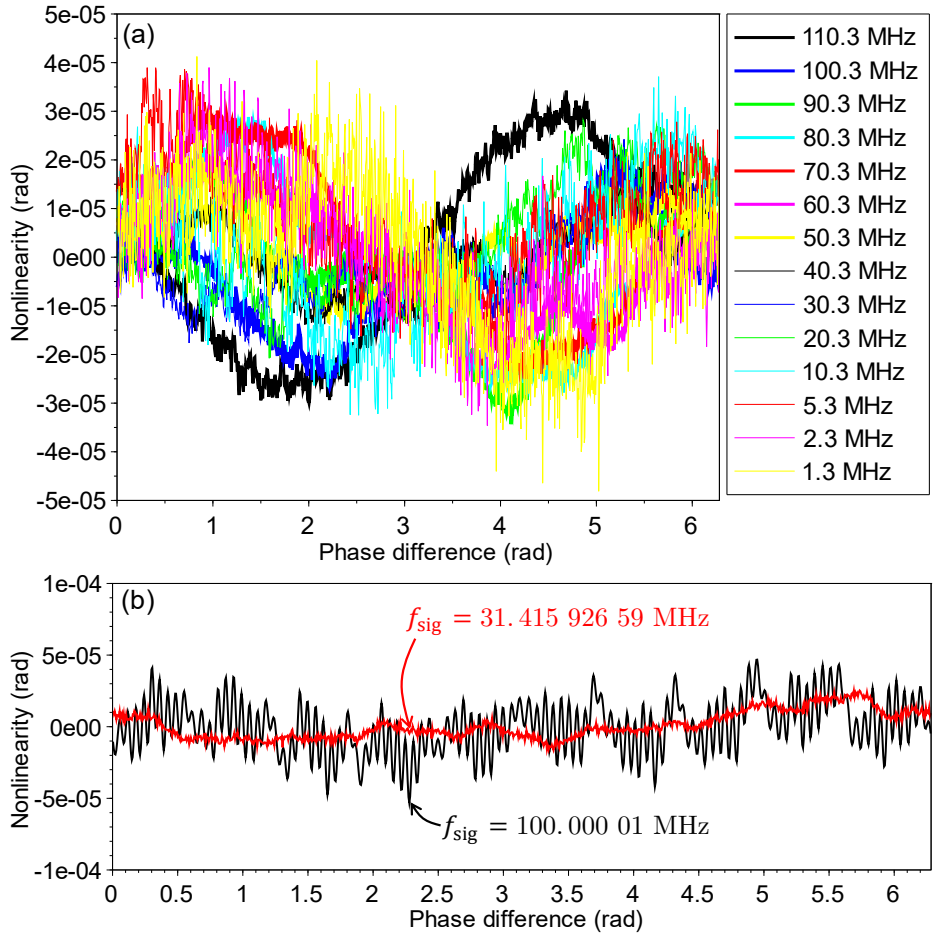

Figure S19. Measured nonlinearity for different input signal frequencies. (a) Results far from singular frequency. (b) Results for the irrational frequency (31.41592659 MHz) and slightly-different frequency (100.00001 MHz).

Here we note that even if the input signal frequency is too close to the singular frequencies, the nonlinearity can still be avoided by two methods. One is lower measurement bandwidth and an LPF with sufficiently cutoff. This method also sacrifices the signals above the measurement bandwidth and generally leads to a longer measurement time.

The other way is that the use of different clock frequencies for the ADC. Even the singular frequencies can be avoided by inputting the frequency-shifted reference clock to the phase meter. For instance, if the input signal frequency is 100 MHz, we can effectively shift the frequency by 1 ppm using a reference clock, which is different from the nominal frequency (10 MHz) by 1 ppm. For such a case, the level of nonlinearity is similar to that for

S44

100.0001 MHz (Figure S19(b)). With this technique, the measurement bandwidth can be increased even when $f_{\text{sig}} = f_{\text{sglr}}$. For example, the fundamental frequency of the ZI error $r|f_{\text{sig}} - f_{\text{sglr}}|$ is shifted by $rf_{\text{sglr}} \times 10^{-7}$ (Hz) when the clock frequency is shifted by 0.1 ppm. Here, $f_{\text{sglr}} > 20$ MHz and $r > 10$, approximately (Table S2). Therefore, we can cut the ZI error out by choosing a low measurement bandwidth; $f_{\text{MBW}} \sim< 20$ Hz.

Note that we can only evaluate an approximate upper limit of the nonlinearity for different frequencies in this evaluation. This is because we need a phase standard that should have a much lower nonlinearity than $1 \times 10^{-4}$ rad to distinguish whether the function generator or phase meter is the source of the nonlinearity for this test setup.

### *2. Nonlinearity for square waves*

We also confirmed excellent-low nonlinearity for low-frequency square waves. In Figure 20(a), the test setup is shown. We used 1-kHz input signal with 1-$V_{\text{p-p}}$ amplitude. Difference from Figure S18(a) is that the two frequencies for Ch. A and Ch. B are the same; the phase difference is made by the function generator. This is because far longer measurement time is needed if we use the same setup as shown in Figure S18(a). The phase difference is set to be 0 rad to $2\pi$ rad with an interval of $2\pi/32$ rad. We selected 1/32 cycle as the phase interval because of the resolution of phase accumulator in the direct digital synthesizer (DDS) system in the function generator; if the target phase offset is irrational number in the DDS, a rounding error may occur. The rounding error cannot be negligible; $2\pi/2^{16}/2 \sim 5 \times 10^{5}$ rad. The phase difference is recorded at 250-kHz sampling rate and 5-s measurement time. Then an LPF with the cutoff frequency of 200 Hz is applied and the average is taken on the time range between 0.1 s and 4.9 s to avoid the ripple.

The difference between the obtained phase and set phase is calculated and plotted in Figure 20(b). An offset of $1.11 \times 10^{-7}$ rad, which comes from cable difference of the two channels, is subtracted. The measured nonlinearity is less than $1 \times 10^{-8}$ rad, which corresponds to $\sim$ 1.6 ps in time difference. In other words, time difference measurement up to $1 \text{ ms} = (1 \text{ kHz})^{-1}$ with $\sim$1.6 ps nonlinearity is achieved with the phase meter.

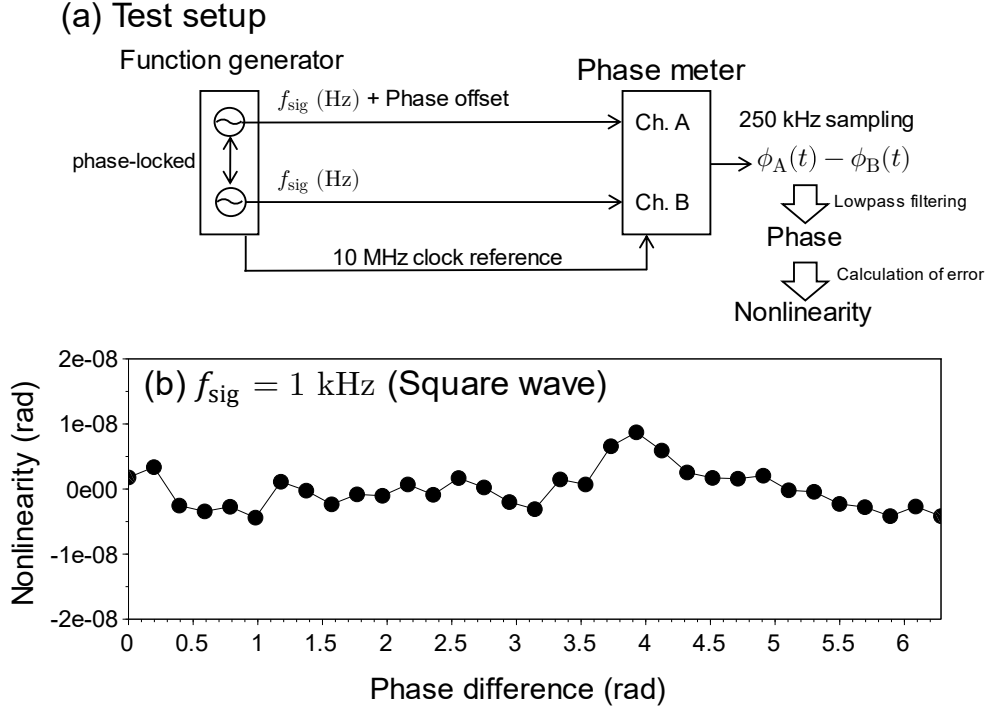

Figure S20. Nonlinearity measurement for 1-kHz square waves. (a) Test setup, (b) Result.

## II.G. Numerical simulation and comparison with real data

Numerical simulation is needed to estimate how large error is present in the high offset frequency range, i.e., when the measurement bandwidth is near the input signal frequency (cf. cases (3)-(ii) and (4)-(iii) in Table S3). Here, we show two examples of numerical simulation and compare them with real data from the phase meter.

Figs. S21 and S22 present a comparison between simulated and measured data. Both figures show that at high offset frequencies, the QA error is dominant. However, at low offset frequencies, the ZI error becomes larger. We calculate the QA and ZI errors for sinusoidal signals and combine them via summation for simplicity. They can be calculated more precisely if we use the complex frequency response. Here, our simulation includes an additional signal processing framework because only data up to 250 MHz are accessible, unlike 1 GHz (the sampling frequency of the ADC) for our hardware. Therefore, we used the following steps to obtain the final measurement from the instantaneous phase estimator at 1 GHz: (i) apply a 1 GHz → 250 MHz boxcar filter, and (ii) resample at 250 MHz. This process comes from a 1:4 parallelization (Figure 1 in the main text) because the FPGA cannot drive with a clock faster than 250 MHz. This parallelization makes the implementation easier; however, it could lead to another aliasing from high-frequency to low-frequency ranges through the resampling.

Figure S21 shows the first case (example 1) with an input signal frequency of 71.429 MHz (1 GHz/14)+10 kHz, i.e., $(s,q,p) = (7,0,1)$ and $\delta f = 10$ kHz . Here, the fundamental frequency becomes $2 \times 7 \times \delta f = 140$ kHz. Therefore, a fundamental peak at 140 kHz and harmonics can be found in both the QA error (Figure S21(a)) and ZI error (Figure S21(b)). The fundamental peak in the ZI error is higher than that in the QA error because the input signal frequency of 71.439 MHz is relatively high, and there is high nonlinearity around the zero-crossings. However, the frequency dependency is different. The ZI error peak exhibits a −60 dB/dec slope, whereas the QA error peaks have a −20 dB/dec slope. Therefore, the QA error exceeds the ZI error above ca. 1 MHz in this case. At high offset frequencies, for instance, above several tens of MHz, large peaks are present in both the QA and ZI errors. The lowest peak is at approximately 35.7 MHz, corresponding to $f_{\text{sig}}/2$, originating from the aliasing in the resampling process from 1 GHz to 250 MHz. Figure S21(d) shows the data obtained using our hardware. Except for the white noise made by the signal-to-noise ratio (SNR) of the ADC, the peaks correlate well with the simulated results.

Figure S22 shows the second case (example 2) with an input signal frequency of $10.000 \text{ MHz} + 20 \text{ kHz}$, i.e., $(s, q, p) = (50, 0, 1)$ and $\delta f = 20$ kHz. The fundamental frequency becomes 1 MHz. For offset frequencies above several MHz, the QA error induces many peaks. The ZI error is negligible compared to the QA error because the input signal frequency is low enough to that nonlinearity in $f_{\text{ADC}}^{-1} = 1$ ns around the zero-crossings is relatively small. As in the first case (example 1), the simulation correlates well with the measured data, except for the white noise.

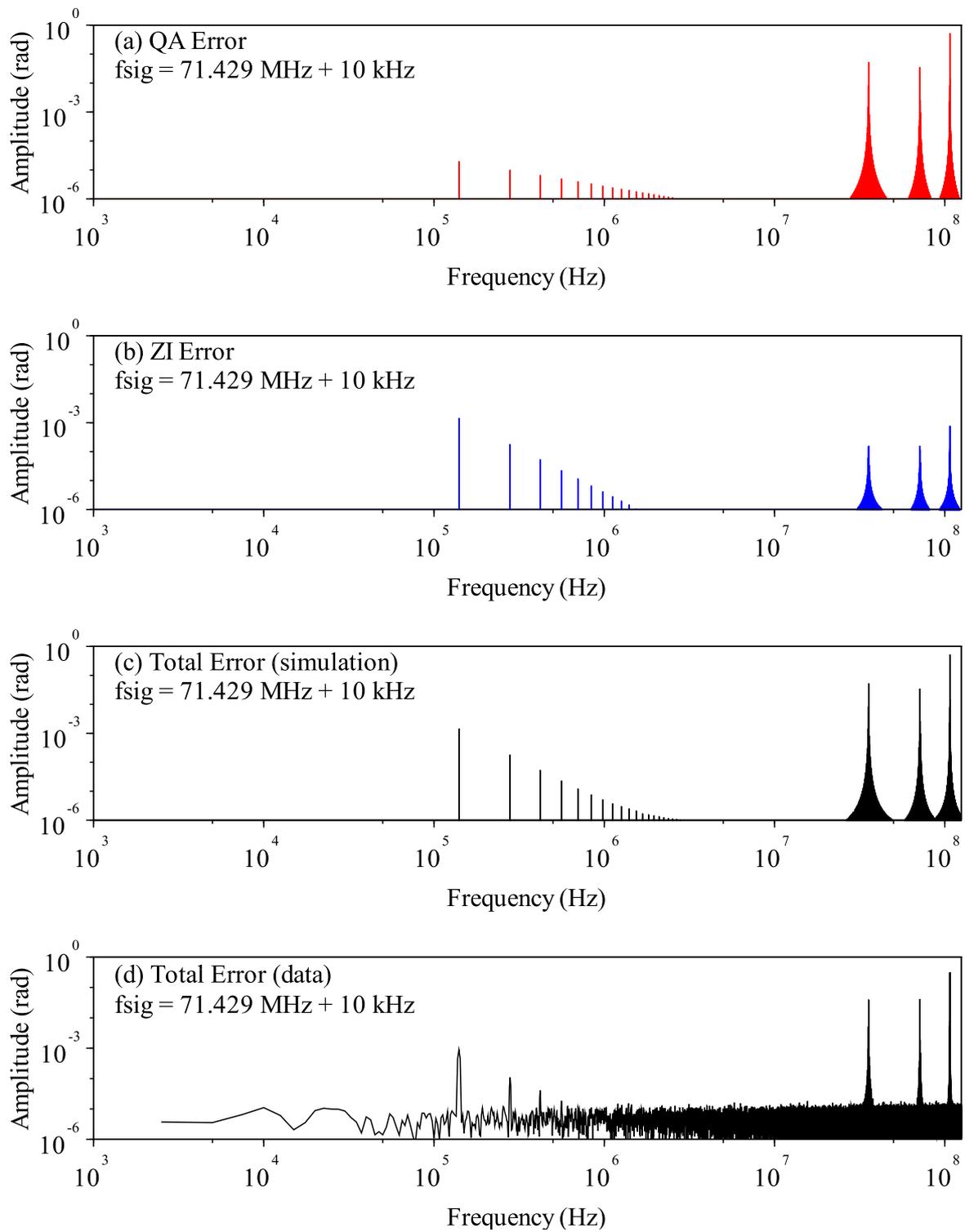

Figure S21. Comparison of simulation and data for example 1. Input signal frequency is 71.429 MHz + 10 kHz. i.e., $(s, q, p) = (7, 0, 1)$ and $\delta f = 10$ kHz.

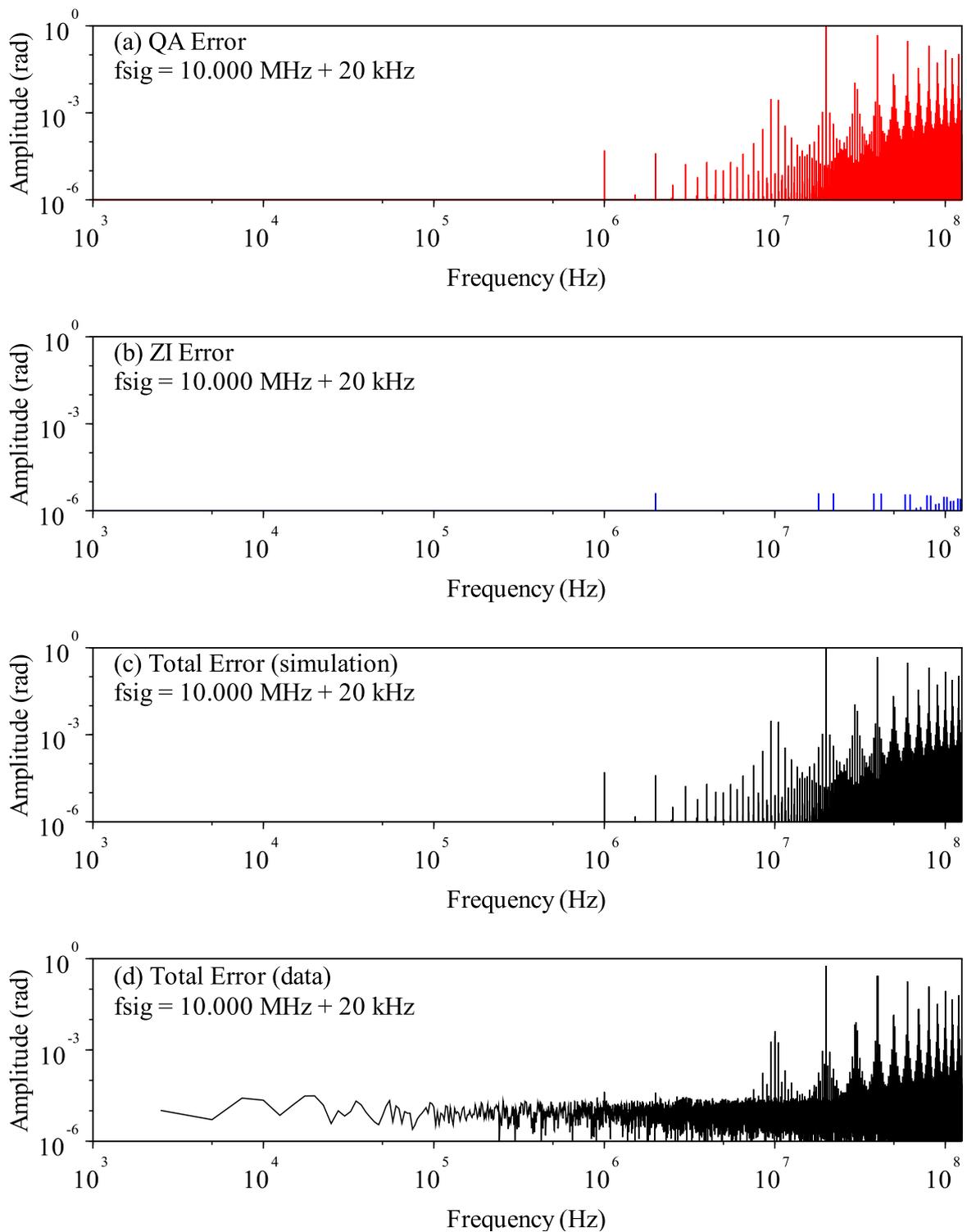

Figure S22. Comparison of simulation and data for example 2. Input signal frequency is 10.000 MHz + 20 kHz. i.e., $(s, q, p) = (50, 0, 1)$ and $\delta f = 20$ kHz.

### II.H. Summary: estimation of total error

Here we summarize the presented error components and show how to roughly estimate the total error. We first show the errors for a general case. Then, we show a special case in which the input signal is limited to low drift. Measurement bandwidth is also assumed to be far lower than the input signal frequency.

#### *1. General case*

We consider general input signals and do not assume any shape, e.g., sinusoidal, square, etc., or frequency. Table S3 lists four error components that are described in detail in the following paragraphs.

(1) TA error can be estimated by using Equation (S48), indicating that TA error is proportional to the average rate of change of the period of the input signal. If the input frequency has small drift, Equation (S48) is approximated to Equation (S49). TA error is generally still smaller than other types of error, even if the drift is large.

(2) FZ error only appears in the first measured or acquired data of the phase measurement. As shown in Equation (S61), the error is less than $(f_{\mathrm{MBW}} f_{\mathrm{sig}}^{-1})^2/4$. In most cases, FZ error is negligible when compared the noise. If FZ error is high, the first measured phase, $\Phi[1]$, can be abandoned.

(3) QA error emerges at low frequencies only when the input signal frequency is close to the singular frequency. The fundamental frequency and its amplitude can be estimated by Equations (S72) and (S73). The fundamental amplitude is proportional to the detuning frequency; therefore, in practice, the peak becomes less than the noise floor when $\delta f \to 0$. See also Figure S10. When the measurement bandwidth is higher, i.e., $f_{\mathrm{MBW}}$ is near $f_{\mathrm{sig}}$, QA error generally dominates the error spectrum in high offset frequencies. In this case, the error spectrum must be estimated by numerical calculation. See the examples in Figure S21 and Figure S22. Note that QA error does not depend on the signal shape; i.e., QA error is the same for sinusoidal, square-wave, and other input signals.

(4) ZI error also appears at low offset frequencies only when the input signal frequency is around the singular frequencies. In general, ZI error makes smaller peaks in high-frequency components than that of QA error because $\gamma_{\mathrm{MAX}}$ is by far smaller than 1 rad, as shown in Figure S14(a); in contrast, QA error has a peak-to-peak amplitude of $\pi$ rad (Figure S9(a)). The most important characteristic of ZI error is that it remains in the DC component when the detuning frequency is zero. The DC phase error also means that there is a secular phase shift in the results. The amplitude depends on the nonlinearity of the input signal at zero-crossings. To remove the error, a relatively large detuning frequency and lower measurement bandwidth are effective. Reducing the input signal frequency is also an effective way to avoid ZI error.

Table S3. Summary of the error components for general input signals.

| Error components | Related parameters | Description | Ref. Eq. |
|---|---|---|---|
| (1) Trapezoidal approximation error $\delta\Phi_{\mathrm{TA}}$ | Input signal frequency $f_{\mathrm{sig}}$ <br> Frequency drift $\dot{f}_{\mathrm{sig}}$ | $\delta\Phi_{\mathrm{TA}} \cong \frac{\pi}{24} \frac{f_{\mathrm{sig(end)}}^{-1} - f_{\mathrm{sig(start)}}^{-1}}{t_{\mathrm{end}} - t_{\mathrm{start}}}$ [large drift] <br><br> $\delta\Phi_{\mathrm{TA}} \cong -\frac{\pi}{24} \frac{\dot{f}_{\mathrm{sig}}}{f_{\mathrm{sig}}^2}$ [small drift] | (S48) <br><br> (S49) |
| (2) First zero-crossing error $\delta\Phi[1]$ | Input signal frequency $f_{\mathrm{sig}}$ <br> Measurement bandwidth $f_{\mathrm{MBW}}$ | $\left(\frac{\delta\Phi[1]}{\Phi[1]}\right) < \left(\frac{f_{\mathrm{MBW}}}{2f_{\mathrm{sig}}}\right)^2$ <br> Note: for the first sample only | (S61) |
| (3) Quantization–aliasing error $\delta\Phi_{\mathrm{QA}}$ | Input signal frequency $f_{\mathrm{sig}}$ <br> Measurement bandwidth $f_{\mathrm{MBW}}$ <br> Detuning frequency $\delta f$ <br> $\left(f_{\mathrm{sig}} = \frac{1}{2\left(s + \frac{q}{p}\right)} f_{\mathrm{ADC}} + \delta f\right)$ | (i) $f_{\mathrm{MBW}} \ll f_{\mathrm{sig}}$ <br> Fundamental peak with amplitude of $\frac{2\delta f}{p f_{\mathrm{ADC}}}$ at the frequency of $2(sp+q)\delta f$ and its harmonics with $-20$ dB/dec dependence appear in the measurement bandwidth only when input signal frequency is around singular frequencies. Maximum amplitude is $\frac{f_{\mathrm{MBW}}}{2p^2 f_{\mathrm{ADC}}^2} f_{\mathrm{sig}}$ when detuning frequency is at maximum, $\frac{f_{\mathrm{MBW}}}{2(sp+q)}$. See also Figure S10. | (S72) <br> (S73) <br> (S74) |
| | | (ii) $f_{\mathrm{MBW}} \sim f_{\mathrm{sig}}$ <br> In high offset frequencies, there are many peaks induced by the QA error. Numerical simulation is needed. See Figure S21 and Figure S22 for example. | |
| (4) Zero-crossing interpolation error $\delta\Phi_{\mathrm{ZI}}$ | Input signal frequency $f_{\mathrm{sig}}$ <br> Measurement bandwidth $f_{\mathrm{MBW}}$ <br> Detuning frequency $\delta f$ <br> $\left(f_{\mathrm{sig}} = \frac{1}{2\left(s + \frac{q}{p}\right)} f_{\mathrm{ADC}} + \delta f\right)$ <br> Input signal nonlinearity | (i) $f_{\mathrm{MBW}} \ll f_{\mathrm{sig}}$ and input signal is sinusoidal <br> Fundamental peak with amplitude shown in Table S2 at the frequency of $2(sp+q)\delta f$ and its harmonics with $-60$ dB/dec dependence appear in the measurement bandwidth only when input signal frequency is around the singular frequencies. When $\delta f = 0$, DC phase error occurs. See also Figure S16 and Table S2. | |
| | | (ii) $f_{\mathrm{MBW}} \ll f_{\mathrm{sig}}$ and input signal is not sinusoidal <br> Fundamental peak amplitude and harmonics dependence may be different from the case (i). They can be calculated numerically by using oneZIEF shown in (S87), which indicates nonlinearity of the input signal around zero-crossings. | (S87) |
| | | (iii) $f_{\mathrm{MBW}} \sim f_{\mathrm{sig}}$ <br> In high offset frequencies, the ZI error is generally less than the QA error. However, numerical simulation is needed. See Figure S21 and Figure S22 for example. | |

## 2. Special case: low-drift input signals with low measurement bandwidth

We consider a special case, in which the input signal is like that for an ordinary phase meter, i.e. the following two conditions are true. (i) The input signal has sufficiently low drift, which results in negligible TA error. (ii) The measurement bandwidth is sufficiently lower than the critical measurement bandwidth, according to Equation (S80), leading to negligible FZ error and QA error when compared with the noise. Note that the input signal frequency does not have to be low under these conditions.

Table S4 summarizes the errors in this special case. We must consider only ZI error. Only when the input signal frequency is around the singular frequency, the error appears at a frequency of $2(sp+q)\delta f$ and its harmonics, according to Table S2. For a sinusoidal input signal, the fundamental peak has an amplitude of $\frac{4f_{\text{sig}}^3}{p^3 f_{\text{ADC}}^3}$ (Equation (S103)) and its harmonics have a −60 dB/dec dependency (Equation (S101)). If the input signal frequency is much lower than $f_{\text{ADC}}$, the maximum amplitude can be approximated with $4.0 \times \left(\frac{f_{\text{sig}}}{f_{\text{ADC}}}\right)^3$ (Equation (S97)). When $\delta f = 0$, there is DC phase error; its largest amplitude is listed in Table S2.

For signals that have other waveforms, fundamental peak amplitude and harmonics dependence can be calculated numerically by using oneZIEF shown in (S87), which indicates nonlinearity of the input signal around zero-crossings.

In terms of the nonlinearity, we estimate that the phase meter has a nonlinearity of less than ~$10^{-4}$ rad. When the input signal frequency be equal to singular frequencies, the clock of the phase meter should be slightly shifted by inputting an external reference clock to avoid the singularities.

Table S4. Summary of the error components for a special case: low-drift input signals with low measurement bandwidth.

| Error components | Related parameters | Description |
|---|---|---|
| (1) Trapezoidal approximation error | | Negligible |
| (2) First zero-crossing error | | Negligible |
| (3) Quantization–aliasing error | | Negligible |
| (4) Zero-crossing interpolation error | Input signal frequency $f_{\text{sig}}$<br>Measurement bandwidth $f_{\text{MBW}}$<br>Detuning frequency $\delta f$<br>$\left(f_{\text{sig}} = \frac{1}{2\left(s+\frac{q}{p}\right)} f_{\text{ADC}} + \delta f\right)$ | Fundamental peak at the frequency of $2(sp+q)\delta f$ and its harmonics appear (Table S2). For sinusoidal signals, the fundamental peak amplitude is $\frac{4f_{\text{sig}}^3}{p^3 f_{\text{ADC}}^3}$ and harmonics dependence is −60 dB/dec. For not sinusoidal signals, fundamental peak amplitude and harmonics dependence can be calculated numerically by using oneZIEF. When $\delta f = 0$, DC phase error occurs. |

## III. Noise analysis

Noise is a random fluctuation in a measurement that originates from random processes in devices and/or references. Generally, noise is one of the most important parameters of a measurement apparatus because it limits how small of a signal can be detected. In this section, we explain in detail how the noise is generated in the phase meter, and we estimate how large the types of noise typically are.

### III.A. Overview: noise sources

To analyze the sources of noise, we consider a block diagram of the phase meter as presented in Figure S23. The two input signals are digitized with different ADCs that use a common sampling clock. The digitized data are converted to the phase with the phase measuring algorithm by digital signal processing. With the phase measuring algorithm, the single-channel phase for each input (expressed by $\phi_A(t)$ and $\phi_B(t)$ in Figure S23) is estimated. We call these single-channel phases. In most applications, such as interferometric displacement measurement, the phase difference of the two input channels is of interest. The difference $\phi_A(t) - \phi_B(t)$, which we call phase difference, is easily obtained by subtraction in the digital signal processing framework.

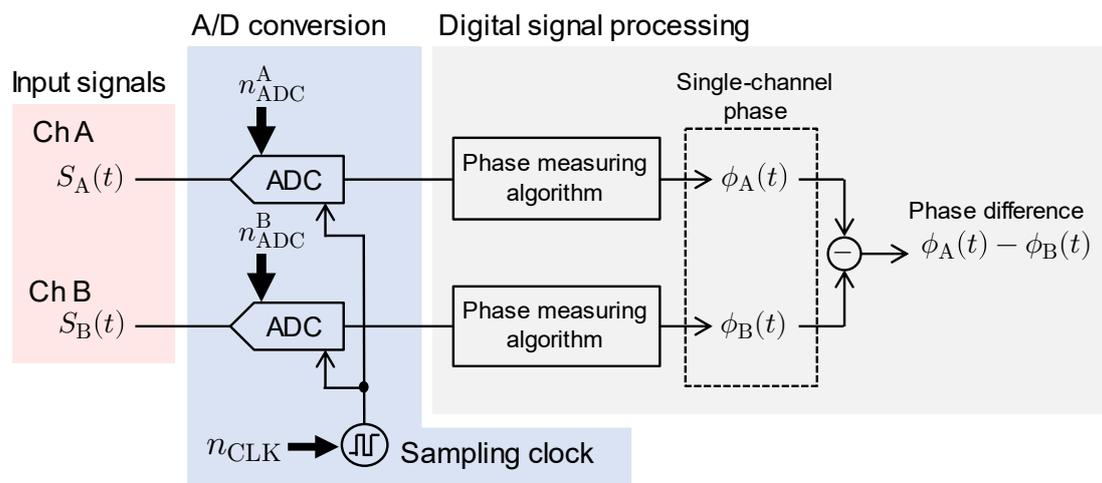

Figure S23. Block diagram of the phase meter for noise analysis. Single-channel and phase difference are shown. $n_{ADC}^A$ and $n_{ADC}^B$ are the noise of the ADC for channel A and channel B, respectively. The phase noise of the sampling clock is shown as $n_{CLK}$.

Throughout the whole phase measuring process, noise is only generated in the analog to digital (A/D) conversion. This occurs because digital signal processing adds no noise as long as it is adequately designed not to produce round-off errors (or quantization errors). There are two types of noise in the A/D conversion process: correlated and uncorrelated noise between the two channels. The correlated noise arises from the sampling clock. On the other hand, uncorrelated noise comes from each A/D conversion process and consists of white phase noise and flicker phase noise. The correlated and uncorrelated noise could include coupling processes to the nonlinear property of the input signals. For instance, amplitude modulation to phase modulation (AM-PM) conversion and DC-offset to phase conversion because of ADC nonlinearity could cause additional noises. However, we suppose these contributions are sufficiently low here, except for the nonlinear conversion for flicker phase noise (Subsection D).

Figure S24(a) clarifies how the noise appears in the spectral domain. In the single-channel phase, the results are generally limited by the sampling clock noise. The phase difference noise consists of only the uncorrelated noise when the frequencies of the two input channels are the same. Levels are the root-sum-square of those for the two ADC channels. Below we describe the three types of noise. Figure S24(b) shows an example of the measurement for the 30-MHz, +7-dBm signal.

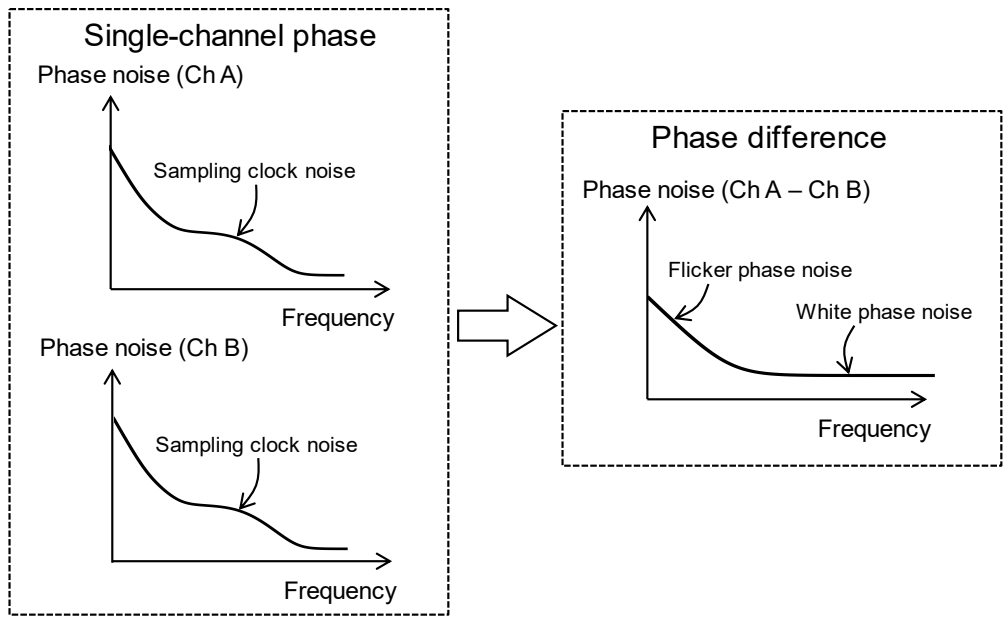

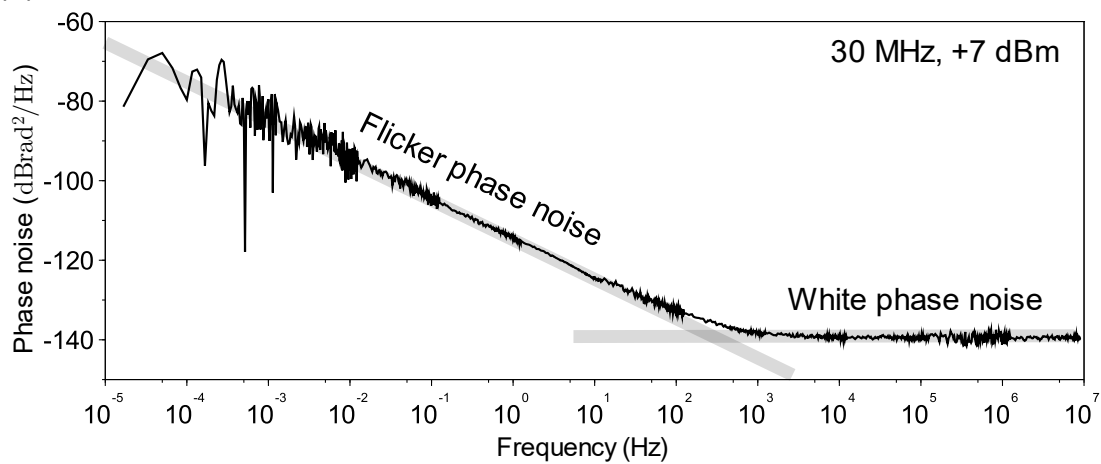

Figure S24. (a) Noise spectra of single-channel and phase differences when the input frequencies for channel A and B are the same. (b) An example of phase difference. Input signal is 30 MHz and +7 dBm.

1. Sampling clock noise. Even if the input signal contains no phase noise, the single-channel phase contains phase noise induced by the reference clock, called sampling clock noise. Here we suppose that electrical noise in analog frontend, such as electronic interference and common-mode drift, is sufficiently avoided. The sampling clock noise is a correlated noise because we assume that both ADCs are driven by a common clock. The noise appearing in both channels is identical when the two input frequencies are the same. In other words, in such a case, this noise will be canceled out upon measuring the phase difference. Note that if the two ADCs use two independent clocks, this noise should be regarded as uncorrelated noise and cannot be canceled out in the differential measurement. In the following Subsection III.B, we explain that this noise is proportional to the input signal frequency.

2. White phase noise. The first component of the uncorrelated noise, which appears flat in the frequency domain of the phase difference, comes from the white voltage noise of the ADC that is usually known as the noise spectral density or SNR. The noise of the two channels is not correlated because the white noise of the ADC is independent of that of other channels. In the following Subsection III.C, we prove that the white phase noise level is determined by the following parameters, (i) SNR of the ADC, (ii) input signal frequency, and (iii) effective power of the input signal. Here, the effective power refers to the equivalent power of the sinusoidal signal that exhibits the same slew rate at zero-crossings as that of the input signal. Based on this relationship, the white noise for low-frequency square waves becomes significantly low since the effective power is far larger (for example, more than +100 dBm) than the maximum input power at the ADCs.

3. Flicker phase noise. The second component of the uncorrelated noise is flicker phase noise, which is dominant in the low offset frequency. The noise is called flicker noise because the noise has a −10 dB/dec dependency. However, it is generally challenging to identify the sources of noise. We suspect that there are two noise sources. One is the flicker noise component of ADC aperture jitter, and the other is the nonlinear conversion from near-DC flicker voltage noise at the ADCs. In Subsection III.D, we show that (i) The level of noise is independent of the power of the input signal and (ii) the level of noise depends on the frequency of the input signal.

### III.B. Sampling clock noise
#### *1. Effect of the sampling clock*
We show how the sampling clock noise affects the phase measurement in Figure S25. Constant input signal frequency is assumed. The ideal sampling point is changed to the actual sampling point due to the timing fluctuation $\delta t$.

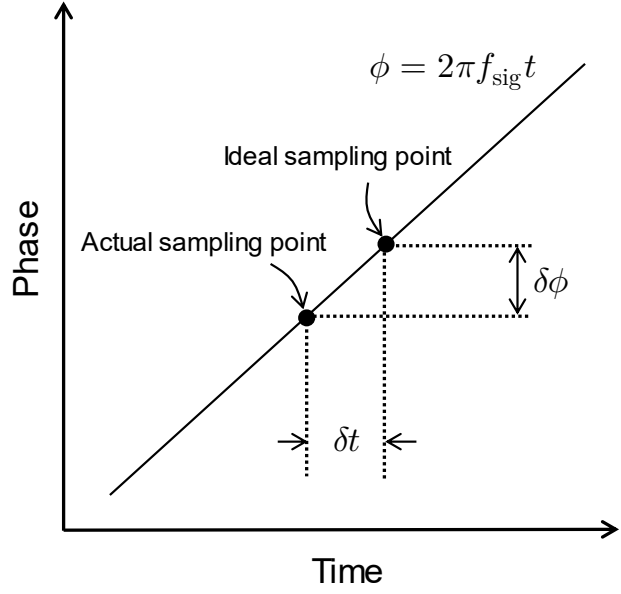

Figure S25. The effect of sampling timing fluctuation on phase measurement. The input signal frequency is $f_{\text{sig}}$. The phase measurement differs by $\delta\phi$, which is induced by the clock noise fluctuation $\delta t$. Ideal sampling point refers to a sampling point without sampling clock noise. The actual sampling point is that with the sampling clock noise.

As shown in Figure S25, the resulting phase fluctuation $\delta\phi$ is related to timing fluctuation $\delta t$ as follows:

$$\delta\phi = 2\pi f_{\text{sig}} \delta t, \tag{S106}$$

and the phase fluctuation of the sampling clock, $\delta\phi_{\text{CLK}}$, is expressed by the fluctuation of the sampling timing $\delta t$ and the clock frequency $f_{\text{CLK}}$.

$$\delta\phi_{\text{CLK}} = 2\pi f_{\text{CLK}} \delta t \tag{S107}$$

Therefore, the phase fluctuation is expressed as

$$\delta\phi = \frac{f_{\text{sig}}}{f_{\text{CLK}}} \delta\phi_{\text{CLK}}. \tag{S108}$$

Note that Equation (S108) also indicates that the timing jitter induced by the clock phase noise is equal to the timing jitter of the clock normalized to the input signal frequency. Converting the phase fluctuation to phase noise PSD, which has units of $\text{rad}^2/\text{Hz}$, the following Equation is obtained for the sampling clock noise, $S_{\phi,\text{CLK}}$.

$$S_{\phi,\text{CLK}} = \left(\frac{f_{\text{sig}}}{f_{\text{CLK}}}\right)^2 S_{\text{CLK}} \tag{S109}$$

$S_{\text{CLK}}$ is the phase noise PSD at the frequency $f_{\text{CLK}}$. Note that the phase fluctuation is proportional to $f_{\text{sig}}$. For the phase difference, the noise contribution becomes

$$S_{\phi,\text{CLK}} = \left(\frac{\Delta f_{\text{sig}}}{f_{\text{CLK}}}\right)^2 S_{\text{CLK}}, \tag{S110}$$

where $\Delta f_{\text{sig}}$ is the frequency difference between the two input channels, hence the noise does not contribute when the frequency difference is sufficiently small.

## 2. Measuring the sampling clock noise

The sampling clock noise is generally not easy to measure because the phase noise of the ADC sampling clock cannot be directly accessed in most cases, i.e., the clock signal driving the ADC generally cannot be extracted from instruments. In addition, in our hardware, we cannot access the single-channel data due to the data stream design of the FPGA. Therefore, we used a phase difference measurement technique, as shown in Figure S26.

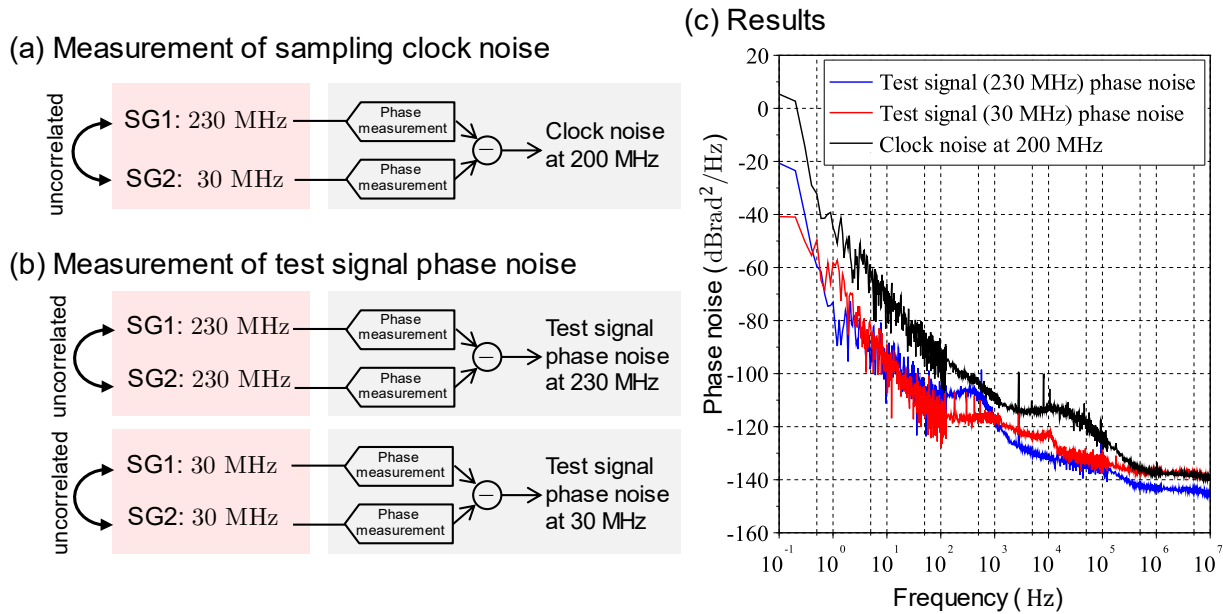

Figure S26. Sampling clock noise measurement: (a) measurement of clock noise; (b) measurement of the phase noise of the test signal; (c) clock noise (black) and phase noise of test signals at 230 MHz (blue) and 30 MHz (red).

We used two different input signal frequencies, 230 MHz and 30 MHz. As shown in Figure S26(a), two signal generators (SGs) (both are the same model, Agilent E4425B) are used to generate the sinusoidal input signals. The data obtained from our hardware represent the clock phase noise at 200 MHz according to Equation (S110). Note that the phase differences are linearly fitted, and the linear trend is subtracted from it because it contains phase evolution with 200 mega cycles per second ($2\pi \times 2 \times 10^8$ rad/s). One concern is that the result may be limited by the phase noise of the test signals. Therefore, we also conducted phase noise measurements for both 230 MHz and 30 MHz with the same input signal frequency, as generated from the two SGs (Figure S26(b)). In all three measurements, the two SGs are not locked with an external reference signal such that the phase noise of the two input signals remains uncorrelated. Furthermore, the sampling clock is also not locked to any external reference. We obtained phase noise of the signal by multiplying the results by $1/\sqrt{2}$ ($-3$ dB).

The results are presented in Figure S26(c). The clock noise at 200 MHz, which is plotted in black, exhibits the typical shape of clock noise. In particular, the bump at several tens of kHz implies that the signal is phase-locked to some internal reference from a noisy seed oscillator. It is also found that the noise of the test signals does not limit this measurement, except above 500 kHz. We infer that in this high frequency range, the clock phase noise should be better than

−140 dBrad²/Hz. As for low offset frequencies (e.g., 1 Hz), the is a 60-dB difference when comparing the noise level of −45 dBrad²/Hz at 1 Hz to the flicker phase noise at the A/D conversion, measured as −105 dBrad²/Hz at 1 Hz. Therefore, if the difference of the input signal frequency is less than ca. 0.2 MHz, the sampling clock noise is less than the noise from the A/D conversion, even at 1 Hz. Note that the sampling clock can be locked to a reference oscillator to reduce the sampling clock noise in such low offset frequencies.

By applying this measurement technique, locking bandwidth to the external reference of the ADC clock can be measured. The measurement can be done as described here. First, two frequencies are input into Ch. A and Ch. B and a 10-MHz reference clock is connected to the reference input of the digitizer. In this setup, the phase modulation in the reference clock appears in the results. Therefore, the locking bandwidth of the clock management of the digitizer can be observed by changing the phase modulation frequency. From the data, we estimate that the −6 dB bandwidth for the digitizer is around 50 kHz – 100 kHz. The value is consistent with the cutoff in the clock noise (the black curve in Figure S26(c)), which shows the loop bandwidth of the internal PLL for clock management.

### III.C. White phase noise

White phase noise is one of the uncorrelated noise components. It originates from the white voltage noise in the A/D conversion process. The noise is characterized by the level (dBrad²/Hz) in the phase noise spectrum. In this subsection, we analyze this noise and estimate how much white phase noise appears in the results.

#### *1. Estimation of white phase noise*

We express the white phase noise with the ADC noise and signal properties and assume some of the parameters. The parameter $s$ (V/s) represents the slew rate of the input signal at zero-crossings. We also express the noise PSD $P_\text{ND}$ (W/Hz) for the white voltage noise spectrum at the ADC. Note that the unit is not dBm/Hz, hence we use a nonlogarithmic scale in this derivation. The white noise generates voltage deviation in each sample; the magnitude of the voltage deviation $\sigma_\text{V}$ (V) is expressed using the noise density $P_\text{ND}$

$$\sigma_\text{V}^2 = \frac{P_\text{ND} f_\text{ADC} R}{2}, \tag{S111}$$

where $f_\text{ADC}/2$ is the Nyquist frequency for the sampling at the ADC, and $R$ is the input impedance (50 Ω). Based on the nature of zero-crossings, the timing noise of the zero-crossings $\sigma_t$ (s) can be converted to voltage noise with the slew rate at the zero-crossings, $\sigma_t = \sigma_\text{V}/s$; therefore, we obtain

$$\sigma_t^2 = \frac{\sigma_\text{V}^2}{s^2} = \frac{P_\text{ND} f_\text{ADC} R}{2s^2}. \tag{S112}$$

For phase measurement with the given time interval of $T$ (s) and assuming the measurement time is longer than the sampling at the ADC, $T \gg 1/f_\text{ADC}$. As shown in Equation (S30), The RMS noise amplitude of the averaged phase estimator, $\delta\Phi$, can be expressed with the timing of the errors $\delta\tau_j$.

$$\delta\Phi = \frac{\pi}{T}\sum_j^L \delta\tau_j = \frac{\pi}{T}\sqrt{2f_{\text{sig}}T}\sigma_t \quad \text{(S113)}$$

Considering that $L$ is the number of the zero-crossing included in one sampling interval, $L = 2f_{\text{sig}}T$, and that the timing noise $\delta\tau_j$ for each zero-crossing is independent but exhibits identical amplitude, $\sum_j^L \delta\tau_j = \sqrt{L}\sigma_t$.

Next, since the phase noise spectrum density $S_\phi$ (rad$^2$/Hz) of the phase measurement result is white noise, it has the following relationship with $\delta\Phi$.

$$S_\phi = \frac{(\delta\Phi)^2}{\text{BW}_\Phi} = 2T(\delta\Phi)^2 \quad \text{(S114)}$$

Here, the measurement bandwidth, $\text{BW}_\Phi$ (Hz), is $1/(2T)$. By applying Equations (S112) and (S113) to (S114), we derive

$$S_\phi = \frac{2f_{\text{sig}}\pi^2 P_{\text{ND}} f_{\text{ADC}} R}{s^2}. \quad \text{(S115)}$$

Then, we define the effective power of the input signal $P_{\text{eff}}$ (W):

$$P_{\text{eff}} \equiv \frac{1}{2R}\left(\frac{s}{2\pi f_{\text{sig}}}\right)^2. \quad \text{(S116)}$$

This effective power is the equivalent power of the sinusoidal signals, which have the same slew rate at zero-crossings as that of the input signal. By definition, for sinusoidal signals, $P_{\text{eff}}$ is equal to its actual power. A sinusoidal signal with the zero-to-peak amplitude $A$ (V) has a slew rate at zero-crossings of $s = 2\pi f_{\text{sig}} A$. Then, $P_{\text{eff}}$ becomes $P_{\text{eff}} = A^2/2R$, which is equal to the actual signal power. In addition, we also define a degradation factor $D(f_{\text{sig}})$ as follows:

$$D(f_{\text{sig}}) \equiv \left(\frac{4f_{\text{sig}}}{f_{\text{ADC}}}\right)^{-1}. \quad \text{(S117)}$$

This factor indicates how much information from the ADC is used for the algorithm. In the algorithm, only two adjacent data of zero-crossings are used for the calculation; i.e., the effective data rate is $4f_{\text{sig}}$, while the total data rate from the ADC is $f_{\text{ADC}}$. From another viewpoint, this degradation factor reflects the aliasing of noise, corresponding to the ratio between the zero-crossing rate ($2f_{\text{sig}}$) and the Nyquist rate of the ADC ($f_{\text{ADC}}/2$).

Figure S27 shows the frequency dependence of the degradation factor. The slopes are $-10$ dB/dec. When the sampling rate of the ADC ($f_{\text{ADC}}$) decreases, the factor also decreases because the information from the signal is used more efficiently to estimate the phase.

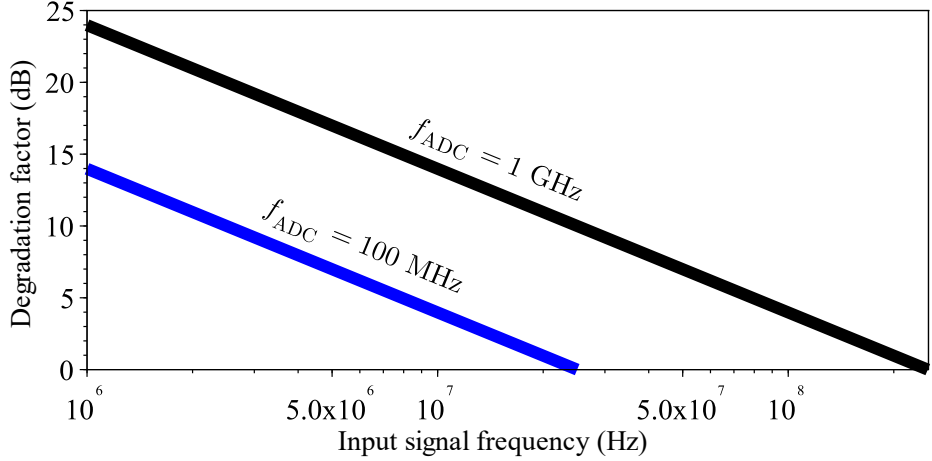

Figure S27. Degradation factor for $f_{\text{ADC}} = 1$ GHz and $f_{\text{ADC}} = 100$ MHz.

From Equations (S115)–(S117), the white phase noise is converted to

$$S_\phi = \frac{P_{\text{ND}}}{P_{\text{eff}}} D(f_{\text{sig}}). \tag{S118}$$

Equation (S118) shows that the phase noise is a product of the degradation factor and $P_{\text{ND}}/P_{\text{eff}}$, thus indicating the effective SNR for the effective power signal.

### *2. Logarithmic expressions and examples*

For convenience, we convert various equations to logarithmic expressions by implementing the following definitions.

$$S_{\phi(\text{dBrad}^2/\text{Hz})} \equiv 10 \log_{10}(S_{\phi,\text{white}}) \tag{S119a}$$

$$P_{\text{ND}(\text{dBm/Hz})} \equiv 10 \log_{10}(P_{\text{ND}}) + 30 \tag{S119b}$$

$$P_{\text{eff (dBm)}} \equiv 10 \log_{10}(P_{\text{eff}}) + 30 \tag{S119c}$$

$$D_{(\text{dB})}(f_{\text{sig}}) \equiv 10 \log_{10}\left(D(f_{\text{sig}})\right) \tag{S119d}$$

Here we show the units of the quantities in the suffix brackets. Taking the logarithm base 10 of both sides of Equation (S118), we obtain

$$S_{\phi \text{ (dBrad}^2/\text{Hz})} = P_{\text{ND}(\text{dBm/Hz})} - P_{\text{eff (dBm)}} + D_{(\text{dB})}(f_{\text{sig}}). \tag{S120}$$

Here, noise density of the ADC, $P_{\text{ND}(\text{dBm/Hz})}$, can be expressed using the parameters that are commonly found in the performance description datasheets of the ADC.

$$P_{\text{ND}(\text{dBm/Hz})} = -\text{SNR}_{(\text{dB})} - \text{BW}_{(\text{dBHz})} + \text{FS}_{(\text{dBm})}, \tag{S121}$$

where $\text{SNR}_{(\text{dB})}$ is the time-domain expression of the SNR for the ADC in dB. $\text{FS}_{(\text{dBm})}$ is the full-scale input of the ADC in dBm. $\text{BW}_{(\text{dBHz})}$ is the logarithmic expression of the Nyquist frequency for the ADC, i.e.,

$$\text{BW}_{(\text{dBHz})} \equiv 10 \log_{10}\left(\frac{f_{\text{ADC}}}{2}\right). \tag{S122}$$

The two independent but identical noises in the two input channels contribute to the measured phase difference; therefore,

$$S_{\phi,\text{diff(dBrad}^2/\text{Hz})} = S_{\phi(\text{dBrad}^2/\text{Hz})} + 3$$
$$= -\text{SNR}_{(\text{dB})} - \text{BW}_{(\text{dBHz})} + \text{FS}_{(\text{dBm})} - P_{\text{eff (dBm)}} \quad (S123)$$
$$+ D_{(\text{dB})}(f_{\text{sig}}) + 3$$

Note that if we should obtain the single-sideband phase noise $\mathcal{L}_{(\text{dBc/Hz})}$ in the unit of $\text{dBc/Hz}$, the following relation can be used:

$$\mathcal{L}_{(\text{dBc/Hz})} = S_{\phi(\text{dBrad}^2/\text{Hz})} - 3. \quad (S124)$$

We present two examples, sinusoidal wave and square wave. For the first example, the ADC in the digitizer, which we also used for the phase meter, has the following parameters[3]: $\text{SNR}_{(\text{dB})} = 67$ dB, $f_{\text{ADC}} = 10^9$ Hz, $\text{BW}_{(\text{dBHz})} = 87$ dBHz, $\text{FS}_{(\text{dBm})} = +9$ dBm. Therefore, when the input signal is sinusoidal with $+7$ dBm power, the white phase noise becomes

$$S_{\phi,\text{diff(dBrad}^2/\text{Hz})} = -145 - 10\log_{10}\left(\frac{f_{\text{sig}}}{100\text{ MHz}}\right). \quad (S125)$$

The white phase noise level has a $-10$ dB/dec dependency on the input frequency.

The other example is square waves. We use the same ADC as for the sinusoidal wave example. Consider the input signal to be a square wave with a fixed slew rate of $s = 0.2$ V/ns. This is the same value we measured with our hardware for $0.7$-$V_{\text{p-p}}$ square wave output from a function generator. Based on this assumption, the effective power becomes

$$P_{\text{eff (dBm)}} = 1 - 20\log_{10}\left(\frac{f_{\text{sig}}}{100\text{ MHz}}\right). \quad (S126)$$

Equation (S126) shows that the effective power can be significantly higher than the actual input power of the ADC. For example, $P_{\text{eff (dBm)}} = +101$ dBm when $f_{\text{sig}} = 1$ kHz. Then, white phase noise is calculated as follows:

$$S_{\phi,\text{diff(dBrad}^2/\text{Hz})} = -139 + 10\log_{10}\left(\frac{f_{\text{sig}}}{100\text{ MHz}}\right). \quad (S127)$$

In this case, the white phase noise level has a $+10$ dB/dec dependency on the input frequency. When the input signal frequency is several tens of MHz, this is not different than that for sinusoidal waves. However, when the input signal frequency becomes much lower, the white noise level decreases significantly. For instance, when $f_{\text{sig}} = 1$ MHz and 1 kHz, the phase noise level becomes $-159$ $\text{dBrad}^2/\text{Hz}$ and $-189$ $\text{dBrad}^2/\text{Hz}$, respectively.

We confirmed the level of white noise in the actual hardware (Figure S28). Figure S28(a) shows the measurement setup. A test signal is generated using a commercial function generator (Keysight 33622A). The test signal is divided by a power splitter (Mini-circuits ZSC-2-1+, 0.1 MHz – 400 MHz) and is input into the two measurement channels of the phase meter. When the input signal frequency is at or below 100 kHz, a resistive power splitter (Mini-circuits ZFRSC-42-S+, DC – 4200 MHz) is used instead, and the power of the test signal is increased by 3 dB to maintain the same power at the input of the phase meter. The measured phase difference is converted to phase noise PSD using the Fourier transformation. The spectral shapes are fitted to calculate the level of white phase noise. Figures S28(b) and S28(c) show the results for sinusoidal waves (+7 dBm at the input of the phase meter) and square waves (0.7 $V_{p-p}$ at the input of the phase meter, slew rate: 0.2 V/ns), respectively.

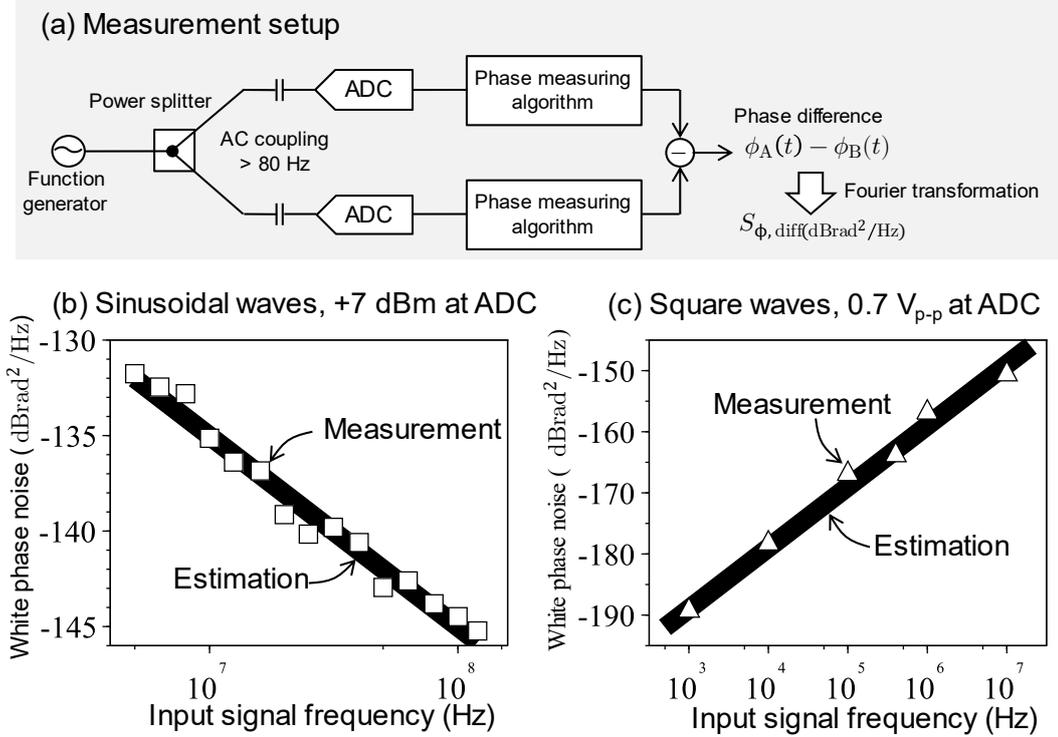

Figure S28. (a) Measurement setup (b) Result for sinusoidal waves with power of +7 dBm (c) Result for a square wave with the amplitude of 0.7 $V_{p-p}$

The measurement results correlate with the theoretical estimations, calculated using Equation (S125) for sinusoidal waves and Equation (S127) for square waves. Note that the input signal frequency for square waves has a lower limit of ~1 kHz because of AC coupling (−3 dB cutoff frequency: 80 Hz) of the phase meter. We consider that much lower white phase noise can be observed using a digitizer with DC-coupled inputs and lower input signal frequencies. Notably, the level of white phase noise can be less than SNR of ADC. In the I/Q demodulation technique, the white noise level is limited by the SNR of the ADC, which is normally around −150 dBFS/Hz for most high-speed, high-resolution ADCs. However, in this phase measurement technique, the white noise level can exceed the limit from the SNR because of the principle of zero-crossing counting.

### III.D. Flicker phase noise
#### 1. Origin of the flicker phase noise

Flicker phase noise is the second type of uncorrelated noise component in phase difference. As its frequency dependence is −10 dB/dec, the flicker phase noise $S_{\phi,\text{Flicker}}(f)$ is characterized using the value at 1 Hz, as shown in Equation (S128).

$$S_{\phi,\text{Flicker}}(f) = S_{\phi,\text{Flicker}}^{(1\text{Hz})} - 10\log_{10}\left(\frac{f}{1\text{ Hz}}\right). \tag{S128}$$

Here, $S_{\phi,\text{Flicker}}^{(1\text{Hz})}$ is the flicker phase noise at 1 Hz, and $f$ is the offset frequency. Figure S24(b) confirmed that the phase noise obeys Equation (S128) at least down to $10^{-5}$ Hz.

There might be a question about the origin of the flicker phase noise: "Why does the phase noise of the RF signal have a close-in flicker noise, albeit that the voltage noise without the RF signal has a flat noise spectrum and the RF signal has no flicker noise?" In other words, the close-in phase noise strangely appears in the measured RF signal, even if the input RF signal has no phase noise at all. The reason is that ADCs generally have imperfections such as fluctuations and nonlinearity. As for RF signal amplifiers, the mechanisms of flicker phase noise are introduced in Enrico Rubiola's book[4]. In the book, the mechanisms are classified into two types. One is the nonlinear mechanism, which is a nonlinear conversion from near-DC flicker noise in voltage. The second is the quasi-linear mechanism, which is the low-frequency gain fluctuation of the amplifier, leading to weak amplitude modulations and, consequently, phase noise in the close-in frequency. For the phase noise in this phase meter, the origin of the noise can be analyzed similarly to the book.

Figure S28 illustrates the noise model for the flicker phase noise. Consider that the input signals are noise-free, the signal in the voltage spectrum has no side lobe; however, after A/D conversion, a side lobe appears because of the imperfection of the ADCs. The uncorrelated component of the side lobe produces the flicker phase noise in the phase difference.

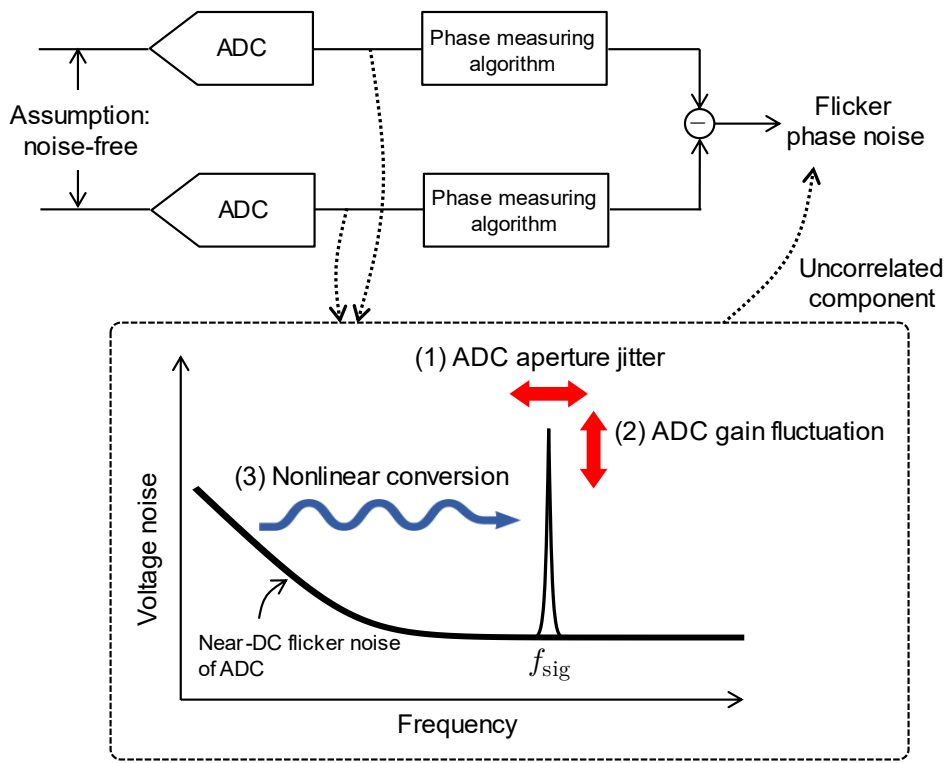

Figure S29. Mechanism of close-in flicker phase noise. (a) Flicker component of aperture jitter of ADC (b) Nonlinear conversion of near-DC flicker voltage noise from nonlinearity of the input signal

We then classified the mechanism for the flicker phase noise into three. One is the aperture jitter of the ADC, referring to the timing fluctuation of the A/D conversion. Regarding modulation, such timing fluctuation corresponds to the phase modulations of the RF carrier signal. Therefore, from the phase modulation, the level of the phase noise is independent of the power of the input signal. Moreover, from the relation $\phi = 2\pi f_{\mathrm{sig}} t$, the level of the phase noise should be proportional to the square of the input signal frequency, i.e., obey Equation (S129).

$$S_{\phi,\mathrm{Flicker}}^{(1\mathrm{Hz})} \propto f_{\mathrm{sig}}^2 \tag{S129}$$

The second is the gain fluctuation of the ADC, referring to the fluctuation of the A/D conversion regarding voltage. Regarding modulation, it is an amplitude fluctuation of the RF carrier signal. Therefore, similar to the aperture jitter of the ADC above, the level of the phase noise is independent of the power of the input signal. However, dependence on $f_{\mathrm{sig}}$ is the opposite. Because it is only amplitude modulation, the level of the phase noise should also be independent of the input signal frequency $f_{\mathrm{sig}}$. Note that the two components, ADC aperture jitter and ADC gain fluctuation, are classified into the quasi-linear mechanism according to Rubiola's book.

The last classification is the nonlinear conversion of near-DC flicker noise. Here we analyze it similar to that in Rubiola's book. Let the input signal be $V_0 e^{-i2\pi f_{\mathrm{sig}} t}$. We use the expression of nonlinearity, as in Equation (S130).

$$V_{\mathrm{out}} = \sum_{i=0}^{\infty} a_i V_{\mathrm{in}}^i \tag{S130}$$

Here, when the ADC has nonlinearity, the ADC output $V_{\mathrm{out}}$ has the second-order term from the ADC input $V_{\mathrm{in}}$; $a_i \neq 0$. If the ADC is perfect, $a_1 = 1$ and the other $a_i = 0$. Furthermore, near-DC flicker (voltage) noise $n(t)$ is added to the input; $V_0 e^{-i2\pi f_{\mathrm{sig}} t} + n(t)$. Then, expanding the right-hand side of Equation (S130), and taking the first order of $e^{-i2\pi f_{\mathrm{sig}} t}$ and $n(t)$, we derive

$$V_{\mathrm{out}} = \left(1 + \frac{2a_2}{a_1} n(t)\right) a_1 V_0 e^{-i2\pi f_{\mathrm{sig}} t}. \tag{S131}$$

Equation (S131) shows that the near-DC flicker voltage noise $n(t)$ couples with the nonlinearity and, consequently, appears as a fluctuation in the output. The coupling constant is $2a_2/a_1$ and independent of the signal power (proportional to the square of $V_0$). Considering that $a_1$ is the A/D conversion gain and is nearly equal to 1, the coupling constant is approximately $2a_2$. Therefore, the coupling is related to the second-order nonlinearity of the A/D conversion. Note that the coefficient $a_2$ generally depends on the input signal frequency; $a_2 = a_2(f_{\mathrm{sig}})$, and thus, the level of the flicker phase noise is not constant to the input signal frequency.

### *2. Measurement of the flicker phase noise*

We could not quantitatively estimate the contribution from the three origins because it is challenging to measure the aperture jitter, gain fluctuation, and nonlinearity of the ADCs separately using our instruments. Therefore, we study the dependence of the level to (i) the power of the input signal and (ii) the input signal frequency and signal shape.

We confirmed that the level of the flicker phase noise is independent to the power of the input signal. Figure S29 shows an example. The test setup for the data is the same as in Figure S28(a), and the input signals used for the test are 30-MHz, 1-$V_{\mathrm{p\text{-}p}}$ square waves. When the power of the input signal changed from −33 dBm to +7 dBm, the level of the phase noise is constant, whereas the level of white phase noise is inversely proportional to the power. Note that this characteristic correlates with the three origins described above.

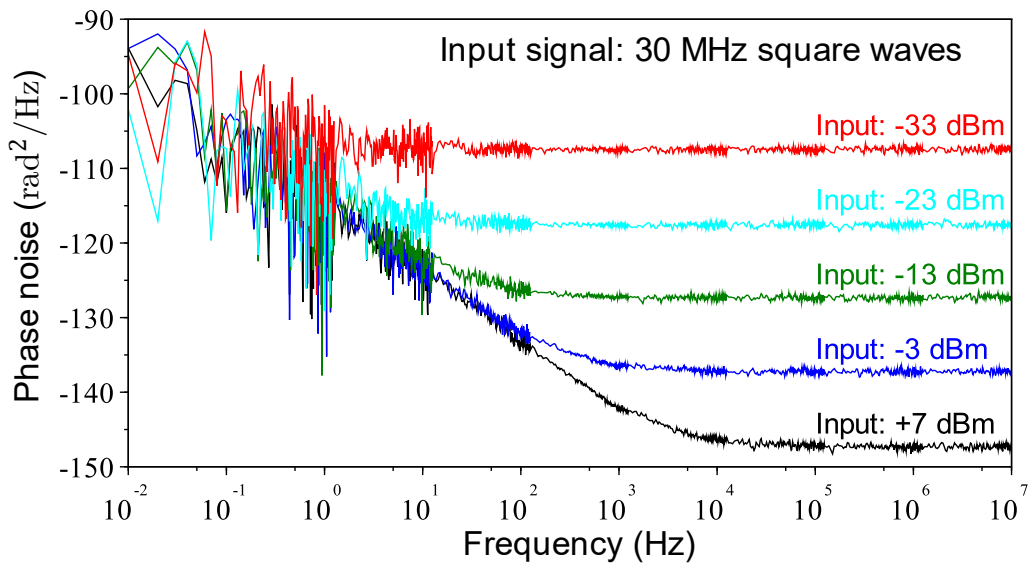

Figure S30. Flicker phase noise vs. input signal power. The level of flicker noise is independent to the input signal power.

Regarding the dependency on the input signal frequency, we measured the level of flicker noise for different input signal frequencies and waveforms (Figure S31), and the test setup is the same as in Figure S28(a). The input signal power is +7 dBm for sinusoidal waves and 1 $V_{p-p}$ for square waves. Here, the phase noise is converted as shown in Equation (S132) to timing jitter, the unit of which is $fs^2/Hz$ for clarity in the plot.

$$S_{t,\text{Flicker}}^{(1\text{Hz})} = \frac{S_{\phi,\text{Flicker}}^{(1\text{Hz})}}{(2\pi f_{\text{sig}})^2} \tag{S132}$$

The lines showing the level of phase noise at $-160 \text{ dBrad}^2/\text{Hz}$, $-130 \text{ dBrad}^2/\text{Hz}$, and $-100 \text{ dBrad}^2/\text{Hz}$ are also plotted.

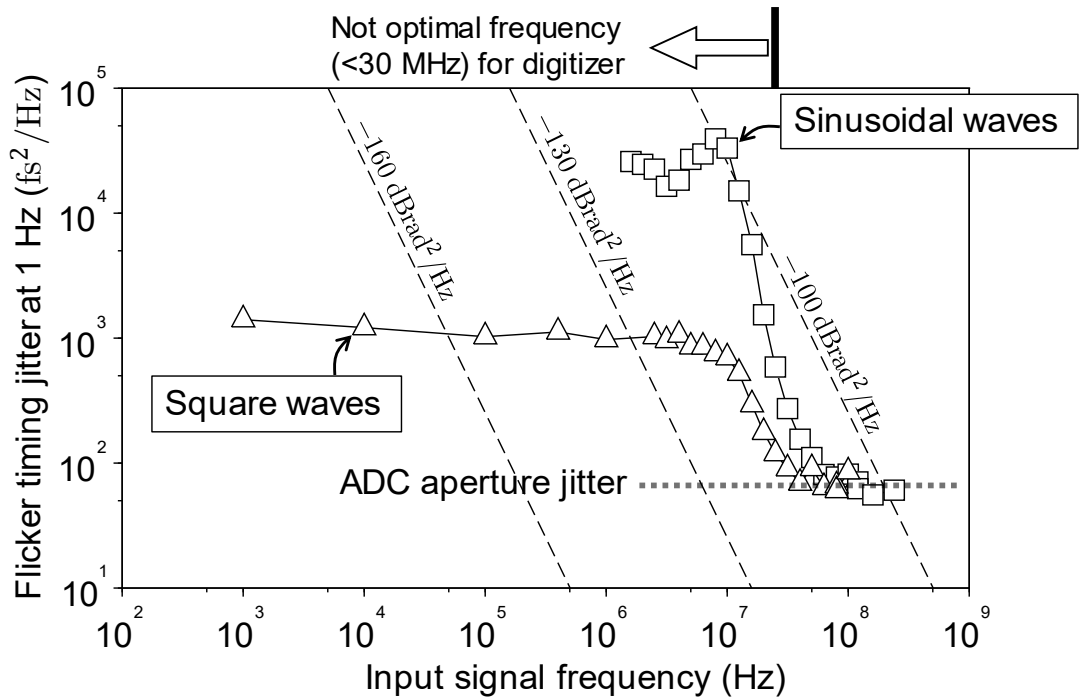

Figure S31. Measured phase noise of the flicker phase noise and its frequency dependence in the unit of timing jitter

For sinusoidal waves (square), the level of the flicker phase noise converges approximately $8 \times 10^1$ fs$^2$/Hz above ~30 MHz. We speculate that the low flat level comes from the differential part of the ADC aperture jitter because the level of timing jitter is independent of the input signal frequency. Therefore, the flicker phase noise at 1 Hz $S_{\phi,\text{Flicker}}^{(1\text{Hz})}$ has a +20 dB/dec dependency in that frequency range. The same level was measured for both sinusoidal and square waves. Such characteristics support our speculation. According to the datasheet[3], aperture jitter of the ADC for the signal channel is specified as 55 fs$_{\text{RMS}}$. However, the integration frequency range for aperture jitter is not described in the datasheet and cannot be compared with our results. Note that the differential jitter between the two inputs for the ADCs is also unspecified.

For lower than 30 MHz input signal frequency, the flicker noise for sinusoidal waves increases more than two orders of magnitude between 10 MHz and 30 MHz. Below 10 MHz, the level does not change so much, between $1 \times 10^4$ fs$^2$/Hz and $5 \times 10^4$ fs$^2$/Hz. For square waves, the characteristics are slightly different, and the level is limited to ~$1 \times 10^3$ fs$^2$/Hz, remaining constant below. According to the datasheet[5], nonlinearity worsens when the input signal frequency is less than 30 MHz. Therefore, for a lower than 30 MHz, the noise could come from some effect regarding nonlinearity at the A/D conversion. We could not find why the level becomes constant at lower than ~10 MHz; further studies are needed. Note that we are considering implementing the same signal-processing framework to the digitizer module with a DC-coupled one, enabling us to measure low-frequency square waves in optimal conditions.

### III.E. Summary: how to estimate phase noise

In the following summary on how to estimate the level of phase noise for a specific measurement device, Table S5 lists the components and how to evaluate them.

1. Clock phase noise. It can be estimated using the frequency difference between the two input signals, as shown in Equation (S110). Clock phase noise can be measured with the differential measurement technique described in Subsection III.B. When the frequency difference is small (typically <MHz), the noise does not need to be considered in most cases.

2. White phase noise. This noise can be estimated theoretically from ADC and input signal parameters. In particular, there are three parameters to consider. One is the noise spectral density of the ADC $P_{\text{ND}}$. The second is the effective power of the input signal $P_{\text{eff}}$, which is derived from the slew rate at the zero-crossings, as given in Equation (S116). The last parameter is the degradation factor, which is proportional to the ratio of $f_{\text{sig}}$ and $f_{\text{ADC}}$; its definition is found in Equation (S117). Note that this level is for a single-channel measurement. In other words, we must take the root-sum-square of the two input channels to obtain it for phase difference.

3. Flicker phase noise. This noise is challenging to estimate deterministically because of the influence of some nonlinearity in the signal and/or ADCs. Generally, the level of the flicker phase noise is independent of the power of the input signal but dependent on the input signal frequency and signal shape. Actual measurements of hardware for use is recommended for as long as possible. In the case of our phase meter, the level of the flicker noise can be approximated in Figure S31.

Table S5. Summary of the noise components for the phase difference measurement. No shape is assumed for the input signals.

| Error components | Related parameters | Description | Ref. Eq. |
|---|---|---|---|
| (a) Clock phase noise | Difference of the input signal frequencies: $\Delta f_{\text{sig}}$<br>Clock phase noise $S_{\text{CLK}}$ at the carrier frequency of $f_{\text{CLK}}$ | $S_{\phi,\text{CLK}} = \left(\dfrac{\Delta f_{\text{sig}}}{f_{\text{CLK}}}\right)^2 S_{\text{CLK}}$ | (S110) |
| (b) White phase noise $S_{\phi,\text{White}}$ | Input signal frequency: $f_{\text{sig}}$<br>Sampling frequency: $f_{\text{ADC}}$<br>Noise spectral density of ADC: $P_{\text{ND}}$<br>Effective power of the input signal: $P_{\text{eff}}$ | $S_{\phi,\text{White}} = \dfrac{P_{\text{ND}}}{P_{\text{eff}}} \dfrac{f_{\text{ADC}}}{4 f_{\text{sig}}}$<br>Note: estimate two noises for each channel and take root-sum-squared. | (S116)<br>(S118) |
| (c) Flicker phase noise $S_{\phi,\text{Flicker}}$ | Input signal frequency: $f_{\text{sig}}$<br>Signal shape | For the phase meter, see Figure S31.<br>Note: measurements for real hardware for use is needed in general case. | — |

## IV. Performance analysis

In this section, we discuss the performance analysis of the phase meter. First, we introduce cycle slips originating from miscounting the zero-crossings when measuring low SNR or low slew rate signals. Second, AM-PM conversion is studied for the phase meter. Third, the phase meter is compared with the phase noise standard at the National Metrology Institute of Japan (NMIJ), validating the phase meter, including software analysis.

### IV.A. Cycle slips

A cycle slip is an undesired jump of the measured phase that is generated by miscounting the zero-crossings when the input signals have a low SNR or low slew rate. In optical experiments, the SNR is limited in most cases, hence it is important to predict or detect cycle slips. Measurements with ordinary frequency counters also suffer from cycle slips. Here, we present a model of the cycle slip in the phase meter, and we analyze how it is determined by the SNR or noise level.

From the nature of zero-crossing counting, the slip rate depends on both the SNR and the slew rate of the input signal. We divide the cases into the two shown in Figure S32; one is the case of low slew rate and high SNR (Figure S32(a)). In this case, the input signal has low frequency compared to the sampling frequency, but the amplitude is sufficiently large. Then there is a miscounting around zero-crossings due to the noise. The other case is high slew rate and low SNR (Figure S32(b)). In this case, the input signal frequency has a much higher frequency than for the case with low slew rate and high SNR. However, it has larger noise or fluctuation, which also leads to a miscounting of the zero-crossings.

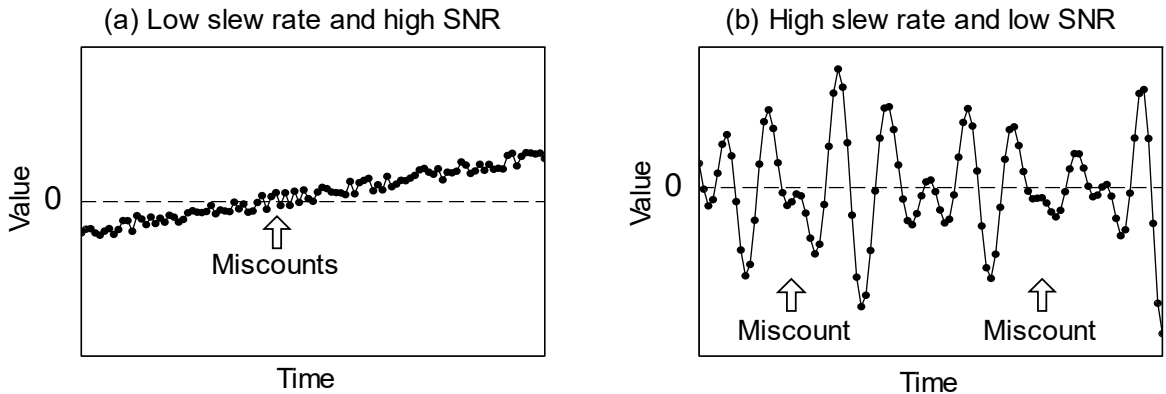

Figure S32. Examples of cycle slips: (a) low slew rate and high SNR; (b) high slew rate and low SNR.

### *1. Case A: low slew rate and high SNR*

We analyze the case shown in Figure S32(a) where the input signal has a high SNR but low slew rate. In Figure S33, an effect that noise can have at a zero-crossing of the input signal is presented; $S(t)$ is the input signal without noise, and the sampled data are shown as $\overline{V_i}$. In this case, input signal increases (decreases, if it is falling edges) monotonically, and there is no miscounting of zero-crossings. On the other hand, for a signal with noise, as shown in Figure S33(b), when $V_{i-1}$ becomes positive and $V_i$ becomes negative due to noise, there are three zero-crossings instead of one; this leads to two additional but fake zero-crossings, thus indicating a cycle slip.

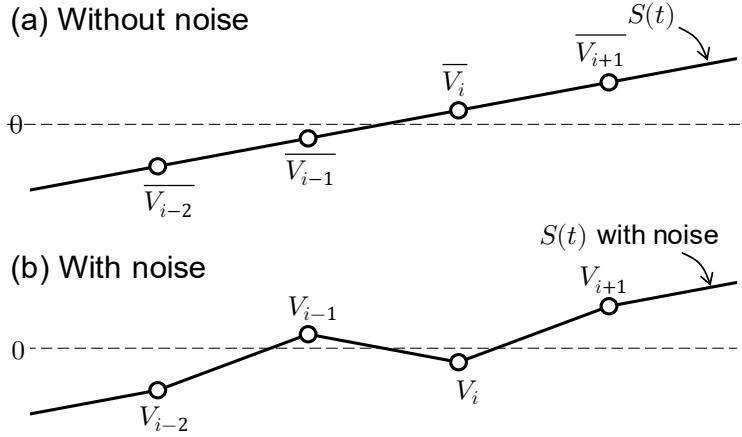

Figure S33. Example of zero-crossing counting when a cycle slip occurs: (a) the input signal without noise; (b) the input signal with noise. Three zero-crossings instead of one are detected due to noise.

We can analyze the probability of a cycle slip in the following way. First, assume the noise follows a normal distribution $N(x)$, which has a standard deviation of $\sigma_V$ and mean value of zero.

$$N(x) \equiv \frac{1}{\sqrt{2\pi}\sigma_V}\exp\left(-\frac{x^2}{2\sigma_V^2}\right) \tag{S133}$$

The cumulative distribution function is defined as

$$C_N(x) \equiv \int_{-\infty}^{x} N(x')dx'. \tag{S134}$$

Here, the sampled data without noise, which we use $\overline{V_i}$ to express, have the following relationship:

$$\overline{V_i} = \overline{V_{i-1}} + st_{\text{ADC}}, \tag{S135}$$

where $t_{\text{ADC}}$ is the sampling interval of the ADC, and the slew rate at the zero-crossings is expressed as $s$. Then, the probability that $V_{i-1}$ becomes positive is same as that of the noise exceeding $-\overline{V_{i-1}}$, i.e.,

$$C_N(\overline{V_{i-1}}) = C_N(\overline{V_i} - st_{\text{ADC}}). \tag{S136}$$

Similarly, the probability that $V_i$ becomes negative is

$$C_N(-\overline{V_i}). \tag{S137}$$

By taking the integration of the normal distribution, the probability of a cycle slip at each zero-crossing is obtained.

$$\int_0^{st_{\text{ADC}}} C_N(\overline{V_i} - st_{\text{ADC}})C_N(-\overline{V_i})d\overline{V_i} \tag{S138}$$

Considering that the zero-crossings occur at $2f_{\text{sig}}$ frequency, the slip rate (SR) becomes

$$\text{SR} = 2f_{\text{sig}}\int_0^{st_{\text{ADC}}} C_N(\overline{V_i} - st_{\text{ADC}})C_N(-\overline{V_i})d\overline{V_i}. \tag{S139}$$

S69

Using the definition of the effective power in Equation (S116), $st_{\mathrm{ADC}}$ can be converted as follows:

$$st_{\mathrm{ADC}} = 2\pi f_{\mathrm{sig}} t_{\mathrm{ADC}} \sqrt{2RP_{\mathrm{eff}}}. \tag{S140}$$

Therefore, the SR depends on three parameters, input signal frequency $f_{\mathrm{sig}}$, effective power $P_{\mathrm{eff}}$, and noise amplitude $\sigma_V$. Note that $P_{\mathrm{eff}}$ and $\sigma_V$ are related to the effective SNR.

For example, we consider the lower limit of the input signal frequency for sinusoidal input. We assume that the noise is generated only by the ADC. Therefore, $\sigma_V$ is only related to the SNR of the ADC. By conducting numerical simulation, we estimate the SR, which is shown as the blue thick line in Figure S34. We assume the input signal is sinusoidal with a power of +7 dBm. We also apply the following parameters based on our hardware; the full scale of the ADC is +9.5 dBm, and the SNR of the ADC is 64 dB. Due to the nature of fast decay of the normal distribution, the estimated SR drops quickly when the frequency of the input signal increases.

We also measured the real SR with the input signal from SGs. Sinusoidal input signals with +7 dBm power were input to channel A of the phase meter, and a +7 dBm square wave was input to channel B for the reference. The result is shown as black filled circles in Figure S34. Below 500 kHz, the measured values follow the prediction well. However, above 500 kHz, the measured SR is higher than the estimation which is attributed to the noise distribution being slightly different from the normal distribution. In other words, the noise creates excess tails or outliers in the error distribution. We suppose that such excess noise comes from the input signals and not from the A/D conversion. Based on our simulation and measurement, we determined the lower limit of the input signal frequency (+7 dBm sinusoidal) of our hardware is 1 MHz, at which the SR drops sufficiently rarely, i.e., less than once per day.

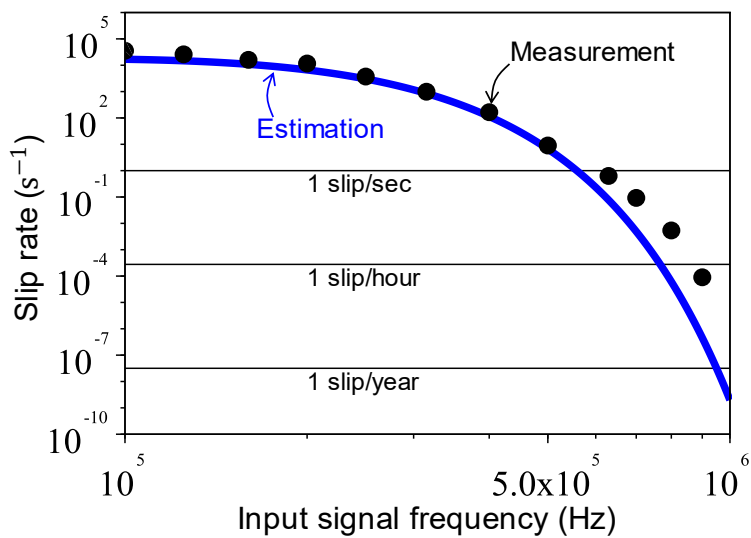

Figure S34. Measurement of the slip rate for low slew rate and high SNR. Input signal is sinusoidal with the power of +7 dBm. Full scale is +9.5 dBm, and SNR is 64 dB.

### *2. Case B: high slew rate and low SNR*

We consider the case where the input signal has a high slew rate and low SNR. Since this phase measuring algorithm is based on zero-crossing counting, the SR depends on not only the SNR but also on the noise amplitude in the time domain. In other words, the SR depends on the bandwidth of the noise.

In Figure S35, we show examples of input signals with high slew rate and low SNR where the noise bandwidth is different. We assume 0 dBm power of the carrier and 100 kHz bandwidth resolution. We make this plot for the signal with an SNR of 20 dB. In Figure S35(a1), the spectrum of the signal without a bandpass filter is shown. The corresponding signal in the time domain (Figure S35(a2)) contains large and fast noise components, and almost all zero-crossings are destructed by the noise. When the noise is filtered by a bandpass filter (BPF), the zero-crossings can be detected. Figure S35(a2) and Figure S35(a3) show the same signal low-pass filtered with cutoff frequencies of 50 MHz and 15 MHz, respectively. When a 50 MHz filter is applied, some of the zero-crossings become detectable (Figure S35(b2)), and by using a 15 MHz filter, the signal becomes cleaner. We call the cutoff of the low-pass filter the noise bandwidth (NBW), which is represented by $f_{\mathrm{NBW}}$. Note that in digital demodulation techniques, the phase noise spectrum simply depends on the SNR of the signal. The phase noise is $-70$ dBrad$^2$/Hz in Figure S35. Therefore, cycle slips do not occur when an appropriate low-pass filter is applied to limit phase fluctuation within $2\pi$ just after the demodulation process.

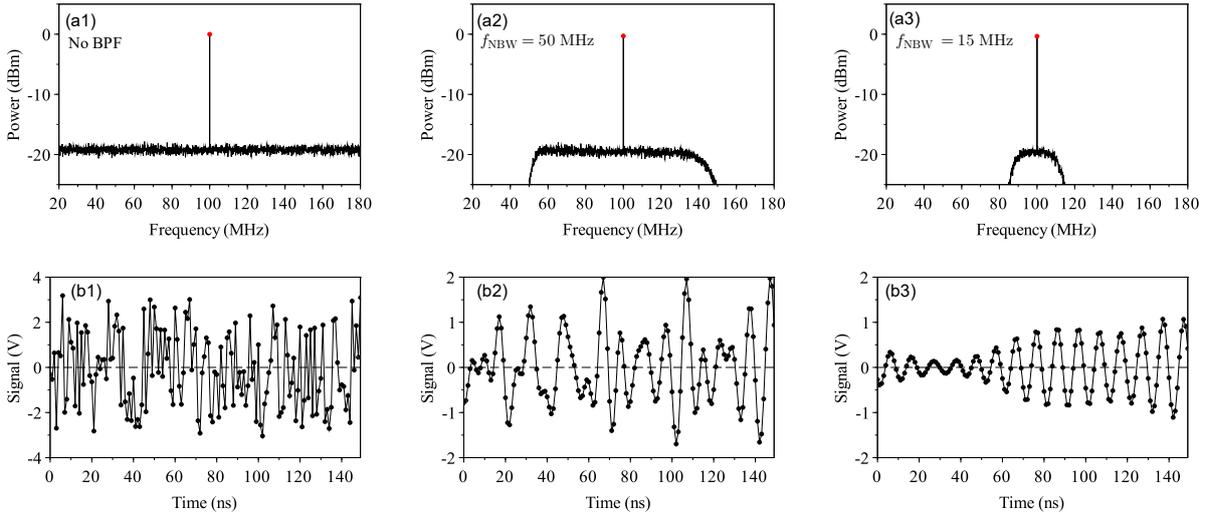

Figure S35. Examples of high slew rate and low signal-to-noise ratio (SNR) input signals with different noise bandwidth (NBW): (a1) signal spectrum with no bandpass filter; (a2) signal spectrum with $f_{\mathrm{NBW}}$ = 50 MHz; (a3) signal spectrum with $f_{\mathrm{NBW}}$ = 15 MHz; (b1) time domain representation of the signal in (a1); (b2) time domain representation of the signal in (a2); (b3) time domain representation of the signal in (a3).

We conducted a simulation to investigate how the SR depends on the SNR and $f_{\mathrm{NBW}}$. Figure S36 shows the signal we used for the simulation. A BPF is inserted before the input of the ADC to make the simulation realistic. First, the noise spectrum is cut between $f_{\mathrm{sig}} - f_{\mathrm{NBW}}$ and $f_{\mathrm{sig}} + f_{\mathrm{NBW}}$ by the BPF. Next, white noise is also added to the signal to representing ADC noise. Then, the phase measuring algorithm is applied to the simulated data, and the result is compared with that obtained from a signal without noise. We obtain the SR in units of s$^{-1}$ by changing the parameters of the SNR and $f_{\mathrm{NBW}}$.

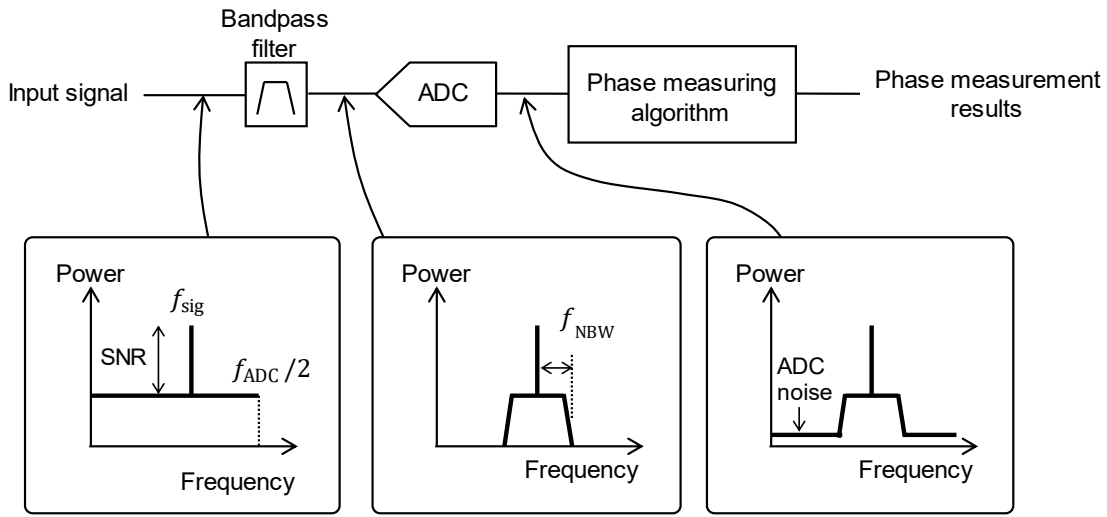

Figure S36. Assumed signal for the simulation.

When the SNR is low and the NBW is large, there is significant noise. The zero-crossings are disturbed by the noise, and cycle slips occur, as shown in Figure S37. Note that $f_{\mathrm{NBW}} = 5$ MHz, $f_{\mathrm{sig}} = 50$ MHz, and SNR = 23 dB for Figure S37.

We found that there are two types of cycle slips. In a positive cycle slip, the measured phase jumps by $+2\pi$ at the slip; a negative slip corresponds to $-2\pi$ jump. Figure S37(a1) shows the time domain signals for a positive slip, and the negative slip is shown in Figure S37(a2). In both plots, the input signal with noise is shown in red, and the reference signal without noise is shown in black. The amplitude of the signal with noise is significantly reduced at the point of cycle slips. Figure S37(b1) and Figure S37(b2) show the measured phase difference corresponding to each cycle slip. The black dots represent the instantaneous phase estimator derived at the same rate as that of the A/D conversion (1 GHz), while the blue lines show the low-pass-filtered instantaneous phase estimator (with the cutoff frequency of 50 MHz). There are obvious jumps of the measured phase, which can be used to detect the cycle slips.

One interesting characteristic is that the ratio of the rate of positive and negative cycle slips converges to $\sqrt{2}:1$ as the simulation time increases. However, we have not yet clarified why it converges to that ratio. We speculate that both phase fluctuation and amplitude fluctuation contribute to positive cycle slips, whereas only amplitude fluctuation contributes to negative slips.

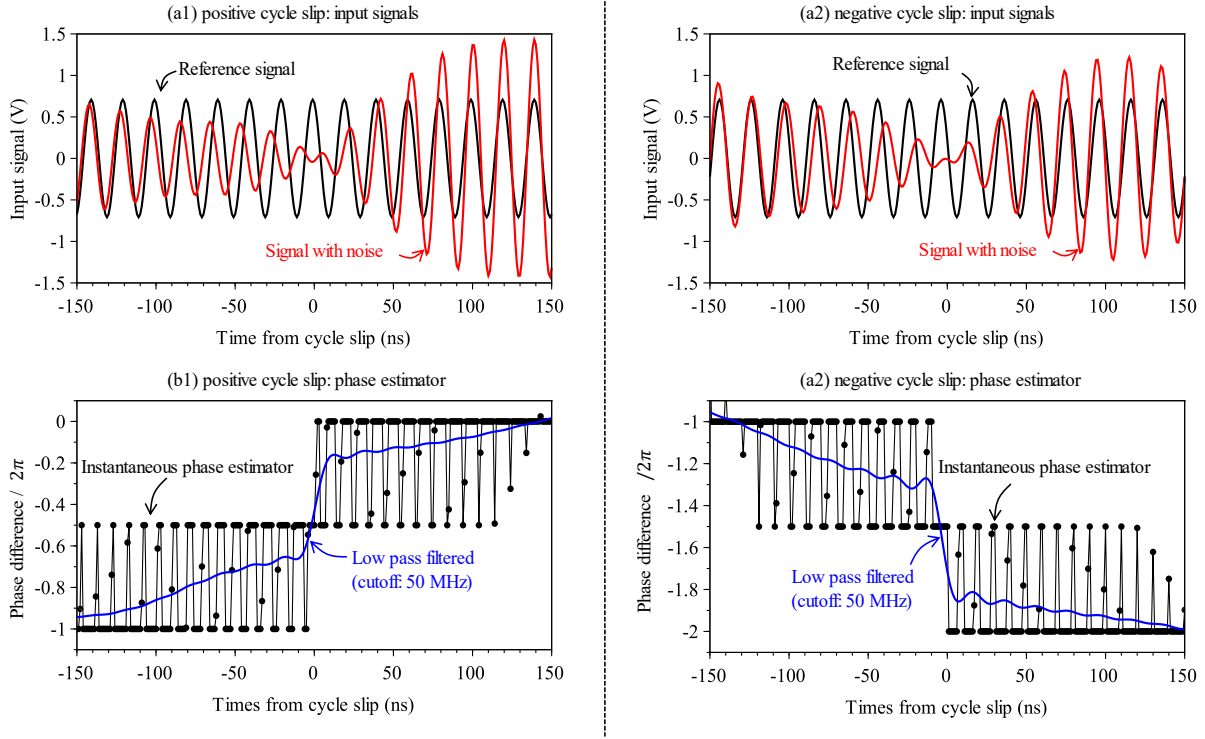

Figure S37. Examples of positive and negative cycle slips for signals with high slew rate and low SNR.

We conducted numerical calculations to obtain the SR for different $f_{\text{NBW}}$ and SNR. The simulation results are shown in Figure S38. Here, we counted the total number of positive and negative slips. The simulation time was set to a maximum of 1 second, which means there are $10^9$ data points for a 1 GHz sampling rate. Therefore, the minimum detection rate is several slips per second. The noise bandwidth is set from 2.5 MHz to 20 MHz; the input signal frequency is assumed to be 50 MHz, and its power $P_{\text{sig}}$ is +7 dBm. The SNR was taken from 0 to 37 dB; the SNR of the ADC is 64 dB, and the full scale of the ADC is +9.5 dBm. Therefore, the noise spectrum of the ADC noise is −141.5 dBm/Hz. The simulation results show two characteristics; (i) the SR converges to a certain level when the SNR becomes low, and (ii) the SR drops quickly when the SNR is high.

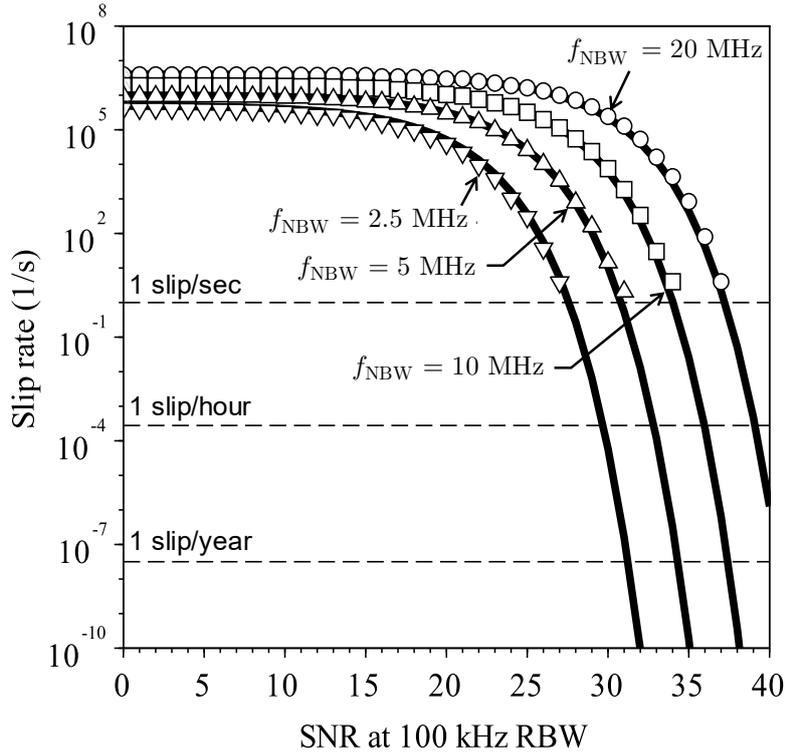

Figure S38. Simulated SR versus SNR (open circles, squares, and triangles) for signals with high slew rate and low SNR: (open circles, squares, and triangles) each shape corresponds to a different noise bandwidth; (black lines) fits the simulated data fitted based on the model.

The results in Figure S38 exhibit two dependencies. One dependency is that the SR at the low SNR becomes proportional to $f_{\text{NBW}}$; the second is that the cutoff of the SR is also proportional to $f_{\text{NBW}}$. Using these trends, we developed a model with a good fit using two constants, $c_1$ and $c_2$, as expressed in Equation (S132).

$$\log_{10}(\text{SR}) = c_1 P_{\text{noise}}^{-1} + \log_{10}(c_2 f_{\text{NBW}}) \tag{S141}$$

Here, $P_{\text{noise}}$ is the noise power integrated within the noise bandwidth.

$$P_{\text{noise}} = 2 f_{\text{NBW}} 10^{(P_{\text{sig}} - \text{SNR} - 50)/10} \tag{S142}$$

Note that 50 in Equation (S142) is the resolution bandwidth; $10 \log_{10} \text{RBW} = 50$. By fitting the data with the model in Equation (S141), we find:

$$c_1 = -2.5 \tag{S143a}$$
$$c_2 = 0.2 \tag{S143b}$$

In Figure S38, the fits are plotted in thick black lines, showing good agreement with the simulation results. We also found that the results remain unchanged when the input signal frequency is differed. Note that the model implies that small white noise introduced by the A/D conversion does not significantly affect the SR; this can likely be explained by $P_{\text{noise}}$ for the ADC noise in this simulation being about $10^{-5.5}$, thus leading negligible SR.

## IV.B. Amplitude-modulation to phase-modulation (AM-PM) conversion

AM-PM conversion is an undesired conversion from amplitude to phase. In some cases, the AM-PM conversion effect should be considered because, in optical measurements, the amplitude modulation is generally large. Moreover, in some precise phase noise measurements, phase noise target sensitivity is significantly low. Therefore, the AM-PM conversion is one of the error sources for phase noise measurement. The AM-PM sensitivity for this phase measuring algorithm is still not well-modeled; therefore, AM-PM sensitivity was evaluated in real hardware.

We estimate the AM-PM conversion factor using the setup shown in Figure S39(a). We use a two-channel function generator (Keysight 33622A) for generating the test signals. One signal is a 30-MHz, +4-dBm sinusoidal wave with amplitude modulation. The amplitude modulation is sinusoidal and $10\ \%_{0-p}$. The other signal is without amplitude modulation, with the same frequency and power as the first signal. We fit the phase difference with a sinusoid to obtain the AM-PM conversion factor. By changing the modulation frequency, we achieve frequency dependence of the AM-PM conversion factor.

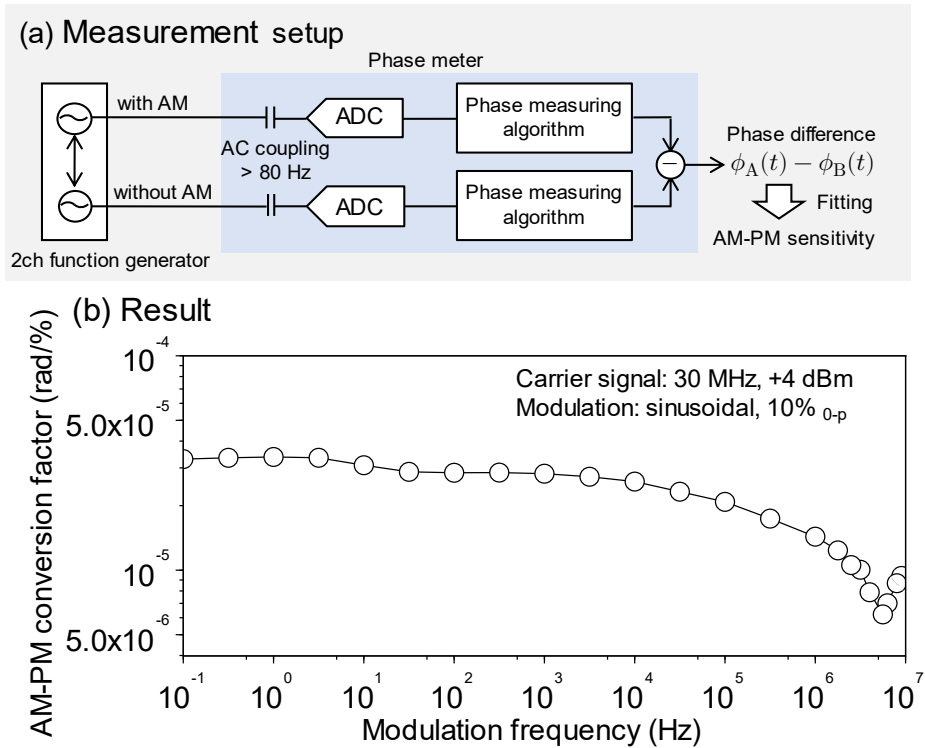

Figure S39. (a) Measurement setup for AM-PM conversion factor. (b) A result for 30-MHz, +4-dBm signal with 10-$\%_{0-p}$ amplitude modulation

Figure S39(b) shows the results. Below several tens of kilohertz, the AM-PM conversion factor converges to a certain value, which is approximately $3 \times 10^{-5}$ rad/%. This value corresponds to approximately −25 dB when the factor is expressed as the ratio between the amplitude modulation power and phase modulation power. In other words, when the AM is −100 dBc/Hz, the phase noise induced by the AM-PM conversion will be −125 dBc/Hz. At approximately 5 MHz, it seems to dip in the frequency response. We do not identify the origin but we speculate that it is because of the different responses to AM and PM of the phase meter.

The problem in measuring the AM-PM conversion factor with such an experiment is that there is no standard AM signal. Pure amplitude-modulated signals without any phase modulation are extremely challenging to generate. Therefore, when a signal from the function generator is amplitude-modulated, the signal might also have undesired phase modulation, shown as phase modulation in the results, but is not from AM-PM conversion. Therefore, the result in Figure S39(b) is only an upper limit for the AM-PM conversion factor, and the AM-PM sensitivity might be smaller than that in Figure S39(b).

We also study the empirical and approximate dependency of the AM-PM conversion factor. Let the input signal $S(t)$ be as expressed in Eq. (S144).

$$S(t) = (1 + A_{\text{AM}}\sin\left(2\pi f_{\text{AM}}t + \phi_{AM}\right))A_{\text{sig}}\sin(2\pi f_{\text{sig}}t) \tag{S144}$$

where $A_{\text{AM}}$, $f_{\text{AM}}$, and $\phi_{AM}$ is the amplitude, frequency, and the initial phase of amplitude modulation, respectively. $A_{\text{sig}}$ is the amplitude of the input signal, and $f_{\text{sig}}$ is the input signal frequency. Then, the AM-PM conversion factor is defined as

$$C_{\text{AM-PM}} \equiv \frac{\partial \phi}{\partial A_{\text{AM}}}, \tag{S145}$$

and we observe that the conversion factor is approximately expressed as

$$C_{\text{AM-PM}} \propto A_{\text{AM}}^0 f_{\text{AM}}^0 \phi_{AM}^0 A_{\text{sig}}^2 g(f_{\text{sig}}). \tag{S146}$$

Therefore, the conversion factor is independent of the amplitude, frequency, and the initial phase of amplitude modulation but is linearly dependent on the power of the input signals ($\propto A_{\text{sig}}^2$). Note that the range of $f_{\text{AM}}$ is limited to 100 kHz in this expression. When $f_{\text{AM}}$ increases to above ~100 kHz, $C_{\text{AM-PM}}$ decreases (Figure S39(b)). Regarding the dependency of the input signal frequency, there is some nonlinear dependency and $g(f_{\text{sig}})$ becomes large when $f_{\text{sig}} < $ ~10 MHz, whereas $g(f_{\text{sig}})$ becomes small when $f_{\text{sig}} < $ ~30 MHz. This strange dependency implies that the AM-PM conversion factor is related to some nonlinearity at the ADC because signals below 30 MHz are not recommended regarding nonlinearity for the digitizer, as also stated by expressing the dependency of the flicker phase noise.

### IV.C. Comparison with the national phase noise standard

To validate the phase meter, we measured calibration signals generated by the national phase noise standard[6] of Japan with the phase meter. Figure S40(a) shows that the 10-MHz calibration signals have standard white phase noise spectra up to several megahertz. The standard noise levels of the signals are assured using a digitizer and discrete-Fourier transformation (DFT). The expanded uncertainty ($k = 2$) of the phase noise standard for a −100-dBc/Hz calibration signal is 0.6 dB (at 1 Hz), 0.26 dB (at 10 Hz), and 0.22 dB (at 1 kHz, 10 kHz and 100 kHz), as shown in the black error bars in Figure S40(c). The standard phase noise is measured using the phase meter by measuring the phase difference between the phase noise standard and a 10-MHz reference signal from a low-noise oscillator. The phase difference measured using the phase meter is recorded on a personal computer and converted to phase noise PSD with our self-made software, which is also used for data analysis of the phase meter.

Figure S40(b) and Figure S40(c) show the results from four calibration signals from −120 dBc/Hz to −90 dBc/Hz. Above 1 kHz, all results agree with the corresponding calibration levels within ~0.3 dB, while in low frequencies, the results increase because of the instrument noise floor. The carrier phase noise of the calibration signal is small enough and does not affect this case (the straight line in Figure S40(b)). Dispersions in the noise spectra in Figure S40(c) originate from insufficient averaging; therefore, a longer measurement time will improve the results. Hence, we verified the measurement reliability within ±1 dB and enough linearity over 30 dB for the phase meter with the national phase noise standard.

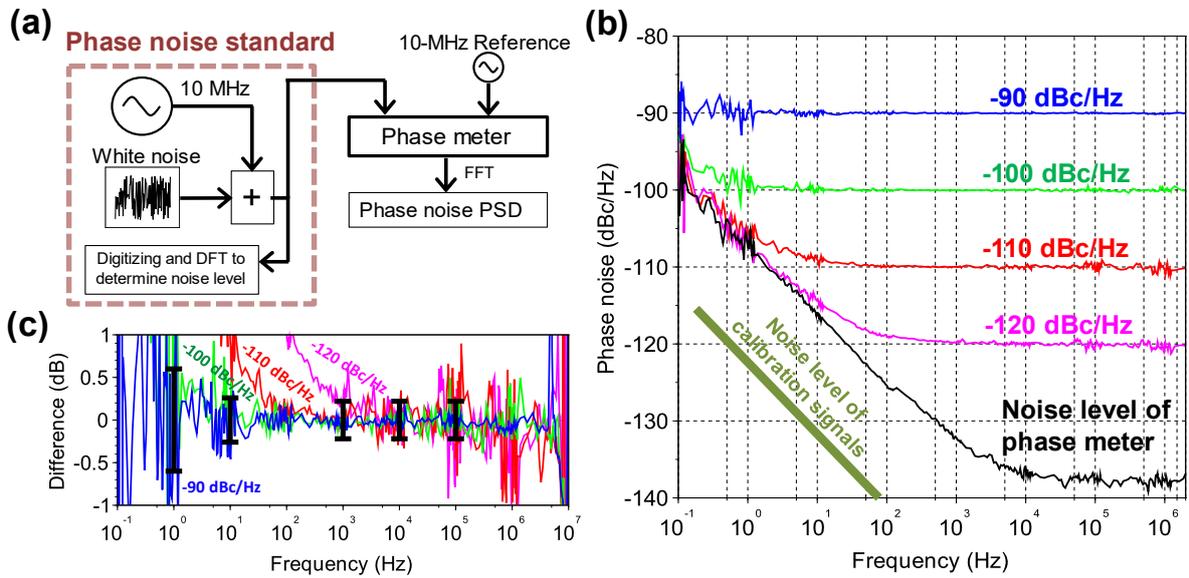

Figure S40. (a) Schematic of the experiment. (b) Results for four calibration signals. The black line shows noise level of the phase meter. The thick line shows phase noise of the calibration signal, which is used to generate the phase noise standard and 10-MHz reference signal. (c) Enlarged view of the difference between the calibration level and measured phase noise.

# V. Improving the phase meter

The phase meter has several limitations, such as the input frequency range and ZI error. In this section, we provide information to improve the phase meter. In subsection V.A, we show a technique to measure and compensate the phase offset. In subsection V.B, a cross-correlation technique is introduced for the phase meter to reduce noise levels. In subsection V.C, we provide how to expand both the upper and lower frequency limits of input signals. In subsection V.D, we discuss the alternative architecture, which is theoretically free from ZI errors, for the phase meter using time-to-digital converters (TDC) instead of ADCs. In subsection E, we introduce ideas on other improvements for the phase meter.

## V.A. Minimizing offset: zero compensation

The phase meter has zero offset in phase difference, resulting from a skew between Ch. A and Ch. B of the phase meter. Furthermore, the difference in characteristics in analog frond-end electronics for the ADC also causes zero offset. To compensate for and measure the zero offset, we use the measurement shown in Figure S41.

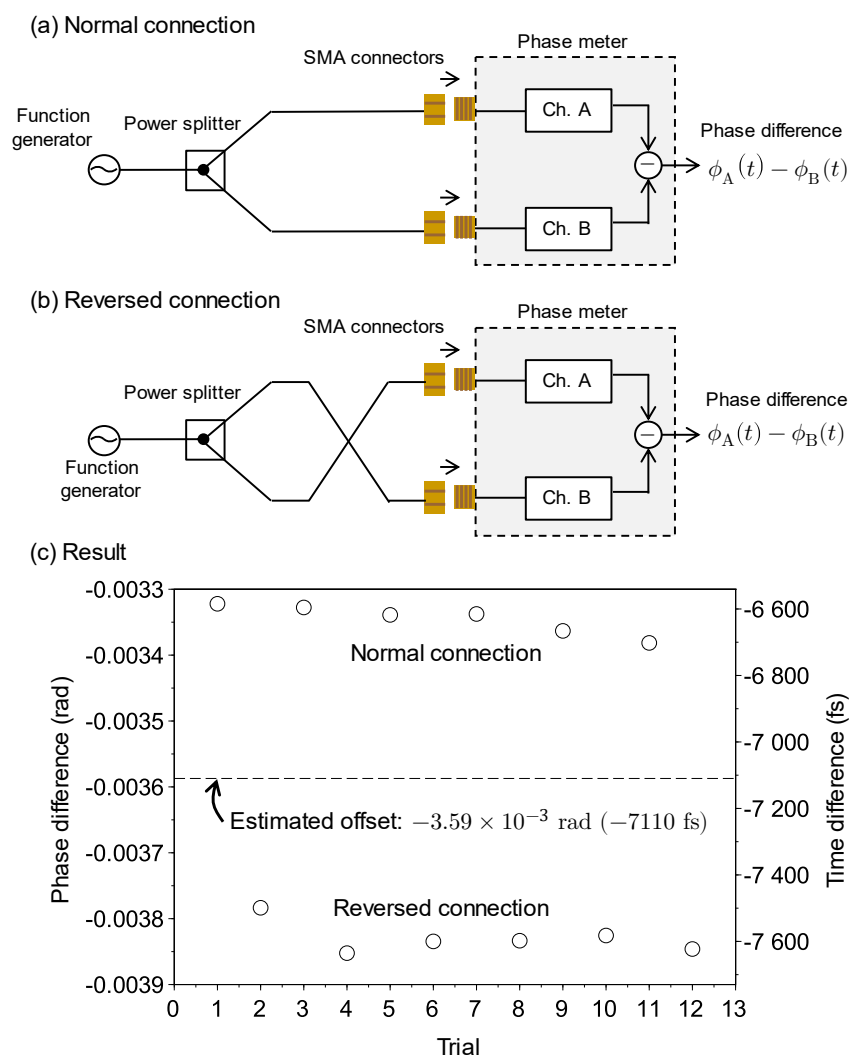

Figure S41. Experiment for measuring phase offset of the phase meter. (a) Normal connection (b) Reversed connection (c) Result

A function generator (Keysight, 33622A) is used to generate a test signal (80.3 MHz, +10 dBm), and the test signal is divided by a power splitter (Mini-circuits Z99SC-62-S+, 0.5 MHz – 600 MHz). The two signals propagate through short semi-rigid cables (Mini-circuits 086-3SM+, 7.6 cm) and input into Ch. A and Ch. B of the phase meter. The connector type is SMA. A calibrated torque wrench (Tohnichi, NSP100CNX8X90cNm, 0.9 Nm, $\pm 5$ % uncertainty) is used to connect the cables, ensuring that the connector is always tightened with the same torque. The phase difference is measured in this configuration, which we call the normal connection (Figure S41(a)). The averaged phase difference from a measurement time of 0.1 s is recorded. Then, the connection is reversed; the correspondence between the cables and connectors in the phase meter becomes the opposite (Figure S41(b)). The same measurement is conducted, and the result is compared to those in normal connections. Because of the phase imbalance of the power splitter and length difference of the cable, the phase difference at the end of the cable has an imbalance. Results in normal connection $\phi_{\text{norm}}$ and reversed connection $\phi_{\text{rev}}$ satisfies

$$\phi_{\text{norm}} = \phi_{\text{imbalance}} + \phi_{\text{offset}} \tag{S147a}$$

$$\phi_{\text{rev}} = -\phi_{\text{imbalance}} + \phi_{\text{offset}} \tag{S147b}$$

where $\phi_{\text{imbalance}}$ and $\phi_{\text{offset}}$ is the phase imbalance at the end of the cable and a zero offset of the phase meter, respectively. Therefore, using the average of the two results, $\phi_{\text{offset}}$ can be estimated as

$$\phi_{\text{offset}} = (\phi_{\text{norm}} + \phi_{\text{rev}})/2. \tag{S148}$$

Note that this measurement is done in the laboratory where the temperature is stabilized at 21.0 °C within $\pm 0.5$ °C in 24 h.

We repeated this measurement set 6 times, resulting in 12 measurements (Figure S41(c)). The results in the normal connection are between –0.0034 rad and –0.0033 rad, whereas for the reversed connection, the results are between –0.0039 rad and –0.0038 rad. Taking the average of the results, the offset is estimated as $-3.59 \times 10^{-3}$ rad, corresponding to –7110 fs in the time domain. Subtracting this offset, the phase differences are within $\pm 200$ fs. These phase differences refer to the path length difference of 0.04 mm. Therefore, we conclude that the offset of the phase meter could be measured and compensated within $1 \times 10^{-4}$ rad for an input signal frequency of 80.3 MHz. Note that the results are different for different input signal frequencies even in the time domain, possibly because of the unbalanced frequency responses in analog frond-end electronics for the ADCs.

### V.B. Improving noise level: cross-correlation technique

A cross-correlation technique is used to improve the noise level of the phase meter. Compared to analog phase noise measurement, the noise level is a disadvantage of digital phase measurement because ADC normally has a worse SNR compared to analog circuits. Therefore, the white noise level is higher than in the analog circuit; typically $-140 \text{ dBrad}^2/\text{Hz}$ to $-150 \text{ dBrad}^2/\text{Hz}$ for several tens of megahertz of input signal frequency. We overcome this limitation using a cross-correlation technique because the noise of different ADCs is not correlated.

Figure S42 shows a schematic of the cross-correlation technique for the phase meter. The two signals carrying the phase difference are split into four signals by two power splitters. The four signals are input into the four channels of the phase meter. Ch. C and Ch. D measure the same phase difference between Ch. A and Ch. B. The noise in the two phase differences are uncorrelated because the noise in phase difference comes from the ADCs. Thus, using the cross-correlation of the two phase differences, the noise level in the phase noise can be suppressed. When the number of $M$ traces are used to take cross-correlation, the noise level is suppressed by $10 \log_{10} \sqrt{M}$ dB. Note that, after the power splitter, the power of the signals is attenuated by 3 dB, then the white noise level for the single phase difference also increases by 3 dB. Therefore, more than four traces of correlation are needed to surpass the sensitivity of the original signal.

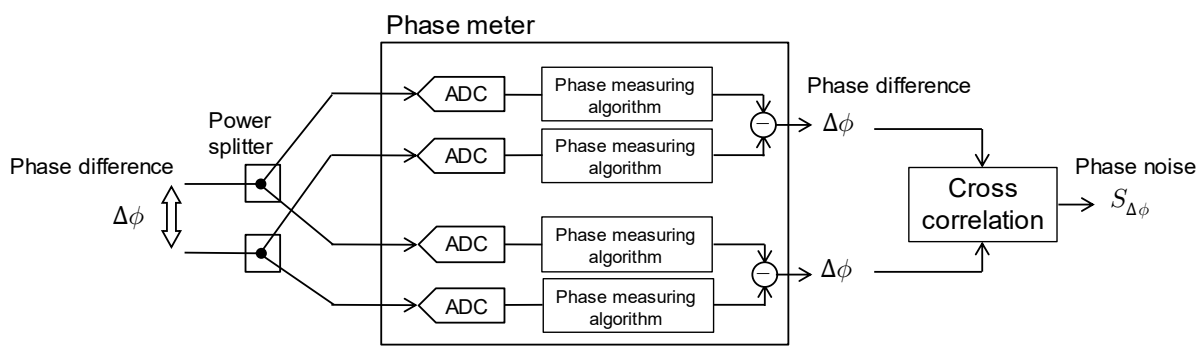

Figure S42. Schematics of cross-correlation technique for phase difference

Here we briefly present the performance of cross-correlation for the phase meter. First, the limit of suppression is measured as in Figure S43(a). A two-channel function generator (Keysight 33622A) is used to generate a pair of signals (80 MHz, +7 dBm) with white phase noise. The phase noises in the two signals are not correlated because of the digital random noise generator inside the function generator. The white noise level is significant ($-76.5$ $\mathrm{dBrad}^2/\mathrm{Hz}$), so that only the limitation of the cross-correlation is measured and not limited by other noise sources, such as common-mode and reference noises. For reference signals, another signal generator (Keysight E4425B) is used to generate 80 MHz, +10 dBm signal. The signal is divided into two with a power splitter, the two signals are used as references, and are input to Ch. B and Ch. D simultaneously. The phase noise of the reference signal is lower than $-130$ $\mathrm{dBrad}^2/\mathrm{Hz}$, which is small enough phase noise for the suppression up to 50 dB. The two phase differences are once recorded on a personal computer, and the cross-correlation is computed by software.

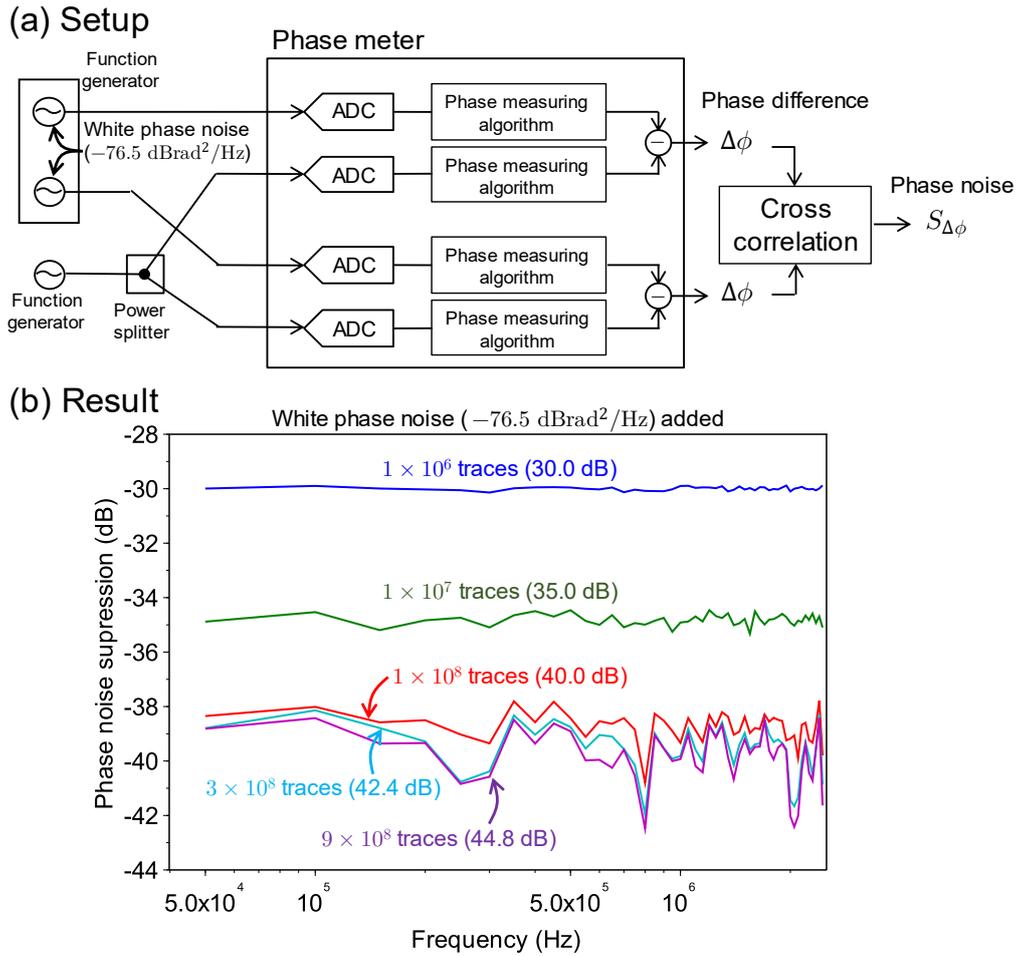

Figure S43. Measurement of limitation of cross-correlation. (a) Measurement setup, (b) Result for phase noise suppression.

In Figure S43(b), the phase noise suppression, which is the difference between the results with and without the cross-correlation is plotted. For less than $1 \times 10^7$ traces, corresponding to the suppression of 35.0 dB, the results correlate well with the theoretical estimation. However, for data of more than $1 \times 10^8$ traces, suppression is limited to approximately –39 dB, which could be from crosstalk between the two channels; crosstalk of $Y$ dB limits the cross-correlation suppression by $Y/2$ dB. Note that this limitation could also come from an unknown correlation between the two outputs in the function generator. Regarding measurement time, the suppression of 39 dB is enough for almost all measurement situations because the 39-dB suppression corresponds to $9 \times 10^7$ traces. When the frequency resolution is $5 \times 10^4$ Hz, which is as in Figure S43(b), the measurement time is as long as 1800 s.

Crosstalk between four input channels of the phase meter can be a fundamental limitation of the cross-correlation. When crosstalk between input Ch. A and Ch. C is $Y_{AC}$ dB, the Ch. C – Ch. D phase difference has an undesired correlation with the Ch. A – Ch. B phase difference by $Y_{AC}/2$ dB. The origin of the crosstalk includes electromagnetic interference in the cable, frond-end electronics, and ADCs. For the real hardware, the crosstalk is measured as Figure S44. We input an 80 MHz, +7 dBm (~–2.2 dBFS) signal into one channel of the phase meter, and measure the signal power in the four channels. In this measurement, the phase-measuring algorithm is not used. The data from the ADCs are transferred to the computer and we apply FFT analysis. Crosstalk between adjacent channels, Ch. A to Ch. B, is approximately –80 dB, whereas crosstalk between nonadjacent channels, Ch. A to Ch. D, is approximately between –110 dB and –120 dB because a common ADC (Analog devices AD9680) is used to convert the adjacent channels. From the measured crosstalk between channels, the suppression should reach –55 dB, surpassing the observed suppression of approximately –39 dB in Figure S43(b). This inconsistency indicates that the phase meter has additional undesired correlations between channels. Above all, the –55-dB suppression is unrealistic because the correlation of $10^{11}$ traces is needed, leading to unfeasibly long measurement time.

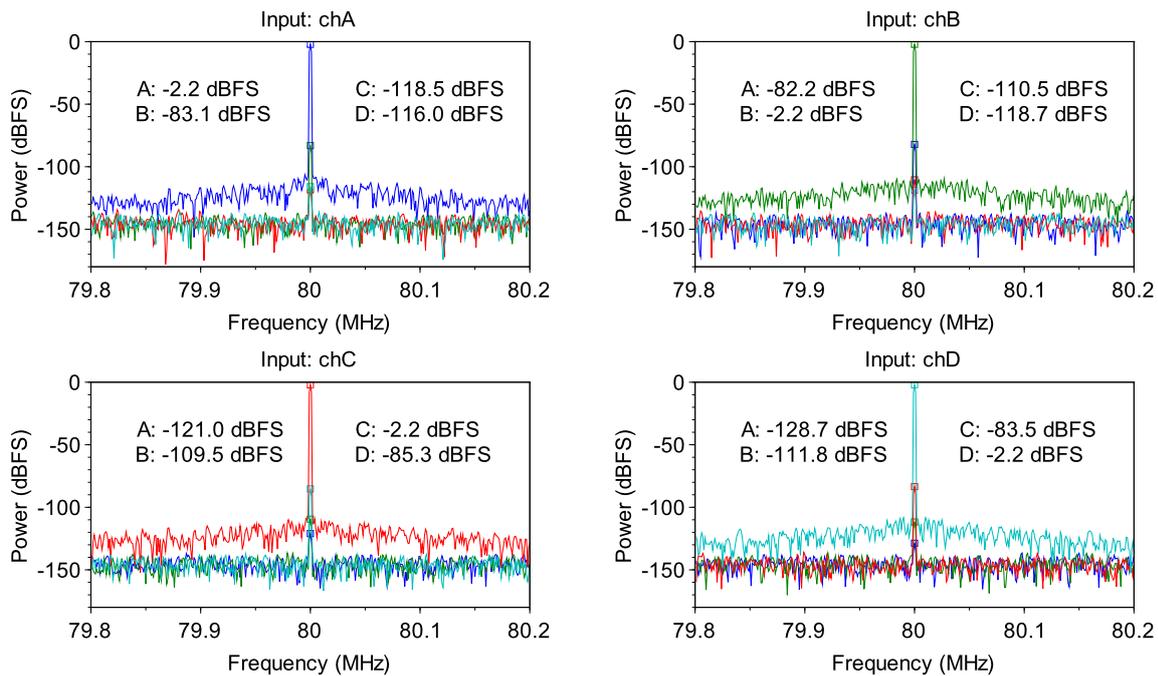

Figure S44. Measurement of crosstalk. Power spectrum in the unit of dB in full scale (dBFS) is shown for each input.

Despite high suppression, any common-mode noise in the two phase differences cannot be suppressed by cross-correlation. We measure the level of common-mode noise in the phase difference by using the setup shown in Figure S45(a). Two function generators (Keysight 33622A and Keysight E4425B) are used to generate test signals (80 MHz, +10 dBm), and the test signals are divided into two, which are two inputs for phase measurement. The phase differences of Ch. A–Ch. B and Ch. C–Ch. D are ideally uncorrelated; however, somehow, the two phase differences correlate. Then, we take the cross-correlation of the two phase differences and evaluate whether the results converge to some level or not, showing the level of common-mode noise.

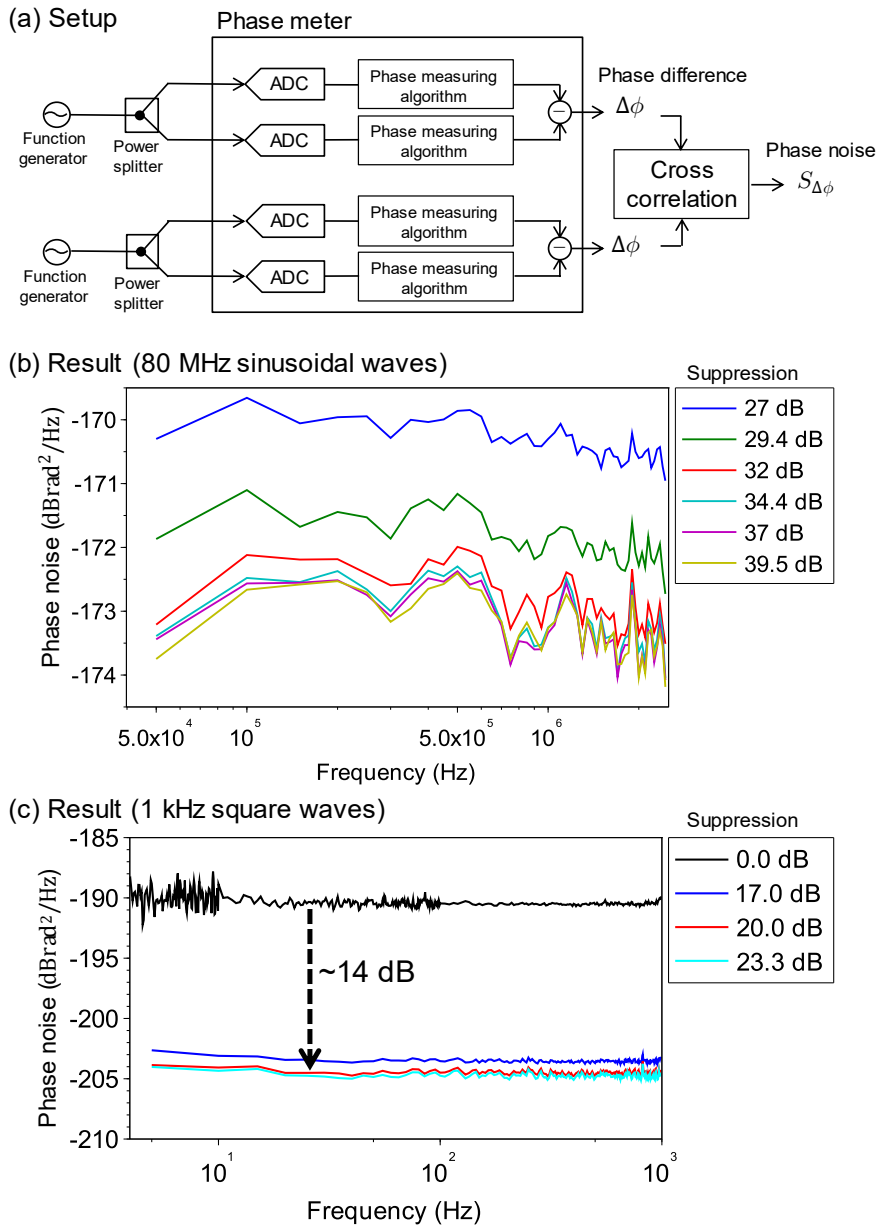

Figure S45. Measurement for common-mode phase difference. (a) Measurement setup (b) Result for 80 MHz sinusoidal waves (c) Result for 1 kHz square waves

Figure S45(b) shows that the suppression limit is approximately −173 dBrad$^2$/Hz. Moreover, the shape of the spectra does not show complete flat spectra, implying two possibilities. One is that the common-mode noise has a structure. The other scenario is that there is some cross-correlation collapse, resulting from undesired anti-correlation components in the two phase differences. Note that we must be careful about cross-correlation collapse because anti-correlation could easily occur in the phase meter. For example, crosstalk from Ch. A to Ch. D results in anti-correlation in the two phase differences.

Notably, the cross-correlation of square waves is possible; no other instruments can do such a measurement. We demonstrate it by performing the measurement shown in Figure S45(c). We use the test signal from the function generators and divide them by two using a T-type adapter (just because availability of two resistive-type power splitters). The signal level is +12 dBm so that the power at the input is near a maximum of +9.5 dBm. Without cross-correlation, the flat phase noise is $-190$ $\mathrm{dBrad}^2/\mathrm{Hz}$ up to 1 kHz (black curve in Figure S45(c)). When we activate cross-correlation, the phase noise is suppressed by 14 dB. Though we have not yet identified why the suppression is limited to such low suppression, we achieve the lowest level of phase noise of $-204$ $\mathrm{dBrad}^2/\mathrm{Hz}$. Note that in this measurement, a 20.0 dB suppression (10000 traces) requires a relatively long measurement time of 2000 s.

In summary, we confirmed the following about the cross-correlation technique: (i) The suppression limit is at least 39 dB when the common-mode phase difference is sufficiently small. (ii) The common-mode phase difference, which also limits the phase noise level of cross-correlation, is approximately $-173$ $\mathrm{dBrad}^2/\mathrm{Hz}$ for 80-MHz sinusoidal input signals and $-204$ $\mathrm{dBrad}^2/\mathrm{Hz}$ for 1-kHz square input signals.

### V.C. Expanding the limits of the input frequency range

The input frequency range for the phase meter has limitations. Specifically, the input frequency is limited to 1/4 of the sampling frequency at ADC to detect all zero-crossings properly. Moreover, the input frequency range has a lower limit of ~1 MHz when the input signal is sinusoidal because of the insufficient slew rate (see Subsection V.A.1). In this subsection, we introduce techniques to expand the limits using preprocess of the input signals.

#### 1. Expansion of the upper frequency limit

The upper-frequency limit, which is a quarter of the sampling frequency of ADC, is a disadvantage of the phase measurement algorithm because the sampling frequency of ADC cannot be easily increased; the higher the sampling frequency, the higher the cost. We overcome the upper-frequency limit by extending the input frequency range of the phase meter by converting the input frequency down (Figure S46).

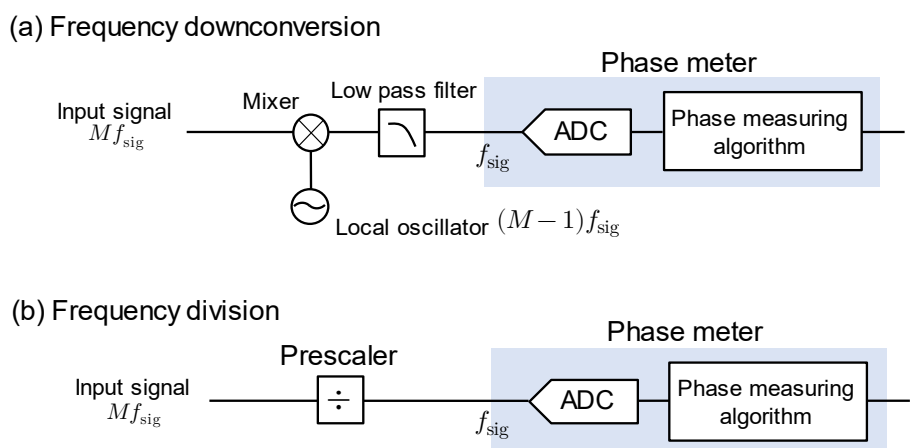

Figure S46. Methods to expand upper limit of input frequency range. (a) Frequency downconversion technique, (b) Frequency division technique.

One technique to down the frequency is heterodyne down-conversion just in front of the ADC. This schematic of a single channel phase is found in Figure S46(a). Here, let the input signal frequency be $Mf_{\text{sig}}$. Here, $M$ is a number representing the down-conversion factor. The input signal is mixed with a local oscillator with a frequency of $f_{\text{LO}} = (M-1)f_{\text{sig}}$. The frequency can also be $f_{\text{LO}} = (M+1)f_{\text{sig}}$. The mixed-signal has a beat-note of $f_{\text{sig}}$. To avoid the harmonics of $f_{\text{sig}}$, the beat-note is filtered with an LPF. The resolution of the phase is maintained though this down-conversion process, i.e., when the phase difference of the two input signals $Mf_{\text{sig}}$ differs by $\delta\phi$, the phase difference between two beat-notes also differs by $\delta\phi$. In exchange for maintaining the resolution, the range of the input frequency is the same as the setup without down-conversion. The input signal frequency should follow Equation (S149).

$$f_{\text{LO}} + f_{\min} < Mf_{\text{sig}} < f_{\text{LO}} + \frac{f_{\text{ADC}}}{4}, \tag{S149}$$

where $f_{\text{LO}}$ shows the frequency of the local oscillator, and $f_{\min}$ is the lower limit of the input frequency.

The other way to decrease the signal frequency is frequency division before the input of digitizers (Figure S46(b)). A prescaler (or frequency divider) is used for this method, which is the same as high-frequency inputs in ordinary frequency counters. The setup is simpler than that for frequency down-conversion. In this technique, the input signal of $Mf_{\text{sig}}$ is directly divided into $f_{\text{sig}}$ with the pre-scaler. The resolution and frequency range for the input signal is different from the case in frequency down-conversion. The resolution is decreased to 1/M and the noise level is increased by $M$, because the phase is also divided by $M$. However, the frequency range for the input signal increases $M$-fold compared to the frequency down-conversion, as shown in Equation (S150).

$$Mf_{\min} < Mf_{\text{sig}} < M\frac{f_{\text{ADC}}}{4}, \tag{S150}$$

Note that the noise added by the pre-scaler might also limit the noise level.

We measured the performances of the two techniques as presented in Figure S47. The performances of frequency down-conversion are measured as the setups shown in Figure S47(a). The test signal (950 MHz, +20 dBm) is made by a vector signal generator (Rohde & Schwarz SMJ100A). The test signal is divided into two with a power splitter (Mini-circuits ZFSC-2-372-S+) and mixed with local oscillator signals. For the local oscillator, a signal generator (Stanford Research Systems SG386, 6 GHz) is used. The power of the signal generator is +13 dBm and frequency is 869.7 MHz. Two frequency mixers (Mini-circuits ZEM-4300+, 300 MHz – 4300 MHz, level 7) are used to down the frequency of the singal. Then, the two mixed-down signals are filtered with LPFs (Mini-circuits BLP-150+) with a cutoff frequency of 150 MHz. The signal at the input of the ADCs has a frequency of 80.3 MHz and power of –3 dBm (–13 dBFS). The signal generator for the test signal, local oscillator, and phase meter are all locked into a common 10-MHz reference clock.

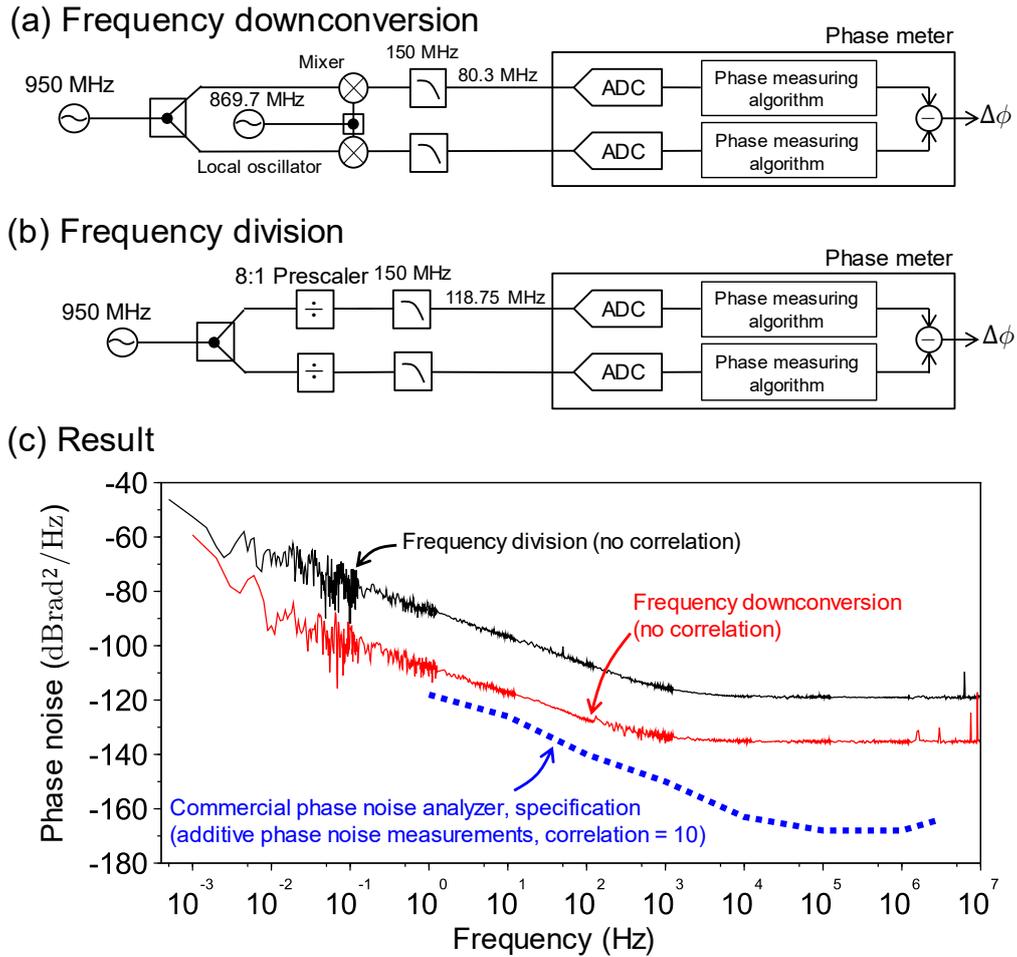

Figure S47. Noise level measurement for the input signal frequency of 950 MHz. (a) Setup for frequency down-conversion (b) Setup for frequency division (c) Results for the two setups. Specification of commercial phase noise analyzer is shown on the dotted line.

Figure S47(b) shows the frequency division test setup. The test signal is generated by the same signal generator (Rohde & Schwarz SMJ100A), and the frequency is the same (950 MHz), but the power is +7 dBm. After being split by the same power splitter (Mini-circuits ZFSC-2-372-S+), the signals are sent to an 8:1 pre-scaler (Analog devices HMC434, 0.2 GHz – 8 GHz). The same LPFs (Mini-circuits BLP-150+) are applied to filter out higher frequency components from the output of the pre-scalers. At the ADC inputs, the signal frequency is 118.75 MHz, and power –2 dBm (–12 dBFS). The obtained phase difference from the phase meter is multiplied by 8 (+18 dB in the power spectrum) to convert the results to those at the input signal of 950 MHz.

Figure S47(c) shows the obtained phase noise spectra for the input signal of 950 MHz. In the frequency conversion technique (red curve), the noise level is as low as –110 $\text{dBrad}^2/\text{Hz}$ at 1 Hz, which is within 10 dB of the specification for additive phase noise measurement of the commercial phase noise analyzer[7] (blue dotted curve) for 1-GHz inputs. Note that the specification is with the cross-correlation of 10 traces, which is equivalent to the suppression of 5 dB. A white noise level above ~1 kHz for the result is not good (~ –135 $\text{dBrad}^2/\text{Hz}$) because of insufficient power at the ADCs. A low-noise power amplifier before ADC inputs would help to reduce the white noise level up to ~10 dB. Moreover, with the cross-correlation technique with $10^5$ traces, the white noise level could be down to –170 $\mathrm{dBrad}^2/\mathrm{Hz}$ at the offset frequency of 100 kHz–1 MHz. Such noise level is comparable to the commercial phase noise analyzer.

By using the frequency division, the noise level is worse in exchange for the wide input frequency range, which is suitable for the signal with large frequency drift. The datasheet shows that the additional phase noise from the pre-scaler is –150 dBc/Hz for 4-GHz input signals. We expect that the noise level for 1-GHz input signals is lower, and therefore, the additional phase noise from the pre-scaler does not worsen the measurement noise.

### *2. Expansion of lower frequency limit*

The phase meter has another limit on the input signal frequency, namely the lower frequency limit in the case of sinusoidal input signals because the phase measuring algorithm does not work well, i.e., results might have cycle slips when the slew rate of the input signal is not high enough. Bandpass filtering to reduce noise in the input signal could help to decrease the rate of the cycle slips. However, even if the input signal is noise-free, there are still limits from ADC noises.

To overcome this issue, a comparator can be used, increasing the slew rate at the zero-crossings of the input signal (Figure S48). The advantage of this technique is that the white noise level calculated from Equation (S118) can be improved because the effective power of the signal is significantly increased after the comparator. Moreover, the increased slew rate is useful to avoid cycle slips because the cycle slip rate depends sharply on the slew rate. However, there are drawbacks. First, the noise of the comparator could limit the performance. Specifically, long-term fluctuation of the threshold in the comparator could worsen low-frequency phase noise. Other noises, such as interference from the power supply, should be minimized to reduce noises. Additionally, the input signal should have enough low noise to avoid multiple triggers at the comparator. Some filtering can be used.

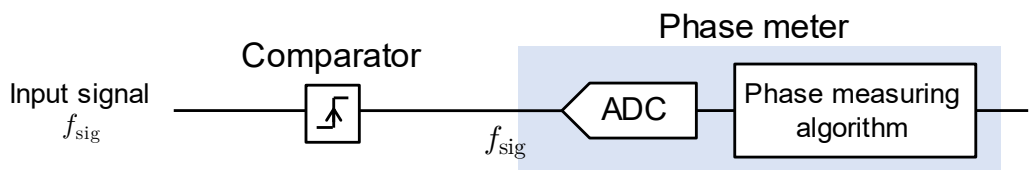

Figure S48. Methods to expand lower limit of input frequency range; increasing slew rate of input signals with comparator

We conducted a performance test for the comparator technique (Figure S49(a)). A test signal of 427.44 kHz is made from a function generator (Keysight 33622A), and the signal power is +10 dBm. The frequency is selected based on other experiments; if the signal frequency is another frequency, e.g., only 400 kHz, there will be no difference. The input signal is applied with LPF (2.5 MHz), divided into two, and sent to comparators. We used a low phase noise logic converter (Analog devices LTC6957-4) as a comparator. The datasheet[8] shows that the additive phase noise for the 100-MHz input signal is –158 $\mathrm{dBrad}^2/\mathrm{Hz}$ above 10 kHz. The output from the comparator is CMOS logic level (0 V–3.3 V) with a rise time of 320 ps. We fit the signal to the input of the phase meter by applying a –5 dB attenuator and 100-MHz LPF.

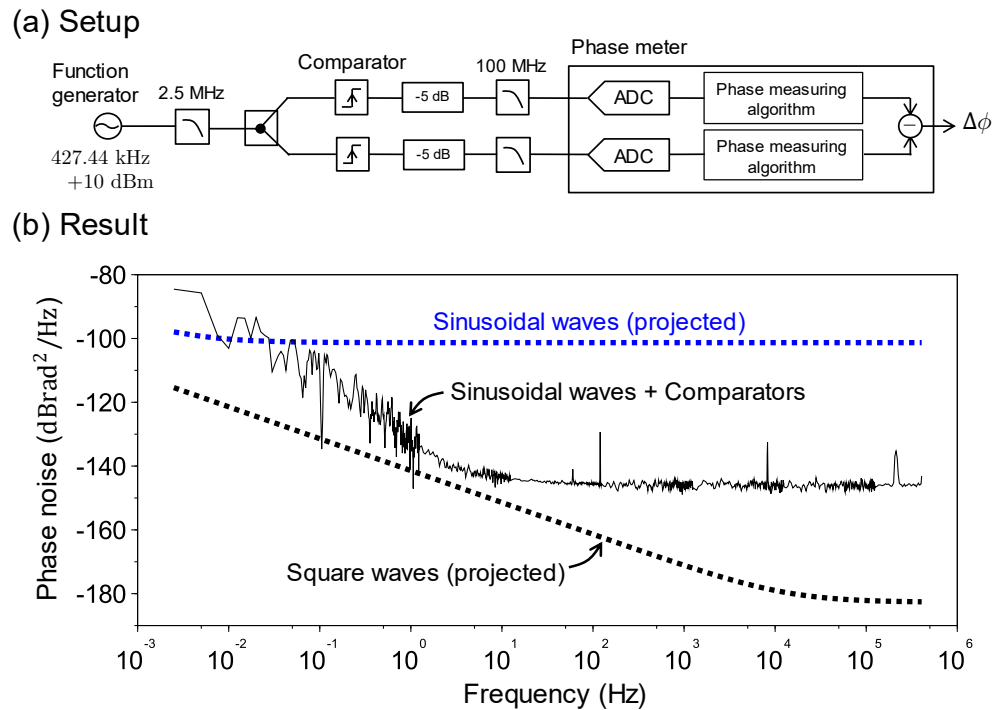

Figure S49. Measurement of noise level for 427.44 kHz input signal using comparators. (a) Setup (b) Result are shown in thin black line

The result is plotted as the black curve in Figure S49(b). For comparison, projected noise levels according to the noise analysis presented in Section III are also plotted as a broken black line for the 427.44-kHz square waves and broken blue line for the 427.44-kHz sinusoidal waves. Above several Hertz, the noise level is limited by white noise of approximately –145 $\mathrm{dBrad}^2/\mathrm{Hz}$. Compared to the datasheet[8], this result has larger noise than expected from the ($\sqrt{2}$-times of) additive noise of the comparator, because the white noise comes from the noise of the input signal, not from the comparator itself. In other words, 430 kHz of the input signal frequency is too low for the comparator, i.e., a 2.5-MHz LPF just after the function generator is not enough.

As for long-term performance, the noise curve shows approximately –20 dB/dec, i.e., a white frequency curve instead of –10 dB/dec (flicker phase) below approximately 3 Hz. We suspect that the temperature fluctuation of both comparators causes such low-frequency noise. A stabilized environment could help to improve performance. Moreover, the performance for the cycle slip is significantly improved. In this test, we observed no cycle slip for more than 1000 s in the measurement because of the increase in the slew rate. On the other hand, a 400-kHz, +7-dBm sinusoidal signal induces approximately 100 cycle slips per second (Figure S34).

### V.D. Eliminating zero-crossing interpolation error

We briefly introduce a TDC-based architecture to implement the phase measuring algorithm without ADCs in an attempt to avoid ZI error. When the input signal is sinusoidal with low-drift and low measurement bandwidth, the only error component that remains in the phase measuring algorithm is ZI error, as shown in Table S4. ZI error (Equation (S24)) essentially comes from sampling, i.e., the coupling of zero-crossing interpolation and nonlinearity of the input signal around zero-crossings. To overcome this issue, a TDC and comparator can be used instead of an ADC. A TDC measures the elapsed time between the zero-crossing timing and the next sampling time, which is expressed by the left side of Equation (S24), without interpolation error. Therefore, the result is free from ZI error induced by nonlinearity. The block diagram of the architecture is presented in Figure S50.

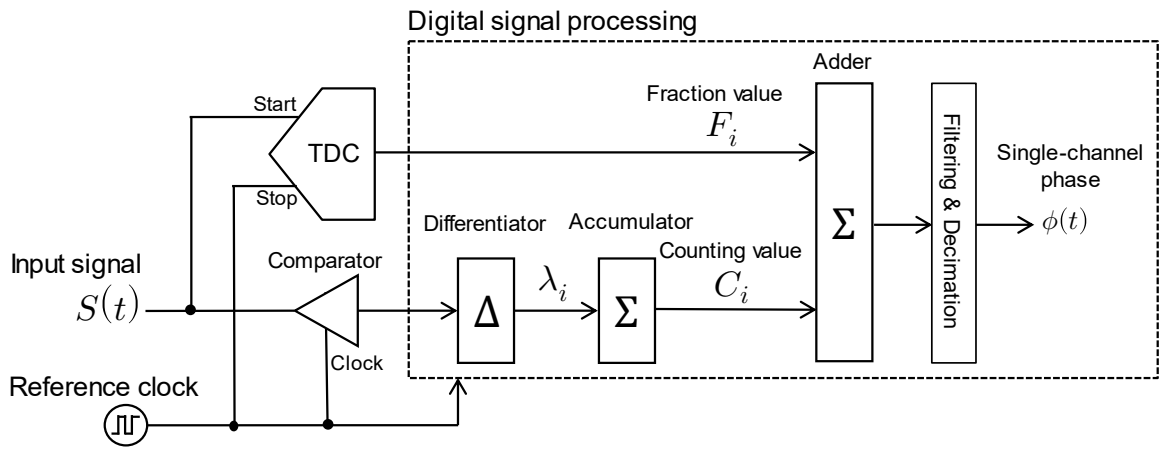

Figure S50. Proposed time-to-digital converter (TDC) architecture.

The TDC architecture works as follows. The input signal is fed into the start signal of a TDC. The stop signal of the TDC is the reference clock signal. Therefore, digital output from the TDC represents the elapsed time between the zero-crossings of the input signal and a clock edge representing the next sampling time. The obtained digital data from the TDC is the fraction value in the phase measuring algorithm. At the same time, the input signal is also input in the comparator, which effectively acts as a 1-bit ADC. The comparator is driven by the same reference clock. The output of the comparator is obtained at the frequency of the reference clock. The output of the TDC and comparator are sent to the digital signal processing unit. The output of the comparator is converted to the zero-crossing detection function, $\lambda_i$, through a differentiator. After accumulating it, the counting value, $C_i$, is obtained. Then, we obtain the two values needed for the phase measuring algorithm, $F_i$ and $C_i$. Finally, single-channel phase measuring can be achieved with an adder, appropriate filtering, and decimation processes afterwards.

The advantage is that the TDC only needs to measure the time interval up to one clock cycle; this improves the requirement for the capture range of the TDC because it means the TDC does not need to exhibit a large measurement range. A drawback is that the TDC is generally sensitive to high-frequency noise in the input signal. Implementing a low-pass filter to limit the bandwidth before the TDC can help resolve this issue. Another point of drawback is that the characteristics of the TDC and the comparator must be matched. Otherwise, there is critical miscounting when the zero-crossing of the input signal occurs with nearly the same timing as that of the edges of the reference clock. Additionally, the phase measurement may contain an error due to the nonlinear response of the TDC, especially around the time difference is around zero.

### V.E. Other possible improvements

Here we present ideas for further improvements of the phase meter briefly.

1. Parallel operation with conventional I/Q demodulation

As stated in Section III, this algorithm has an intrinsic measurement error. For example, there is a DC phase error from zero-crossing interpolation. Moreover, the phase noise has large peaks from errors in the high-offset frequencies (Figure S21 and Figure S22). To overcome this issue, a conventional phase measuring algorithm (digital I/Q demodulation) could be operated in parallel with this phase measurement algorithm. Combining the two results, the measurement error in the phase measuring algorithm can be found. Moreover, this parallel operation enables us amplitude measurement, which is impossible with the phase-measuring algorithm.

2. Digital filtering after ADC to improve the white phase noise level

One way to further improve the white noise level is to use digital LPF between the ADC and the phase-measuring algorithm. The low-pass filter effectively collects the information around a certain data point and reduce the noise, making the waveform smoother and keeping it away from the noise. This technique effectively reduces the degradation factor at the same input signal frequency. The other way is to reduce the sampling rate of the ADC, because the lower the sampling frequency, the higher the SNR. For example, a market-available ADC with a 105-MHz sampling rate has an SNR of 82 dBFS, which is better than the ADC that we used by more than 15 dB. Consequently, the white noise level can be improved with a low cost, in exchange for the limited input frequency range.

3. Managing cycle slips

We consider several ways to avoid the cycle slips in high slew rate and low SNR signals, because measurement of low-SNR signal is desired in some optics experiments. One is the bandwidth limitation, such as digital bandpass filtering after ADC, which is shown effective (subsection VI.A). Another way is post-processing; when a cycle slip occurs, the loss or appearance of zero-crossing can be detected by monitoring the time difference between adjacent zero-crossings and consequently, can be corrected. Note that, generally, cycle slips cannot be detected with an ordinary frequency counter. For low slew rate signals, hold-off could also be effective to avoid cycle slips.

4. Measuring pulses as an input signal

When pulses, the duty cycle that is not 50 %, is measured, the measurement algorithm still works; however, the noise property is different from those of square waves. The change of the duty cycle contributes to a rapid phase error like the QA error, causing higher frequency components in the phase measurement results. This can be removed when appropriate LPF is applied to extract low-frequency components. Another way to deal with the signal without a 50-% duty cycle is to use only the rising (or falling) edges for the phase measurement algorithm.

In this case, the white phase noise level increases by 3 dB because the degradation factor increases by the same, corresponding that the amount of data used for calculation becomes half.

5. Multiple thresholds for enhancement of resolution

Resolution can be enhanced with several thresholds instead of only zero for zero-crossings. As shown in the derivation of the principle, zero-crossings can be generalized to different thresholds. In that case, the effective data rate increases by $M$-times if we use $M$ thresholds instead of only two zero-crossings in one signal cycle. Then, the white noise level becomes $10 \log M$ (dB) lower. However, there is a drawback; the measurement becomes vulnerable to signal properties other than phase, including amplitude fluctuations, nonlinearity and signal shape.